%% file: main.tex
\RequirePackage{fix-cm}
\documentclass[smallextended]{svjour3}       
\smartqed  

\usepackage{textcomp}
\usepackage{graphicx}
\usepackage{xspace}
\usepackage{cite}
\usepackage{amsmath,amssymb,amsfonts}
\usepackage{hyperref}
\usepackage{algorithmic}
\usepackage{graphicx}
\usepackage{textcomp}
\usepackage{pgfkeys} 
\usepackage{listings}
\usepackage{float}
\usepackage{seqsplit}
\usepackage[utf8]{inputenc}
\usepackage[table,xcdraw]{xcolor}
\def\BibTeX{{\rm B\kern-.05em{\sc i\kern-.025em b}\kern-.08em
    T\kern-.1667em\lower.7ex\hbox{E}\kern-.125emX}}

\usepackage{xcolor}   
\usepackage{xspace}   

\usepackage{listings}
\lstdefinestyle{customCallStack}{
  numbers=none,
  language=CallStack,
  commentstyle=\color{gray},
  basicstyle=\ttfamily\footnotesize,
  breaklines=true,
  rulecolor=\color{black},
  captionpos=b,
  belowcaptionskip=\baselineskip,
  moredelim=[is][\color{blue}]{~}{^},
  moredelim=**[is][\color{black}]{^}{^},
  moredelim=**[is][\color{red}]{^}{\$},
  moredelim=**[is][\color{black}]{\$}{)},
}
\def\code{\lstinline}
\usepackage[most]{tcolorbox} 
\usepackage{multicol}
\usepackage{multirow}
\usepackage{tabularx}
\usepackage{booktabs}
\usepackage{pgfplots}
\pgfplotsset{compat=1.18}
\usepackage{subcaption}
\usepackage{listings}
\usepackage[inline]{enumitem}
\usepackage{colortbl}
\usepackage{makecell}
\usepackage{hyperref}
\usepackage[authoryear,round]{natbib}
\setcitestyle{aysep={} }

\lstdefinestyle{diff}{
    escapechar=\%
}

\newcommand{\GHilight}{\makebox[0pt][l]{\color{green}\rule[-4pt]{1\linewidth}{10pt}}}
\newcommand{\RHilight}{\makebox[0pt][l]{\color{pink}\rule[-4pt]{1\linewidth}{10pt}}}

\newcommand{\decl}[2]{%
  \noindent\textbf{#1} #2\par\vspace{0.8em}
}

\newcommand\byamurl{\href{https://github.com/chains-project/bacardi}{https://github.com/chains-project
/bacardi}\xspace}

\newcommand{\toolname}{Byam\xspace}
\newcommand{\bumptotal}{571}
\newcommand{\bumptotalcompile}{243}
\newcommand{\totalcompileper}{43}
\newcommand{\bumpjavaver}{78}
\newcommand{\bumpjwerror}{8}
\newcommand{\finaldata}{103}

\input{file_error_level}
\input{rq1-tab}

\input{rq1_pgfkeys}
\input{rq3-values}
\input{rq3-tab2}

\begin{document}

\title{\toolname: Fixing Breaking Dependency Updates with Large Language Models}


\author{Frank Reyes \and
        May Mahmoud \and
        Federico Bono \and
        Sarah Nadi \and
        Benoit Baudry \and
        Martin Monperrus
       }


\institute{F. Reyes, KTH Royal Institute of Technology, Stockholm, Sweden \at\email{frankrg@kth.se} \and
           M. Mahmoud, New York University Abu Dhabi, United Arab Emirates \at\email{m.mamhoud@nyu.edu}\and
           F. Bono, KTH Royal Institute of Technology, Stockholm, Sweden \at\email{fbono@kth.se}\and
           S. Nadi, New York University Abu Dhabi, United Arab Emirates \at\email{sarah.nadi@nyu.edu}\and
           B. Baudry, Université de Montréal, Canada \at\email{benoit.baudry@umontreal.ca}\and
           M. Monperrus, KTH Royal Institute of Technology, Stockholm, Sweden \at\email{monperrus@kth.se}\and
}

\date{}


\maketitle
\vspace{-2cm}
\begin{abstract}
Application Programming Interfaces (APIs) facilitate the integration of third-party dependencies within the code of client applications. However, changes to an API, such as deprecation, modification of parameter names or types, or complete replacement with a new API, can break existing client code. These changes are called \textit{breaking dependency updates}; It is often tedious for API users to identify the cause of these breaks and update their code accordingly.
In this paper, we explore the use of Large Language Models (LLMs) to automate client code updates in response to breaking dependency updates. We evaluate our approach on the BUMP dataset, a benchmark for breaking dependency updates in Java projects. Our approach leverages LLMs with advanced prompts, including information from the build process and from the breaking dependency analysis.
We assess effectiveness at three granularity levels: at the build level, the file level, and the individual compilation error level.
We experiment with five LLMs: Google Gemini-2.0 Flash, OpenAI GPT4o-mini, OpenAI o3-mini, Alibaba Qwen2.5-32b-instruct, and DeepSeek V3. 
Our results show that LLMs can automatically repair breaking updates. 
Among the considered models, OpenAI's o3-mini is the best, able to completely fix \pgfkeysvalueof{o3_P_8_BUILD_SUCCESS_percent}\% of the builds when using prompts that include contextual information such as the erroneous line, API differences, error messages, and step-by-step reasoning instructions. 
Also, it fixes 78\% of the individual compilation errors.
Overall, our findings demonstrate the potential for LLMs to fix compilation errors due to breaking dependency updates, supporting developers in their efforts to stay up-to-date with changes in their dependencies. 

\end{abstract}

\section{Introduction}
\label{sec:intro}
Modern software development heavily relies on third-party software libraries to improve developers' productivity and time-to-market.
The set of libraries used by a client project is referred to as the \textit{dependencies} of that project.
Developers access a library's functionality through its provided \textit{Application Programming Interfaces} (APIs).
These APIs evolve over time to introduce new functionality, fix bugs, or address security vulnerabilities~\citep{apievolve}. 
While such evolution is necessary, it often comes at a cost: updates may introduce breaking changes that prevent client projects from compiling or running ~\citep{apievolve,bump,breakinggood}.
Failures resulting from API updates are referred to as \textit{breaking dependency updates}.
When a new version of a dependency results in a breaking dependency update, the developers using the library (i.e., \textit{the client developers}) have to figure out how to update their code to fix the errors they face.
This is known to be a tedious and time consuming maintenance task~\citep{semanticvers, breakingbad}.

To address this problem, we propose \toolname, a pipeline that leverages Large Language Models (LLMs) to update client code in response to breaking dependency updates \citep{vaswani2017attention, chang2024survey}.
We focus on breaking dependency updates that lead to compilation errors, which are the most common types of causes for breakage \citep{bump,breakinggood}.
\toolname analyzes compilation errors caused by a dependency update, formulates prompts that incorporate relevant build and dependency information, and queries an LLM to generate updated code.
We assess the effectiveness of \toolname using the BUMP dataset ~\citep{bump}, which provides reproducible cases of breaking dependency updates in Java projects.
Our experiments systematically evaluate different LLMs, prompt design strategies, and information granularity levels.

We evaluate \toolname with five different LLMs: Google Gemini-2.0 Flash, OpenAI GPT4o-mini, OpenAI o3-mini, Alibaba Qwen2.5-32b-instruct, and DeepSeek V3. We assess \toolname's effectiveness at three levels of granularity: build, file, and individual error.
At the build level, we measure the number of fully fixed builds, those still failing due to compilation errors, and cases where compilation was resolved but test failures emerged.
At the file level, we track the number of files with compilation errors before and after the fixes, identifying which files were successfully fixed and which ones still include compilation errors.
Finally, at the individual error level, we analyze the number of original compilation errors per file, comparing them before and after the LLM-generated fixes to determine fixed, unresolved, and newly introduced errors.
To quantify the impact of the corrections generated by LLMs, we define a set of metrics to evaluate the performance of LLMs.
We analyze the capability of the models to fully resolve the build process, file-level success rate, and performance in resolving individual errors. 
Since LLMs can also introduce new errors in the process, we measure the relative error fixed as the proportion of previously existing errors that were successfully resolved, while accounting for any new errors introduced after the code update.

\begin{figure}[ht!]
    \centering
        \vspace{0.5em}
     \begin{subfigure}{\columnwidth}
        \centering
        {\footnotesize\sffamily
            \begin{minipage}{\linewidth}
                \includegraphics[width=.95\textwidth]{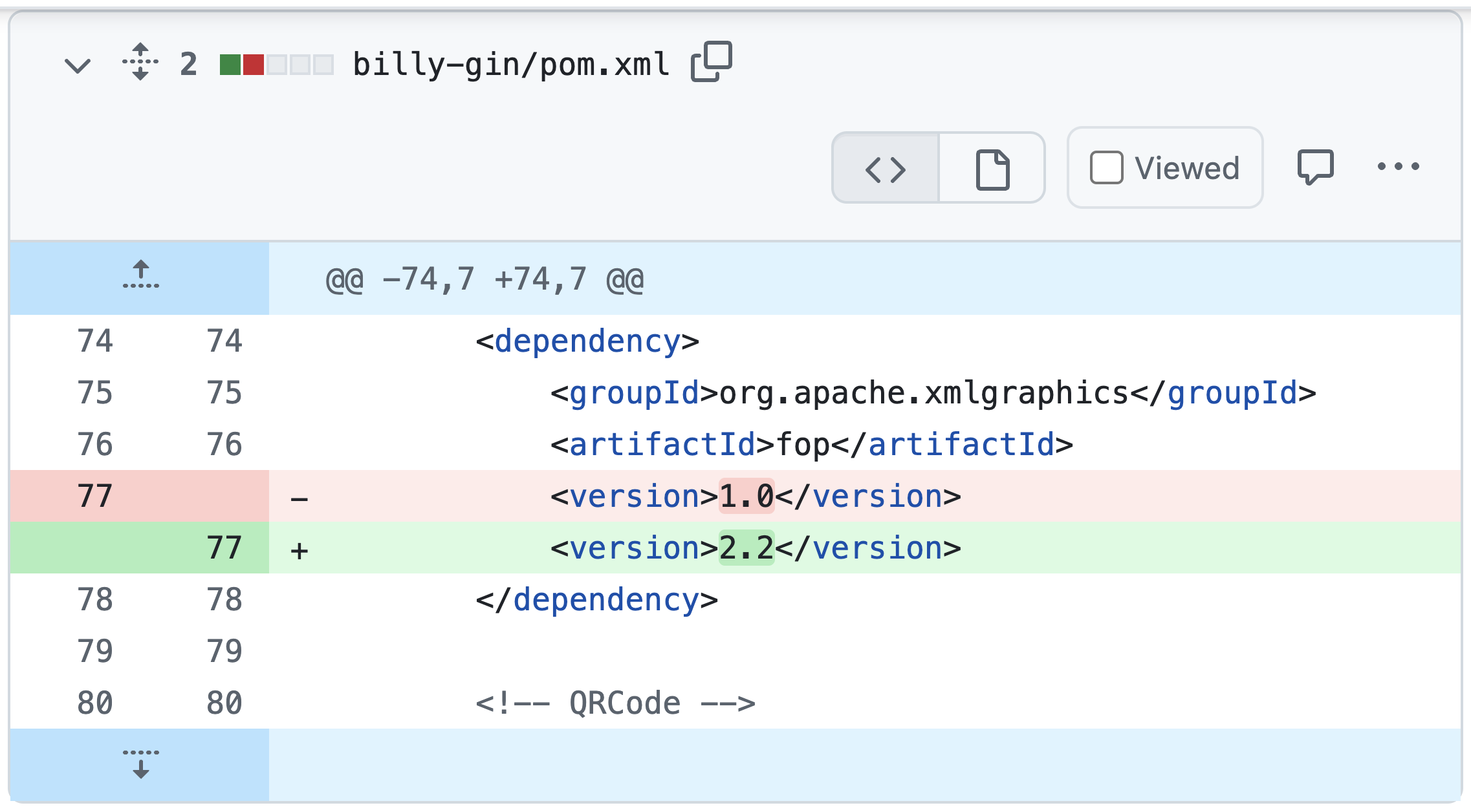}
            \end{minipage}
        }
        \caption{Difference in Maven build file, causing a breaking update.}
        \label{fig:pom_diff}
    \end{subfigure}
    \vspace{0.5em}
    
    \begin{subfigure}{\columnwidth}
        \centering
        \begin{lstlisting}[
            firstnumber=113,
            numbers=left,         
            numberstyle=\tiny\color{gray},
            numbersep=5pt,         
            frame=leftline,        
            xleftmargin=12pt       
        ]
...
// create an instance of fop factory
FopFactory fopFactory = FopFactory.newInstance();
// a user agent is needed for transformation
FOUserAgent foUserAgent = fopFactory.newFOUserAgent();
...
\end{lstlisting}
        \label{fig:buggyLine_example}
        \caption{The broken code with compilation failure after breaking update.}
    \end{subfigure}

  \vspace{0.7em}
    \begin{subfigure}{\columnwidth}
        \centering
        {\footnotesize\sffamily
            \begin{minipage}{\linewidth}
                \begin{verbatim}
[ERROR] /billy/billy-gin/src/main/java/com/premiumminds/billy/gin/services/
impl/pdf/FOPPDFTransformer.java:[115,43] no suitable method found for
newInstance(no arguments)
  method org.apache.fop.apps.FopFactory.newInstance(org.apache.fop.apps
  .FopFactoryConfig) is not applicable
   (actual and formal argument lists differ in length)
  method org.apache.fop.apps.FopFactory.newInstance(java.io.File) is 
   not applicable (actual and formal argument lists differ in length)
  method org.apache.fop.apps.FopFactory.newInstance(java.net.URI) is
   not applicable (actual and formal argument lists differ in length)
  method org.apache.fop.apps.FopFactory.newInstance(java.net.URI,java.
  io.InputStream) is not applicable (actual and formal argument lists 
  differ in length)
                \end{verbatim}
            \end{minipage}
        }
        \caption{Compilation error information from the logs.}
        \label{fig:log_info}
    \end{subfigure}
    \caption{Example of a breaking dependency update when updating \texttt{org.apache.xmlgraphics} (dependency) from version 1.0 (old version) to 2.2 (new version) in the project \texttt{billy}.}
    \label{fig:breakingexample}
\end{figure}

Our results show that LLM can automatically repair breaking updates. OpenAI's o3-mini model achieves the best performance with a build success rate of  \pgfkeysvalueof{o3_P_8_BUILD_SUCCESS_percent}\% when using the best prompt that integrates erroneous line, information regarding the differences between the dependency versions (APIDiff), and structured reasoning techniques  (CoT).
At the file level o3-mini fixes 41\% of the originally erroneous files in unsuccessful builds and
at the error level, o3-mini fixes 79\% of compilation failures.
Overall, our experimental results demonstrate the potential for LLMs to reduce developer effort and time in managing breaking dependency updates, helping developers more effectively stay up-to-date with changes in their dependencies.

To summarize, our contributions in this paper are as follows:
\begin{enumerate}
    \item An approach, called \toolname, for fixing compilation errors from breaking dependency updates with LLMs.
    Its advanced prompting strategies relies on a comprehensive analysis of the breaking update problem space.  In particular, we use API differencing of the breaking library and a Chains-of-thought reasoning command dedicated to breaking updates.

    \item A large scale experiment with five different LLMs and eight prompts, over \finaldata~breaking dependency updates. Our results show that o3-mini achieves the best outcomes when using a prompt that highlights the erroneous line causing the compilation error(s), that provides information about the API differences between the old and updated dependency versions, and that includes Chains-of-thought instructions. 
    
    \item We provide a full replication package that ensures the reproducibility of our study and fosters future research on this topic. The replication package is available at \byamurl 
    \end{enumerate}

\section{Background}
\label{sec:background}

Breaking dependency updates are common in software projects~\citep{brito2018and}, and often discourage developers from upgrading their dependencies \citep{dependencywild,breakingbad,venturini2023depended}.
However, keeping dependencies up to date is crucial for security, as outdated libraries may contain vulnerabilities ~\citep{brito2018and,xavier2017we,brito2020you,larios2020selecting,salza2018developers}.
A breaking dependency update occurs when a library update introduces incompatibilities that cause client code to fail.
This can occur either when developers manually update dependencies or when they use automated tools such as Dependabot or Renovate, which submit pull requests for updates.
Such breakages are often changes in method signatures, deprecations, or class removals can cascade into compilation errors across the client project.
While breaking updates can also trigger test failures, dependency resolution issues or lock conflicts~\citep{bump}, in this paper we focus specifically on compilation errors, which are the most prevalent form~\citep{breakinggood}.

Figure~\ref{fig:breakingexample} shows an example of a breaking dependency update in project \code{billy}\footnote{\href{https://github.com/premium-minds/billy/pull/300}{https://github.com/premium-minds/billy/pull/300}}, where the \code{org.apache.xmlgraphics} dependency was updated from version 1.0 to 2.2 in commit \code{36859167815292f279e570d39dd2ddbcf1622dc6}.
The \code{org.apache.xmlgraphics} library has a method \code{newInstance()}, which has zero parameters in version 1.0 and was changed in the new version 2.0.
This led to a compilation error in \code{billy}, as shown at the bottom of the figure.

Prior work has extensively studied the prevalence and causes of breaking changes and methods to identify them in library code~\citep{brito2018and,xavier2017we,brito2020you,brito2018apidiff,mujahid2020using,breakinggood}.
Migration support has also been explored through rule-based transformations and recommendation systems~\citep{xing2007api,dagenais2009semdiff,brito2018apidiff}.
However, such approaches typically rely on predefined rules, mined examples, or static analysis heuristics, which limit their ability to handle the diversity of real-world breaking updates and the project-specific context required to repair client code.
In contrast, we investigate whether large language models (LLMs) can directly generate client-side repairs when provided with appropriate context.

To formalize our scope, we adopt standard terminology.

\begin{definition}
    \label{depupdate}
    A \textbf{dependency update} is a change made in a build specification file where the version of a specific dependency is updated to a new version. In the experimentation of this paper, we focus on the Maven build file (\texttt{pom.xml}) in Java.
\end{definition}

\begin{definition}
    \label{breakdepupdate}
    A \textbf{breaking dependency update} occurs when updating a dependency version introduces incompatibilities that cause the build to fail.  
\end{definition}

For systematic evaluation in this study, we examine breaking dependency updates as represented in the BUMP dataset \cite{bump}: a pair of commits for a project, consisting of a pre-breaking commit with a passing build and a breaking commit that updates the version of a single dependency, causing the build to fail.

\begin{definition}
    \label{pre-breakingcommit}
    A \textbf{pre-breaking commit} is the commit before the dependency update, where the project is built successfully.
 \end{definition}

\begin{definition}
    \label{breakingcommit}
    A \textbf{breaking commit} is the project commit where the only change is an update to a dependency version that leads to the build failing.
\end{definition}

\section{\toolname: An Approach to Repairing Breaking Updates}
\label{sec:approach}

Existing solutions for breaking dependency updates have limited scalability and adaptability.
They require predefined rules or abundant historical data, which makes them weak in the face of diverse or novel API changes.
Large language models (LLMs) offer a promising alternative due to their ability to generalize across different contexts and reason about unknown patterns.
To be effective in automating the repair of breaking dependency updates, LLMs must be guided with rich prompts.

We propose a novel approach, \toolname, based on generating a code patch with LLMs to fix breaking dependency updates.
\toolname examines error messages, locates the erroneous line, and consults API differences before querying an LLM and applying a fix.
\toolname enables LLMs to operate as context-aware repair assistants, making them more efficient and accurate at addressing breaking changes than prior approaches.

\subsection{Overview}
\begin{figure}[t!]
    \centering
    \includegraphics[width=1\textwidth, trim=0 10 0 10, clip]{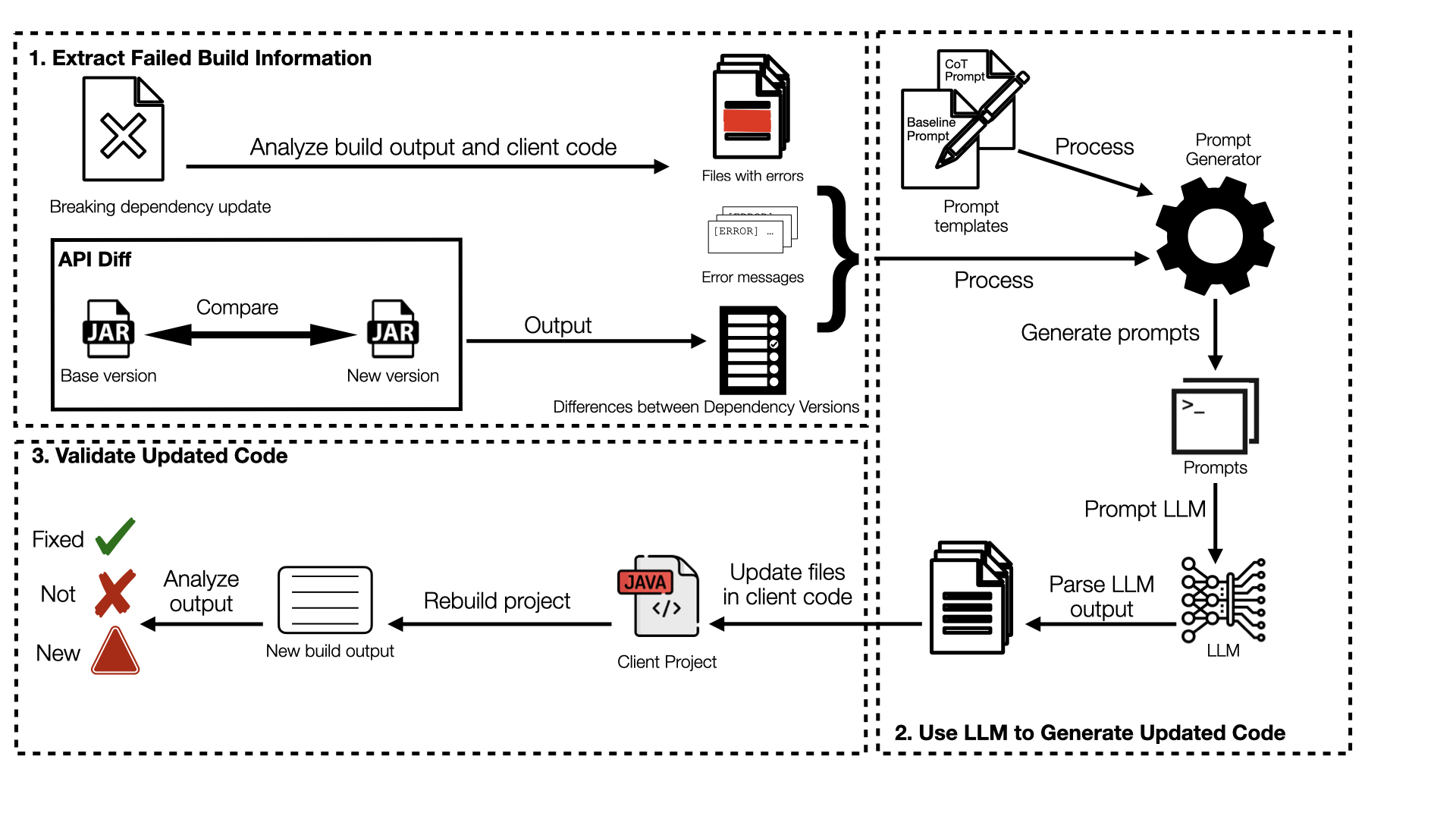}
    \caption{Overview of our the \toolname pipeline using LLMs to fix breaking dependency updates.} 
    \label{fig:process}
\end{figure}

Figure \ref{fig:process} shows an overview of our approach, which we refer to as \textit{\toolname}.
The pipeline consists of three steps.
The first step extracts build information regarding the breaking dependency update, including the error-causing code and the compiler error message(s). It also provides information regarding the differences between the two versions of the dependency. 
The second step focuses on generating a fixed version of the client code using an LLM. We engineer different prompts, which include different types of contextual information, such as API differences, error messages, and affected code lines; we discuss this further below. \toolname~ prompts the LLM, augmented with the extracted build and the previously mentioned contextual details, to generate a fix to address the compilation failure.
The third and final step focuses on validating the generated fix by integrating it into the client code and re-running the build.

We detail each stage of the pipeline in the following subsections.

\subsection{Step 1: Extract Failed Build Information}
This step isolates the parts of the code that are causing a compilation failure after the dependency update and identifies the API differences involved in the failure.
\toolname uses this information to provide the LLM with various contextual data.
To collect this information, \toolname analyzes the build output log after a breaking update.
The pipeline automatically extracts a list of files causing the compilation error(s) and the list of error messages for each file.
Additionally, \toolname analyzes the changes between both versions of the dependency. 

Figure \ref{fig:log_info} shows an excerpt of a build output.
Analyzing this output, \toolname determines that the file \code{/billy/billy-gin/src/main/java/com/premiumminds/billy/gin/services/impl/pdf/FOPPDFTransformer.java} is causing a compilation error.
\toolname also extracts the error message, \textit{``no suitable method found for
\code{newInstance(no arguments)}"}, as well as the error location, at line 115, column 43. 
From the error location, \toolname identifies the specific constructs responsible for the compilation failure.
In the previous example, the error is caused by a method \texttt{newInstanc()}.
Next, \toolname extracts the API differences between the dependency versions to determine whether any changes in the new version are related to the construct that triggers the error.
We term this output the API Difference (APIDiff).
For this example, the APIDiff shows that the signature of the method \texttt{newInstance()}has been changed in the new version 2.0 of the dependency.
\autoref{fig:api_diff_extract} shows an excerpt of the APIDiff corresponding to the example in \autoref{fig:breakingexample}, highlighting the change associated with the construct identified as the root cause of the breaking dependency update.
This APIDiff information is later incorporated into the LLM prompts, as discussed in \autoref{sec:use_llm}.
\begin{figure}[t]
\centering
\begin{lstlisting}[basicstyle=\ttfamily\small, breaklines=true, numbers=none, linewidth=\columnwidth, keywordstyle={}, commentstyle={}, stringstyle={}]
---! REMOVED METHOD: PUBLIC(-) STATIC(-) org.apache.fop.apps.FopFactory newInstance()
	+++  NEW METHOD: PUBLIC(+) STATIC(+) org.apache.fop.apps.FopFactory newInstance(org.apache.fop.apps.FopFactoryConfig)
	+++  NEW METHOD: PUBLIC(+) STATIC(+) org.apache.fop.apps.FopFactory newInstance(java.io.File)
		+++  NEW EXCEPTION: org.xml.sax.SAXException
		+++  NEW EXCEPTION: java.io.IOException
	+++  NEW METHOD: PUBLIC(+) STATIC(+) org.apache.fop.apps.FopFactory newInstance(java.net.URI)
	+++  NEW METHOD: PUBLIC(+) STATIC(+) org.apache.fop.apps.FopFactory newInstance(java.net.URI, java.io.InputStream)
		+++  NEW EXCEPTION: org.xml.sax.SAXException
		+++  NEW EXCEPTION: java.io.IOException
\end{lstlisting}
\caption{API Difference used in \toolname. It captures the relation between the construct that triggers the error and the changes in the new version of the dependency in \autoref{fig:breakingexample}}
 \label{fig:api_diff_extract}
\end{figure}

At the end of this first step, \toolname~has detailed information about: the list of files causing compilation errors; the list of compilation errors for each file, and APIDiff that triggers the error. 

In the next step, we use this output to engineer several prompts for an LLM to generate updated code.

\subsection{Step 2: Use LLM to Generate Repairs}
\label{sec:use_llm}
In this step, we use an LLM to attempt to fix the files with compilation errors.
For each file causing compilation errors, we construct a prompt based on a prompt template. 
Once a prompt is created, we use it to query the LLM to generate fixed code for one buggy file. We iterate over each file in the list to completely fix the project.
To improve LLM performance, we explore different prompt design variations that enrich the context provided to the model.
\autoref{tab:prompt_design} provides a list of the different prompt design decisions we consider. 

\begin{table}[t]
    \centering
    \caption{Prompt Design Space for Fixing Breaking Updates}
    \label{tab:prompt_design}
    \rowcolors{2}{gray!10}{white}
    \begin{tabular}{ll}
        \hline
        \textbf{Variation} & \textbf{Description} \\
        \hline
        \textbf{Baseline Prompt} & \makecell[l]{Includes client code and error message but \\ excludes additional context.} \\
        \textbf{Erroneous Line Inclusion} & \makecell[l]{Adds the specific line of code causing the \\ compilation error.} \\
        \textbf{API Differences (API Diff)} & \makecell[l]{Includes details of API differences between \\ dependency versions.} \\
        \textbf{Chain of Thought (CoT) Prompting} & \makecell[l]{Guides LLM reasoning by incorporating \\ structured reasoning steps.} \\
        \hline
    \end{tabular}
\end{table}

\textbf{Baseline prompt.} Our baseline prompt includes the client code causing the error and the compilation error message.
The client code provides the Java class where the failure occurs.
The error message, extracted from the build log file, helps pinpoint the exact issue.

\begin{figure}[t!]
    \centering
    \begin{tcolorbox}[boxrule=1pt, left=2pt, right=2pt, top=2pt, bottom=2pt, width=\columnwidth]
    \footnotesize  
Act as an Automatic Program Repair (APR) tool, reply only with code, without explanation.\\
You are specialized in breaking dependency updates, in which the failure is caused by an external dependency.\\
To solve the failure you can only work on the client code.\\\\

the following client code fails:\\
```java\\
\texttt{\textless client\_code\textgreater}\\
```\\
\\\\
with the following error message:\\
\texttt{\textless error\_message\textgreater}\\

\begin{itemize}
 \item[-] Propose a patch that can be applied to the code to fix the issue.
 \item[-] Return only a complete and compilable class in a fenced code block.
 \item[-] You CANNOT change the function signature of any method but may create variables if it simplifies the code.
 \item[-] You CAN remove the @Override annotation IF AND ONLY IF the method no longer overrides a method in the updated dependency version.
 \item[-] If fixing the issue requires addressing missing imports, ensure the correct package or class is used in accordance with the newer dependency version.
 \item[-] Avoid removing any existing code unless it directly causes a compilation or functionality error.
 \item[-] Return only the fixed class, ensuring it fully compiles and adheres to these constraints.
\end{itemize}
\end{tcolorbox}
    \caption{Baseline Prompt}
    \label{fig:base_prompt}
\end{figure}

Figure \ref{fig:base_prompt} shows our baseline prompt template, where the placeholders for the different information, with the entire Java class that triggers the compilation error (\textit{client\_code}), the error message extracted from the build log file (\textit{error\_message}).

Based on the baseline, we evaluate specific combinations of prompt design choices rather than generating all possible variations, focusing on erroneous lines, API version differences (\textit{APIDiff}), and Chain of Thought (CoT) reasoning.
Instead of an exhaustive exploration, we systematically test selected configurations that provide meaningful information on their impact on LLM performance.
Our approach considers the incremental addition of each factor to a baseline prompt, ensuring a balanced assessment of each factor's contributions.
This approach examines two key dimensions: the prompting strategy (zero-shot vs. CoT) and the type of contextual information included in the prompt (error information, erroneous line, APIDiff).
This structured evaluation allows us to isolate the effects of each component while maintaining a manageable number of experimental setups. 

\textbf{Providing Erroneous Line Information.} 
Developers, when debugging, typically start by identifying the exact line of code causing the issue, as this helps narrow down the scope of the problem and focus efforts on the relevant section of the code.
Including the erroneous line in the prompt replicates this natural debugging process, potentially improving code understanding for the LLM (the compiler error message does not include the actual line by default).
We include this information by extending the base prompt with the actual line of the client code responsible for the compilation error, as shown in~\autoref{fig:buggy_line}, after the error message.

\begin{figure}[t!]
    \centering
    \begin{tcolorbox}[boxrule=1pt, left=2pt, right=2pt, top=2pt, bottom=2pt, width=\columnwidth]
    \footnotesize  
 the error is triggered in the following specific lines in the previous code:
\texttt{\textless erroneous\_line\textgreater}\\
\end{tcolorbox}
    \caption{Adding Erroneous Line to the Prompt. This helps the LLM to identify and focus on the problem to be fixed.}\vspace{-0.2cm}
    \label{fig:buggy_line}
\end{figure}

\textbf{Providing API Difference Information (\textit{APIDiff})}
When a dependency update breaks a project, developers often compare the previous and new versions of the API to identify relevant changes, such as method renaming or modifications to the parameters~\citep{brito2018apidiff}.
Such changes help developers determine how to update their client code to fix the problem.
Following this practice, our intuition is that by explicitly providing these API differences (APIDiff) to the LLM, we can allow it to reason about the likely source of the error and suggest a more accurate solution.

To extract these differences, we rely on \textit{japicmp} \citep{Japicmp-baseJapicmp}, a widely used tool to detect API changes in Java libraries.
\textit{japicmp} performs a comprehensive structural comparison, allowing it to detect modifications across public, internal, and even deprecated parts of an API.
This tool identifies modifications such as changes in method signatures, parameter types, deprecated methods, or new functionality.
For example, when we use \texttt{japicmp} to analyze the changes between versions \texttt{6.18.1} and \texttt{6.19.1} of the \texttt{jasperreports} dependency, it detects that the \texttt{setLinewith} method was removed from the new version of the dependency. 
We augment the prompt with that information, as in the example shown in \autoref{fig:api_diff_prompt}.
\autoref{fig:apidiffexample} shows the concrete example of APIDiff for \texttt{jasperreports}.

\begin{figure}[t!]
    \centering
    \begin{tcolorbox}[boxrule=1pt, left=2pt, right=2pt, top=2pt, bottom=2pt, width=\columnwidth]
    \footnotesize  
 The error is caused by a change in the API of the dependency. The new library version includes the following changes:
\texttt{\textless api\_diff\textgreater}\\
\end{tcolorbox}
    \caption{Adding API Differences to the Prompt. It helps to the LLM to identify possible replacements to missing constructs.}\vspace{-0.5cm}
    \label{fig:api_diff_prompt}
\end{figure}

\begin{figure}[tp]
    \centering
    {\footnotesize\sffamily
    \begin{minipage}{\columnwidth}
    \begin{lstlisting}[basicstyle=\ttfamily\small, breaklines=true, numbers=none, frame=none, linewidth=\columnwidth, keywordstyle={}, commentstyle={}, stringstyle={}]
The error is caused by a change in the API of the dependency.
The new library version includes the following changes:  
- Method net.sf.jasperreports.engine.JRPen.setLineWidth(float) 
  has been removed in the new version of the dependency.  
- Method net.sf.jasperreports.engine.base.JRBasePen.setLineWidth(float) 
  has been removed in the new version of the dependency.
    \end{lstlisting}
    \end{minipage}
    }
    \caption{APIDiff example included in the prompt to repair the breaking update of the dependency \texttt{jasperreports} from version \texttt{1.18.1} to version \texttt{1.19.1} in project biapi}
    \label{fig:apidiffexample}
\end{figure}

\textbf{Applying Chain of Thought (CoT) Prompting.}
Developers often solve complex errors by following a structured step-by-step reasoning process.
This structured approach helps them to analyze the problem, identify possible causes and systematically find a solution. This structured approach is replicated in prompting in what is referred to as \textit{Chain of Thought (CoT)} prompting \citep{cotmainpaper}.
Previous work has shown that CoT instructions improve performance in code generation and automated repair tasks~\citep{le2019automated,monperrus2018automatic}.
Consequently, we consider CoT prompting for fixing breaking dependency updates.
In our approach, CoT prompting explicitly guides the LLM to analyze the error message, identify the affected code, infer the underlying issue, and reason about possible fixes before proposing a corrected version.
This approach guides the LLM to produce a more structured and coherent response \citep{cotmainpaper}. 
We incorporate the CoT technique in the prompt as shown in Figure \ref{fig:base_anthropic}.

\begin{figure}[t!]
    \centering
    \begin{tcolorbox}[boxrule=1pt, left=2pt, right=2pt, top=2pt, bottom=2pt, width=\columnwidth]
    \footnotesize 
Act as an Automatic Program Repair (APR) tool, reply only with code, without explanation.\\
You are specialized in breaking dependency updates, in which the failure is caused by an external dependency.\\
To solve the failure you can only work on the client code.\\\\

the following client code fails:\\
```java\\
\texttt{\textless client\_code\textgreater}\\
```\\
the error is triggered in the following specific lines in the previous code:\\
\texttt{\textless erroneous\_line\textgreater}\\

with the following error message:\\
\texttt{\textless error\_message\textgreater}\\

 The error is caused by a change in the API of the dependency. The new library version includes the following changes:\\
\texttt{\textless api\_diff\textgreater}\\

Before proposing a fix, please analyze the situation and plan your approach within \texttt{\textless repair\_strategy \textgreater} tags:\\

\begin{itemize}
    \item Identify the specific API changes that are causing the failure in the client code.
    \item Compare the old and new API versions, noting any changes in method signatures, return types, or parameter lists.
    \item Determine which parts of the client code need to be updated to accommodate these API changes. 
    \item Consider any constraints or requirements for the fix (e.g., not changing function signatures, potential import adjustments).
    \item Plan the minimal set of changes needed to fix the issue while keeping the code functional and compliant with the new API.
    \item Consider potential side effects of the proposed changes on other parts of the code.
    \item Ensure that the planned changes will result in a complete and compilable class.
    \item If applicable, note any additional imports that may be needed due to the API changes.
\end{itemize}

\begin{itemize}
 \item[-] Propose a patch that can be applied to the code to fix the issue.
 \item[-] Return only a complete and compilable class in a fenced code block.
 \item[-] You CANNOT change the function signature of any method but may create variables if it simplifies the code.
 \item[-] You CAN remove the @Override annotation IF AND ONLY IF the method no longer overrides a method in the updated dependency version.
 \item[-] If fixing the issue requires addressing missing imports, ensure the correct package or class is used in accordance with the newer dependency version.
 \item[-] Avoid removing any existing code unless it directly causes a compilation or functionality error.
 \item[-] Return only the fixed class, ensuring it fully compiles and adheres to these constraints.
\end{itemize}
\end{tcolorbox}
    \caption{Chain of Thought Prompt for Breaking Update. LLMs benefit from detailed Chain of Thought instructions, esp. for reasoning tasks, such as bug fixing.}\vspace{-0.5cm}
    \label{fig:base_anthropic}
\end{figure}

\begin{table}[ht]
\centering
\caption{The 8 Studied Prompt Configurations in our Experiments.} 
\rowcolors{2}{gray!10}{white}
\label{tab:exhaustive_prompt_list}
 \resizebox{\textwidth}{!}{%
\begin{tabular}{lcccccc}
\toprule
\textbf{\makecell[l]{Prompt\\ID}} & \textbf{\makecell[c]{Prompt\\Name}} & \textbf{\makecell[c]{Client\\Code\\}} & \textbf{\makecell[c]{Error\\Message}} & \textbf{\makecell[c]{Erroneous\\Line}} & \textbf{APIDiff} & \textbf{\makecell[c]{CoT\\Prompting}} \\ \midrule
$P_1$ & Baseline Prompt & \checkmark & \checkmark & & & \\ 
$P_2$ & Erroneous Line & \checkmark & \checkmark & \checkmark & & \\
$P_3$ & APIDiff & \checkmark & \checkmark & &  \checkmark & \\ 
$P_4$ & Erroneous Line + APIDiff & \checkmark & \checkmark & \checkmark & \checkmark & \\ 
$P_5$ & CoT Prompt & \checkmark & \checkmark & & & \checkmark \\ 
$P_6$ & CoT + Errouneous Line & \checkmark & \checkmark & \checkmark & & \checkmark \\ 
$P_7$ & CoT + API Diff& \checkmark & \checkmark & &\checkmark & \checkmark \\ 
$P_8$ & CoT + Erroneous Line + APIDiff & \checkmark & \checkmark & \checkmark & \checkmark & \checkmark \\ \bottomrule
\end{tabular}}
\end{table}

To summarize, Table \ref{tab:exhaustive_prompt_list} provides an overview of all the prompt configurations we experiment with. The table outlines eight distinct prompt variations, detailing the information included in each prompt and whether we utilize the Chain of Thought (CoT) prompting technique. 
Our experiments will study the performance of each prompt (see \autoref{sec:eval}).

\subsection{Step 3: Update Code, Rebuild Project and Analyze the Build Outcome}
In this stage, we replace the original files in the client project, with the LLM-generated files.
Then, we rebuild the project.
We use the default build command of the project, typically \texttt{mvn test} command to build the project and run all existing tests.
If the build is successful, the updates to the code successfully fixed the breaking dependency update.
If not, we identify which files/errors are fixed, which are not fixed, and which new errors might have appeared.
We note the build where the compilation failure is fixed but results in a different failure category, such as a test failure.

\begin{figure*}[t!]
\begin{lstlisting}[backgroundcolor=\color{gray!10}, style=diff,language=Java,belowskip=\baselineskip]
%\RHilight%-        // create an instance of fop factory
%\RHilight%-        FopFactory fopFactory = FopFactory.newInstance();
%\GHilight%+        // create an instance of fop factory with a base URI
%\GHilight%+        FopFactory fopFactory = FopFactory.newInstance(new File(".").toURI());
         // a user agent is needed for transformation
         FOUserAgent foUserAgent = fopFactory.newFOUserAgent();
\end{lstlisting}
\caption{LLM-generated code for fixing the code example from \autoref{fig:breakingexample}}. \label{fig:fixingexample1}
\end{figure*}

\subsection{Example of a Repair}
 \toolname is able to fully fix the breaking dependency update we introduced in \autoref{fig:breakingexample}, where the code on line 3 causes a build breakage. When we process this case using \toolname~with o3-mini, the LLM generates the code in Figure \ref{fig:fixingexample1}.
We can see that \toolname~successfully updated the method signature, fixing the break in the project build.

\section{Experimental Methodology}\label{sec:eval}

In this section, we describe our research questions, the data and LLMs we used in our study.

\subsection{Research Questions}
\label{researchQuestions}

In this section, we present our three research questions designed to evaluate complementary aspects of \toolname's effectiveness to repair compilation errors caused by breaking dependency updates.


\newcommand\rqbuildsuccess{\textbf{(Repair Success): }
How effective is \toolname~in fixing compilation errors due to breaking dependency updates?}

\newcommand\rqFileError{\textbf{(Partial Repair):} To what extent does \toolname~partially fix compilation errors at the file and error level?\xspace}

\newcommand\rqIntroducedErrors{\textbf{(New Error Introduction):}  To what extent do language models tend to introduce new errors during the breaking update repair process?\xspace}

\begin{enumerate}[label=\textbf{RQ\arabic*}, ref=RQ\arabic*]

    \item \label{rq: exploration}{\rqbuildsuccess}
    
    This question aims to measure \toolname's capability to fully repair projects with breaking dependency updates by fixing all files and errors to reach a build success. This metric indicates how effective \toolname is as a fully automated solution without developer intervention.
    
    \item \label{rq: goldenPrompt}{\rqFileError} 

    This question evaluates the extent to which \toolname can repair projects partially, even when the build is not fully fixed.
    While RQ1 focuses on complete build repair, there are cases where \toolname cannot resolve all errors but still fixes a substantial portion of them.
    Such partial repair is valuable in practice, as it reduces the number of remaining errors and allows developers to concentrate their effort on a smaller scope.
    In this RQ, we therefore measure partial repair at both the file level and the individual error level.
    
    \item \label{rq: introducedErrors}{\rqIntroducedErrors} 
    
    In this question, we address the balance between fixing existing errors caused by the breaking dependency update and introducing new ones when applying patches generated by \toolname.
    Minimizing the introduction of new errors with generated fixes is essential for the practical adoption of any automated repair tool, as newly introduced errors will increase debugging effort and affect the developers' confidence in the automated system.

\end{enumerate}

\subsection{Study Subjects}

Our experiments are based on the real-world dependency updates collected in previous research \cite{bump}. We use BUMP, a benchmark for reproducible breaking dependency updates.
BUMP includes \text{\bumptotal} breaking dependency updates, of which \text{\bumptotalcompile} (\text{\totalcompileper\%}), are broken builds due to compilation failure.

BUMP classifies compilation failures into four distinct categories: \textit{Direct compilation errors}, \textit{Indirect compilation errors}, \textit{Java version incompatibility} and \textit{Werror failure} \citep{breakinggood}.
We discard the \text{\bumpjavaver} failures classified under the \textit{Java version incompatibility} category.
Java versioning failures are failures that require a different Java version than the one used in the project. Resolving Java versioning failures requires changing the Java version, rather than updating the code.
BUMP also contains \text{\bumpjwerror} compilation errors of type \textit{Werror failure}, which is a failure due to activating the \code{failOnWarning} option in the configuration file after the dependency update.
Since this type of failure relates to the linter configuration rather than the code, we disregard these data points.
We discard cases in which, when resolving dependencies, the new version of the dependency conflicts with the dependencies declared in the Maven configuration file, causing errors when invoking API calls~\citep{bono2024javaclasshijacksoftwaresupplychain}.
These failures appear in the \textit{Direct compilation errors} and \textit{Indirect compilation errors} categories.
Such cases are related to classpath issues, they are impossible to solve by updating the Java code.
This leaves us with a total of \text{\finaldata} breaking dependency updates for our evaluation. 

\begin{figure}[h!]
    \centering
    \includegraphics[width=1\linewidth]{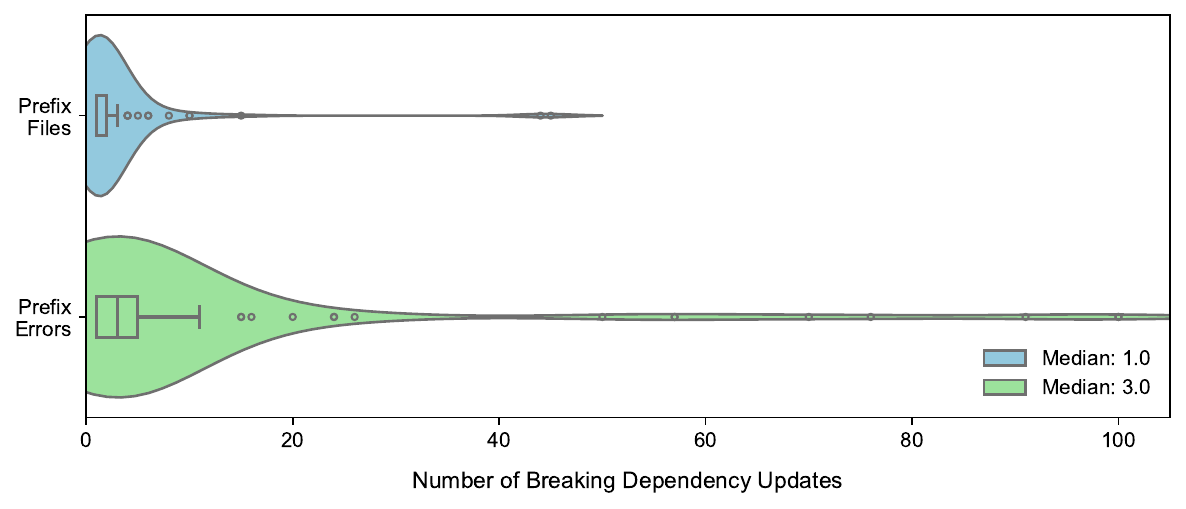}
    \caption{Distribution of initial error files and initial compilation errors on the \finaldata~breaking dependency updates. Most breakages consist of one file, and there are 3 errors to fix on average.}
     \label{fig:file_error_distribution}
\end{figure}

\autoref{fig:file_error_distribution} shows the distribution of the initial files with errors and the initial errors over the \text{\finaldata} breaking updates we consider.
The median number of initial files with errors per breaking updates is \text{1} and the median number of initial errors is \text{3}.
The figure shows that the distribution is right-skewed.
Specifically, \text{70\%} of the updates affect only \text{1} file.
The highest number of initial files with errors is found in the project \texttt{billy} when updating the dependency \texttt{jaxb2-basics-runtime} from version \texttt{0.13.1} to version \texttt{1.11.1}, affects \text{45} files.
The \text{75\%} of the \text{\finaldata} breaking updates result in between \text{1} and \text{5} compiler errors.
For example, in project \texttt{ChangeSkin}, where the update of dependency \texttt{spongeapi} from version \texttt{7.4.0} to \texttt{8.0.0} causes a compilation failure, resulting in errors in \text{15} different files and a total of \text{91} compilation errors.
In general, the more the errors, the harder it is to fix a breaking update. 

\subsection{Experimental Protocol}

\subsubsection{Protocol for RQ1}

This research question aims to investigate how effective is \toolname~in fixing build failures that are caused by breaking dependency updates. 
We run \toolname~with the prompt configurations of \autoref{tab:exhaustive_prompt_list} on the .\text{\finaldata} breaking dependency updates of our experiments.

As success metric for answering RQ1, we define the \textbf{Build Success Rate (BSR)} as the proportion of initially failing projects that successfully build after applying \toolname fixes:
\begin{equation}\label{bsr}
BSR = \frac{N_{\text{fixed\_builds}}}{N_{\text{initially\_failing\_builds}}}
\end{equation}
where \( N_{\text{fixed\_builds}} \) is the number of projects that successfully build after applying fixes, and \( N_{\text{initially\_failing\_builds}} \) is the total number of projects that originally failed to build due to a dependency update.

\subsubsection{Protocol for RQ2}

\toolname may partially fix the build, while leaving some errors unfixed.
We want to know how well it performs when it does not succeed to completely repair the build.
Hence, we consider two additional levels of granularity that are finer than the build success level, and are specifically measured on builds that remain failing after applying the generated fixes: (1) the \textit{file level}, which examines the number of source files that were fixed and no longer contain compilation errors, and (2) the \textit{compilation error level}, which evaluates the fixed compilation errors across the project.
We define the following success metrics to answer RQ2: 

\textbf{File Fix Success Rate (FFSR)} – The percentage of files originally containing compilation errors that were fixed, from failed repairs:
\begin{equation}
\label{ffrs}
FFSR = \frac{N_{\text{fixed\_files}}}{N_{\text{initially\_erroneous\_files}}}
\end{equation}
where \( N_{\text{fixed\_files}} \) is the number of Java files that no longer have compilation errors, and \( N_{\text{initially\_erroneous\_files}} \) is the number of Java files that contained compilation errors before applying fixes.

\textbf{Compilation Error Fix Rate (CEFR)} – The percentage of fixed compilation errors, from failed repairs:
\begin{equation}
CEFR = \frac{N_{\text{fixed\_errors}}}{N_{\text{initial\_errors}}}
\end{equation}
where \( N_{\text{fixed\_errors}} \) is the total number of compilation errors successfully fixed, and \( N_{\text{initial\_errors}} \) is the total number of compilation errors originally present.

We experiment with the different configurations as detailed in Section \ref{sec:use_llm}.

\subsubsection{Protocol for RQ3}

An LLM may be able to fix some errors at the cost of introducing many additional errors.  Those errors are bad for developers: they have to fix them manually, potentially needing more work than the initial breaking update errors.
If the number of introduced errors is too high, automated correction may impose an additional debugging effort, making manual updates more preferable. 
To quantify this trade-off, we define a metric that focuses on the extent to which new errors arise due to LLM-generated patches. 
We define the \textbf{Relative Error Fixed Ratio (REF)} metric in \autoref{eq:saved_effort}.
The metric is a percentage that ranges from 100\% to infinite negative percentages since the best the LLM can do is fix all errors leading to 100\% relative error fixes efforts, but it can introduce any number of new errors leading to an infinite number of negative percentages of relative error fixes effort (i.e., added effort). 

\begin{equation}\label{eq:saved_effort}
    REF = \frac{N_{\text{fixed\_errors}} - N_{\text{new\_errors}}}{N_{\text{initial\_errors}}}
\end{equation}

where \( N_{\text{fixed\_errors}} \) is the number of errors successfully fixed by \toolname, 
\( N_{\text{new\_errors}} \) is the number of new errors introduced after the code updates, and
\( N_{\text{initial\_errors}} \) is the number of errors originally caused by the breaking dependency update.

\subsection{Language Models for Experimentation}\label{llmused}

We select the models we experiment with based on the results of RepairBench \citep{repairbench} and LiveCodeBench~\citep{jain2024livecodebenchholisticcontaminationfree}, which are established evaluation frameworks for LLMs in program repair.
Based on these frameworks, we select the following five LLMs to use in our evaluation:

 \begin{itemize}
     \item Google Gemini-2.0 Flash \footnote{\href{https://ai.google.dev/gemini-api/docs/models/gemini\#gemini-2.0-flash}{https://ai.google.dev/gemini-api/docs/models/gemini\#gemini-2.0-flash}} because it provides a strong balance between speed and accuracy.
     \item OpenAI GPT4o-mini \footnote{\href{https://openai.com/index/gpt-4o-mini-advancing-cost-efficient-intelligence/}{https://openai.com/index/gpt-4o-mini-advancing-cost-efficient-intelligence}} because it offers a lightweight yet powerful alternative to larger models, maintaining high-quality code generation while being more cost-efficient.
     \item OpenAI o3-mini \footnote{\href{https://openai.com/index/openai-o3-mini/}{https://openai.com/index/openai-o3-mini}} because it leads the results in RepairBench at the time of writing.
     \item Alibaba Qwen2.5-32b-instruct
     \footnote{\href{https://huggingface.co/Qwen/Qwen2.5-32B-Instruct}{https://huggingface.co/Qwen/Qwen2.5-32B-Instruct}} because it is the leading open source model in the 32B parameter range at the time of writing in LiveCodeBench.
     \item DeepSeek V3
     \footnote{\href{https://github.com/deepseek-ai/DeepSeek-V3/}{https://github.com/deepseek-ai/DeepSeek-V3}} because it has great cost-efficiency and robust performance.
 \end{itemize}

 \autoref{tab:llm_comparison} shows a comparison of the five models according to different aspects like their provider, number of parameters, and input and output token limits.
Our study deliberately focuses on small and medium-sized, cost-effective models to enable large-scale, reproducible experimentation across many real breaking dependency updates and projects.
To reduce sampling noise, we set the temperature to 0, which minimizes randomness; the remaining limitations of this choice are discussed in \autoref{sec:internal-validity}. This design choice aims to balance scalability, cost, and replicability in our evaluation.

 \begin{table}[t]
    \centering
    \caption{LLMs used in our experiments}
    \resizebox{\columnwidth}{!}{%
    \rowcolors{2}{gray!10}{white}
    \begin{tabular}{@{}lccccc@{}}
      \toprule
        \textbf{Feature} & \textbf{\begin{tabular}[c]{@{}c@{}}Gemini\\2.0 Flash\end{tabular}} & \textbf{\begin{tabular}[c]{@{}c@{}}GPT-4o\\Mini\end{tabular}} & \textbf{\begin{tabular}[c]{@{}c@{}}o3-\\mini\end{tabular}} & \textbf{\begin{tabular}[c]{@{}c@{}}DeepSeek\\V3\end{tabular}} & \textbf{\begin{tabular}[c]{@{}c@{}}Qwen2.5-\\32B-instruct\end{tabular}} \\
        \midrule
        \textbf{\begin{tabular}[c]{@{}l@{}}Model\\ Provider\end{tabular}} & Google & OpenAI & OpenAI & DeepSeek & Alibaba \\
        \textbf{\begin{tabular}[c]{@{}l@{}}Inference\\ Provider\end{tabular}} & Google & OpenAI & OpenAI & OpenRouter & OpenRouter \\
        \textbf{\begin{tabular}[c]{@{}l@{}}Model\\ Open Source\end{tabular}} & No & No & No & Yes & Yes \\
        \textbf{\begin{tabular}[c]{@{}l@{}}Number\\ of Parameters \end{tabular}}& \begin{tabular}[c]{@{}c@{}}Not\\Disclosed\end{tabular} & \begin{tabular}[c]{@{}c@{}}Not\\Disclosed\end{tabular} & \begin{tabular}[c]{@{}c@{}}Not\\Disclosed\end{tabular} & 671B & 32.5B \\
        \textbf{\begin{tabular}[c]{@{}l@{}}Input\\ Token Limit\end{tabular}} & 1,048,576 & 128,000 & 200,000 & 131,000 & 131,000 \\
        \textbf{\begin{tabular}[c]{@{}l@{}}Output\\ Token Limit\end{tabular}} & 8,192 & 16,000 & 100,000 & 131,000 & 131,000 \\
        \textbf{\begin{tabular}[c]{@{}l@{}}Pricing (per\\ 1M tokens)\end{tabular}} & \text{\begin{tabular}[c]{@{}c@{}}In: \$0.10\\ Out: \$0.40\end{tabular}} & \text{\begin{tabular}[c]{@{}c@{}}In: \$0.15\\ Out: \$0.60\end{tabular}} & \text{\begin{tabular}[c]{@{}c@{}}In: \$1.10\\ Out: \$4.40\end{tabular}} & \text{\begin{tabular}[c]{@{}c@{}}In: \$0.90\\ Out: \$0.90\end{tabular}} & \text{\begin{tabular}[c]{@{}c@{}}In: \$0.79\\ Out: \$0.79\end{tabular}} \\
        \textbf{\begin{tabular}[c]{@{}l@{}}Knowledge\\ Cutoff\end{tabular}} & \begin{tabular}[c]{@{}c@{}}August\\2024\end{tabular} 
        & \begin{tabular}[c]{@{}c@{}}October\\2023\end{tabular}
        & \begin{tabular}[c]{@{}c@{}}October\\2023\end{tabular}
        & \begin{tabular}[c]{@{}c@{}}Not\\Disclosed\end{tabular}
        & \begin{tabular}[c]{@{}c@{}}Not\\Disclosed\end{tabular} \\
         \noalign{\smallskip}\hline
    \end{tabular}
    }
    \label{tab:llm_comparison}
\end{table}


\section{Experimental Results}
\label{sec:results}

We now answer our three research questions introduced above.

\subsection{\textbf{RQ1}\label{rq1} \textit{\rqbuildsuccess}}

To answer RQ1, we evaluate the build repair success of \toolname~using the different LLMs listed in Section \ref{llmused} and the different prompt configurations defined in Section \ref{sec:use_llm}. We calculate the \textbf{Build Success Rate(BSR)} as defined in Equation~\ref{bsr}. We summarize the results in Table \ref{tab:build_success_prompt}. 

\buildsuccesstab

\autoref{tab:build_success_prompt} presents the~\textbf{BSR}, indicating the percentage of builds that are successfully repaired by each LLM under different prompt configurations.
Each row of the table represents a different configuration ($P_1$ to $P_8$ shown in \autoref{tab:exhaustive_prompt_list}), while each column corresponds to one of the five evaluated LLMs presented in \autoref{tab:llm_comparison}.
Each cell indicates the corresponding build success rate by \toolname~after applying the generated code, we represent that value in the following format \textit{total number of fixed builds / total number of broken builds (build success rate)}.

The highest success rate in this evaluation is achieved by o3-mini using $P_8$ (CoT + Erroneous Line + API Diff), with \text{ \pgfkeysvalueof{o3_P_8_BUILD_SUCCESS_percent}\%}, meaning \text{\pgfkeysvalueof{o3_P_8_BUILD_SUCCESS}} out of the \finaldata~ original failing builds were completely repaired, with compilation and test execution succeeding. Overall, o3-mini had the highest build success rate across most prompts.
Relatively high performance is also achieved by Deepseek V3 performing second-best in cases $P_1$, $P_5$, $P_6$ and Gemini-2.0-flash performing second-best in  $P_2$, $P_3$, $P_4$, $P_8$

In contrast, Qwen2.5-32b-instruct has the lowest success rate across most prompts. This is because this is the smallest model considered, orders of magnitude smaller than the frontier models considered.

Meanwhile, Gemini‐2.0‐flash shows an increase when API Diff is introduced: for example, it improves from $P_1$ at \text{ \pgfkeysvalueof{gemini_P_1_BUILD_SUCCESS_percent}\%} to $P_3$ at \text{ \pgfkeysvalueof{gemini_P_3_BUILD_SUCCESS_percent}\%}. Interestingly, the performance of Gemini‐2.0‐flash improves when adding more context to the baseline prompt in $P_2$ and $P_3$. 

The introduction of Chains-of-thought (CoT) in the prompt is beneficial for some models (o3-mini) and detrimental for others (Gemini‐2.0‐flash)
In the latter case, the lowest performance for Gemini with $P_7$ at \text{ \pgfkeysvalueof{gemini_P_7_BUILD_SUCCESS_percent}\%} contains CoT. 
Looking at the open-source model, Qwen2.5‐32b‐instruct, it benefits a little from CoT and API Diff, improving from \text{ \pgfkeysvalueof{qwen_P_1_BUILD_SUCCESS_percent}\%} in $P_1$ to \text{ \pgfkeysvalueof{qwen_P_7_BUILD_SUCCESS_percent}\%} in $P_7$.

\begin{figure*}[ht!]
    \centering
    \includegraphics[width=\textwidth]{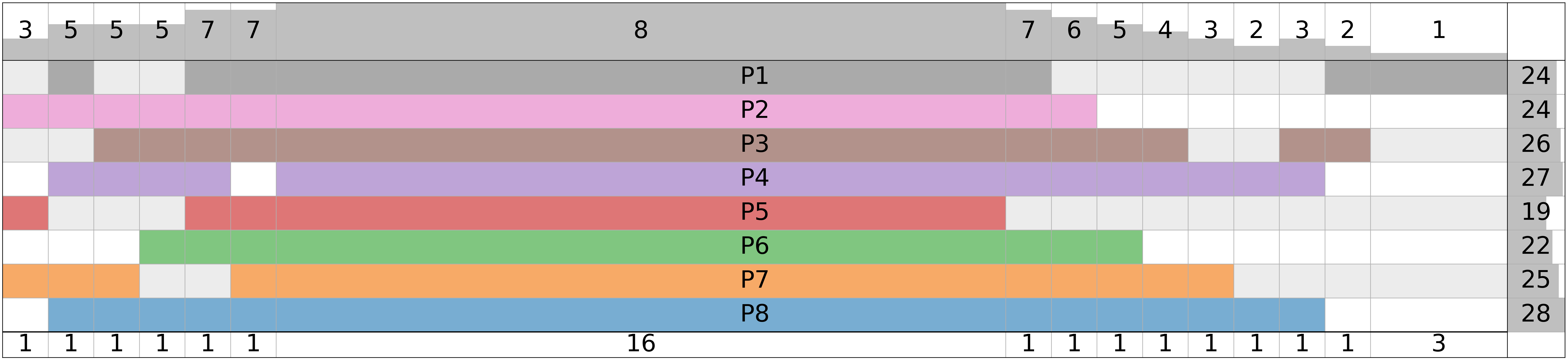}
    \caption{Visualization of Build Success by o3-mini Across Prompts. Each row represents the set of commits repaired using the prompt labeled on that row. The numbers on the right indicate the total number of commits repaired by each prompt. The bottom number represents intersections (“chunks”) of commits that were repaired by multiple prompts. The number at the top of each column indicates how many prompts are involved in that intersection. Overlapping areas correspond to the set intersections among the prompts.}
    \label{fig:o3successven}
\end{figure*}

Given that o3-mini has the highest build success rate, we further analyze its fixes in \autoref{fig:o3successven}. The figure illustrates the number of breaking commits successfully fixed by o3-mini across prompts, and the overlap between prompts. It shows each prompt ($P_1$ through $P_8$) as a horizontally colored row. The right-side number indicates how many commits each prompt successfully fixes in total. Along the upper part of the diagram, the numbers indicate how many prompts overlap in fixing a particular set of breaking updates, and the numbers at the bottom of the diagram are the actual count of these commits. The width of each 'chunk' represents how many breaking updates fall into those intersections between prompts. For example, the chunk labeled `8' at the top and `16' at the bottom indicates that there are 16 breaking updates successfully fixed by all eight prompts. 
Overall, $P_8$ fixed the most commits (28), while $P_4$ fixed the least breaking updates (20), confirming the importance of API Diff and Cot.

The diagram shows that out of the highest number of fixed breaking updates (28) by $P_8$, 16 of those were commonly fixed across every prompt (15\% of total builds). Notably, some smaller chunks (e.g, labeled `3' at the top and `1' at the bottom) show that some breaking updates were only fixed by a couple of prompts. 
Oddly enough, three breaking updates were only fixed by the baseline simple prompt $P_1$.

\begin{figure*}[ht!]
    \centering
    \includegraphics[width=\textwidth]{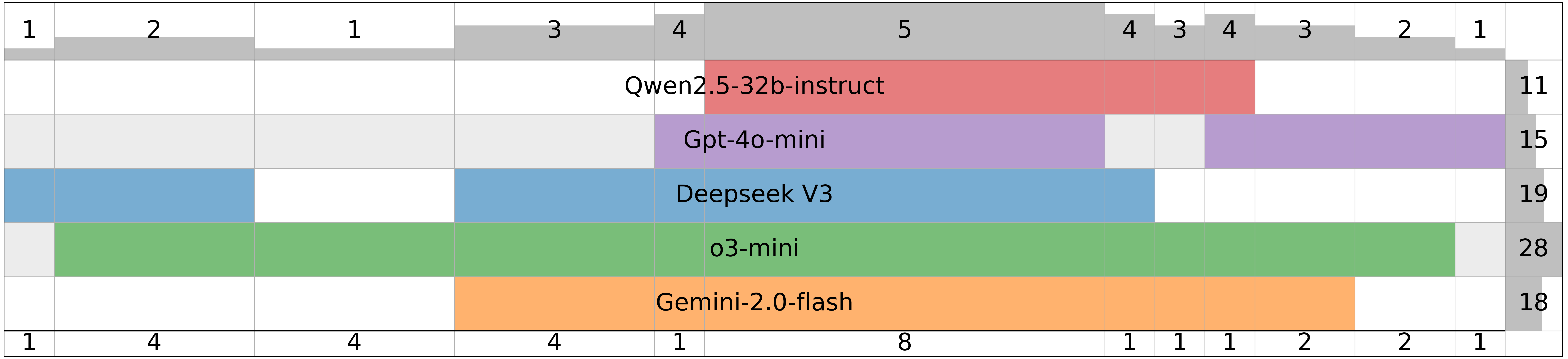}
    \caption{Visualization of Build Success by $P_8$ Across LLMs. Each row represents the set of commits repaired by the LLM labeled on that row. The numbers on the right indicate the total number of commits repaired. The bottom number represents intersections (“chunks”) of commits that were repaired by multiple LLMs. The number at the top of each column indicates how many LLMs are involved in that intersection. Overlapping areas correspond to the set intersections among the LLMs.}
    \label{fig:p8successven}
\end{figure*}

Given that $P_8$ is the prompt that provides the highest number of build success, we now consider the output of $P_8$ across different LLMs in \autoref{fig:p8successven}. Similar to \autoref{fig:o3successven}, the figure illustrates the number of breaking updates successfully fixed by $P_8$ across LLMs. It shows the number of breaking updates the LLMs successfully fix with the prompt configuration $P_8$, as a horizontally colored rows. The right-side numbers indicate how many commits each LLM successfully fixes using $P_8$. The numbers in the upper part of the diagram indicate how many LLMs overlap in fixing a particular set of commits. The numbers at the bottom of the diagram are the actual count of these commits.

The chunk labeled `5' at the top and `8' at the bottom indicates that there are 8 commits successfully fixed by all five LLMs using the $P_8$ prompt. It also shows that there is one breaking update only fixed by Deepseek V3, and another one only fixed by Gpt4o-mini. There are four breaking updates that are only fixed by o3-mini which contribute to its top performance. Overall, \autoref{fig:p8successven} confirms the superiority of o3-mini, which is capable of overlapping with other good models (DeepSeek-v3, gpt4o-mini, fixing unique breaking updates.

\begin{tcolorbox}[boxrule=1pt, left=2pt, right=2pt, top=2pt, bottom=2pt, width=\columnwidth]
    \normalsize
    \textbf{Answer to RQ1}:
    It is possible to fix real-world breaking updates with LLMs.
    o3-mini with prompt $P_8$ (CoT + Erroneous Line + APIDiff) achieves the highest build success rate, fixing \text{\pgfkeysvalueof{o3_P_8_BUILD_SUCCESS_percent}} out of the \finaldata~original failing builds (\pgfkeysvalueof{o3_P_8_BUILD_SUCCESS}\%). 
    The addition of API Diff and Chains-of-thought makes a difference, yet a relatively small one. 
    Open-source model Qwen2.5-32b-instruct, has the lowest success rate, as expected. 
    Overall, the inherent capability of the model is more important than the prompting itself for fixing breaking updates.
 \end{tcolorbox}

\subsection{\textbf{RQ2 }\textit{\rqFileError}}

To address this research question, we evaluate the extent to which \toolname~can partially fix compilation errors in builds that still fail after applying the generated fixes.
We analyze \toolname's performance at two levels of granularity: the file level, which considers the number of completely fixed files, and the error level, which considers the number of individual compilation errors fixed.
\input{rq2-pgfkeys_file-level}

\autoref{tab:file_error_level} presents the \textbf{File Fix Success Rate (FFSR)}, indicating the percentage of files that initially contained compilation errors and were successfully repaired by each LLM under different prompt configurations on builds that failed after applying the generated code.
Each row in the table refers to a different prompt configuration shown in \autoref{tab:exhaustive_prompt_list} and each column refers to one of the five LLMs presented in \autoref{tab:llm_comparison}.
Each cell contains the respective success rate of the files fixed by \toolname~after applying the generated fixes code.

\fileerrorleveltab

It is clearly that \toolname gets lots of partial fixes, in the double digits range.
For example, when it fails to completely repair the build, o3-mini with $P_4$ (Erroneous Line + API Diff) is able to fix \text{\pgfkeysvalueof{P_4_o3-mini-2025-01-31_fixed}} of the \text{\pgfkeysvalueof{P_4_o3-mini-2025-01-31_total_files}} original files with compilation failures (\text{\pgfkeysvalueof{P_4_o3-mini-2025-01-31_fixed_percentage}\%}). 
All models are able to fix dozens of files.

At the prompt level, adding erroneous lines and APIDiff information improves partial repair, confirming RQ1. 
For instance, Gemini-2.0-flash improves from \text{\pgfkeysvalueof{P_1_gemini-2.0-flash-001_fixed_percentage}\%} (\text{\pgfkeysvalueof{P_1_gemini-2.0-flash-001_fixed}/\pgfkeysvalueof{P_1_gemini-2.0-flash-001_total_files}}) with the baseline prompt ($P_1$) to \text{\pgfkeysvalueof{P_4_gemini-2.0-flash-001_fixed_percentage}\%} \text{(\pgfkeysvalueof{P_4_gemini-2.0-flash-001_fixed}/\pgfkeysvalueof{P_4_gemini-2.0-flash-001_total_files})} with $P_4$, which shows that the addition of APIDiff and erroneous lines in the configurations improves the performance of the model.

This is more evidence that the inclusion of APIDiff in the prompts assists the model to understand the nature of the breakage more clearly and to generate a more appropriate fix.
On the other hand, GPT-4o-mini shows low values with the CoT prompt, achieving only \text{\pgfkeysvalueof{P_5_gpt-4o-mini_fixed_percentage}\%} for $P_5$. However, its performance improves notably when using APIDiff and erroneous line information. This highlights how reasoning plays a significant role in influencing model performance.
However, the addition of CoT improves performance in Qwen2.5-32b-instruct, from \text{\pgfkeysvalueof{P_1_qwen-qwen2.5-32b-instruct_fixed_percentage}\%} in $P_1$ to \text{\pgfkeysvalueof{P_5_qwen-qwen2.5-32b-instruct_fixed_percentage}\%} in $P_5$ and $P_6$.
This indicates that CoT positively impacts the performance for fixing error files in specific models.
\input{rq2-pgfkeys_error-level}
\input{error_level}
\errorleveltab

At the compilation error level, \autoref{tab:error_level} shows the \textbf{Compilation Error Fix Rate (CEFR)} over failed repairs.
Each row represents prompts, columns represent LLM, and each cell represents the fixed error rate.
Strikingly, up to 78\% of errors are repaired in unsuccessful builds.
In other words, for those cases, the models are almost able to fix the build, a promising result for practitioners. 
o3-mini is again the best with $P_8$ prompt design, with \text{\pgfkeysvalueof{P_8_o3-mini-2025-01-31_fixed_errors_percentage}\%}, fixing \text{\pgfkeysvalueof{P_8_o3-mini-2025-01-31_errors_fixed}/\pgfkeysvalueof{P_8_o3-mini-2025-01-31_total_errors}} errors, demonstrating high efficiency in fixing individual compilation failures.

For Qwen2.5-32b-instruct, the behavior is inconsistent.
Despite achieving its best result with $P_3$, $P_5$ and $P_6$ (\text{\pgfkeysvalueof{P_5_qwen-qwen2.5-32b-instruct_fixed_errors_percentage}\%}), it drops to \text{\pgfkeysvalueof{P_8_qwen-qwen2.5-32b-instruct_fixed_errors_percentage}\%} in $P_8$, suggesting that prompts with multiple elements may overfit the model, decreasing its effectiveness.
It even shows marked drops in advanced configurations such as $P_6$ and $P_7$.

Consider the case of the \texttt{openfire-hazelcast-plugin} project.
Updating the \texttt{hazelcast} dependency from version \texttt{7.4.0} to \texttt{8.0.0} introduced a total of 15 buggy files and 91 compile errors.
After applying the code generated by \toolname (using o3-mini with $P_8$), 2 of the 15 buggy files are completely repaired, and 86 of the 91 compilation errors are successfully fixed. 
Although not all affected files are restored, \toolname manages to resolve 94\% of the individual errors, demonstrating a great ability to perform fine repairs even when a complete compilation fix is not reached.
For example, in the \texttt{ChangeSkingSpponge} file, \toolname correctly adapts the class signature, changing the generic type from \texttt{CommandSource} to \texttt{Audience} as shown in \autoref{lst:error_fixed}
In another case, \toolname fails to resolve the error \texttt{can't find symbol symbol: class Plugin}.
\autoref{lst:error_fail} shows the change proposed by \toolname, which fails to fix the original error. 

\begin{lstlisting}[backgroundcolor=\color{gray!10}, style=diff,caption={Changes introduced by \toolname correctly fixing the individual error}, label=lst:error_fixed,language=Java,belowskip=\baselineskip, numbers=none]
%\RHilight%-   public class ChangeSkinSponge implements PlatformPlugin<CommandSource> {
%\GHilight%+   public class ChangeSkinSponge implements PlatformPlugin<Audience> {
\end{lstlisting}

\begin{lstlisting}[backgroundcolor=\color{gray!10}, style=diff, caption={Changes introduced by \toolname failing to fix an error}, label=lst:error_fail,language=Java,belowskip=\baselineskip, numbers=none]
%\RHilight%-   @Plugin(id = ARTIFACT_ID, name = PomData.NAME, version = PomData.VERSION,
%\GHilight%+   @Plugin(id = PomData.ARTIFACT_ID, name = PomData.NAME, version = PomData.VERSION, {
\end{lstlisting}

\begin{tcolorbox}[boxrule=1pt, left=2pt, right=2pt, top=2pt, bottom=2pt, width=\columnwidth]
     \normalsize
    \textbf{Answer to RQ2}:
    Our results show that, even when the build is not fully repaired, a large number of errors are actually fixed.
    o3-mini, combined with prompts that include APIDiff and CoT, fixes \pgfkeysvalueof{P_4_o3-mini-2025-01-31_fixed_percentage}\% of files and an impressive \pgfkeysvalueof{P_8_o3-mini-2025-01-31_fixed_errors_percentage}\% of individual compiler errors. This suggests that the top-of-the-line models are close to perfectly repair many more builds, a trend we expect to happen with the next generation of models, especially the reasoning models.
\end{tcolorbox}

\subsection{\textbf{RQ3} \textit{\rqIntroducedErrors}}

\input{rq3_pgfkeys_output}
To answer this question, we evaluate the relative error fixed by \toolname,  considering the number of fixed errors as well as any new errors introduced by the LLM generated fix. We use the different LLMs listed in Table \ref{llmused} and the different prompt configurations defined in Section \ref{sec:use_llm}. We calculate
the \textbf{Relative Error Fixed Ratio (REF)} as defined in Equation \ref{eq:saved_effort} per breaking build for each configuration (Prompt and LLM). For this metric, the higher the better, meaning more fixed errors than introduced ones.

We summarize the results in Table \ref{tab:releative_error_prompt}, where we show the median value of the \textbf{REF}. Across all breaking builds, o3-mini performs the best, delivering the highest median REF on every prompt, peaking at \pgfkeysvalueof{P4_o3-mini-2025-01-31_median_relative_fixed}\% with $P_4$ (Erroneous Line and APIDiff). It also shows promising stability, never dropping below \pgfkeysvalueof{P3_o3-mini-2025-01-31_median_relative_fixed} \%, which shows that the model is robust enough that providing it with basic information ($P_1$-$P_5$) or more advanced reasoning and contextual information ($P6$-$P8$ ) will not result in new errors. 

In contrast, Qwen-2.5-32B-instruct fails according to this metric (with a median of 0\% for all prompts), indicating that the model either does not help in fixing errors or introduces more new errors than it fixes. 

DeepSeek-v3 benefits the most from the full context prompt $P_8$, with an increased median REF of \pgfkeysvalueof{P8_deepseek-deepseek-chat_median_relative_fixed}. 
These results are consistent with the results of RQ1 we present in Section \ref{rq1}. 

When we consider the results from the prompts viewpoint, we see that the improvement of $P_2$ (Erroneous Line) is consistently better, boasting the results of four of the five models.
On the contrary, $P_5$ (CoT) is the least consistently effective, helping only o3-mini. This indicates that adding the exact erroneous line is a valuable prompt addition to most LLMs.
CoT has an added value only when the model supports reasoning, such as o3-mini. \\

\relativeerrortab

Next, we investigate why new compilation errors appear. To do so, we manually analyze a random sample of new errors.
Our first example is from the \texttt{WorldwideChat} project, where the dependency update of \texttt{XSerie}s from version \texttt{8.5.0.1} to \texttt{8.6.0} fails because the method \texttt{parseEnchantment()} was removed in the new version of the dependency.
o3-mini with $P_8$ introduces a new error when generating the patch to fix the error.
The model replaces the call to the \texttt{parseEnchantment()} method with \texttt{getEnchantment()}.
The method proposed by LLM does not exist in the new version of the dependency.
\autoref{lst:patch-diff} shows the change applied by LLM that triggers the new compilation error.

\begin{figure}
\begin{lstlisting}[backgroundcolor=\color{gray!10}, style=diff,caption={Changes introduced By \toolname to fix a real breaking dependency updates}, label=lst:patch-diff,language=Java,belowskip=\baselineskip]
%\RHilight%-   currentLangMeta.addEnchant(XEnchantment.matchXEnchantment("power").get().
%\RHilight%     parseEnchantment(), 1, false);
%\GHilight%+   currentLangMeta.addEnchant(XEnchantment.matchXEnchantment("power").get().
%\GHilight%     getEnchantment(), 1, false);
\end{lstlisting}
\end{figure}

\begin{tcolorbox}[boxrule=1pt, left=2pt, right=2pt, top=2pt, bottom=2pt, width=\columnwidth]
    \normalsize
    \textbf{Answer to RQ3}:
    Our results show that o3-mini is the model that introduces the least new errors.  Prompt-wise, providing the erroneous line ($P_2$) yields the most gains across all LLMs to avoid new error introduction.
    Models with reasoning abilities are the ones which introduce the least new errors, benefiting more from the added contextual information.
 \end{tcolorbox}

\section{Discussion}\label{sec:discuss}

In this section, we reflect on our findings and their implications for both future research and practical use.
\autoref{comparison} compares \toolname with prior work, situating our results in relation to existing approaches.
\autoref{unsuccessful} analyzes the main causes of unsuccessful repairs , highlighting the limits of fixes.
Finally, \autoref{future-work} discusses how \toolname can be integrated into developer workflows and outlines promising directions for future wrk.

\subsection{Comparison with Prior Work}
\label{comparison}

In \cite{funkte2025}, the authors introduce two new approaches to fix breaking dependency updates using LLMs: a zero-shot prompting method and an agent system.
Their agent iteratively attempts up to 30 fixes per case.
The authors process the Bump dataset to produce two distinct evaluation data sets, the first set is considered as ``light slice" with 65 projects with breaking dependency updates caused by only one file, and a ``full slice" of 140 projects that has multiple files causing breaking in build.
The authors test the zero-shot approach on the light-slice dataset while they test the agent system on both datasets.
The zero-shot baseline uses no prompt engineering or design and is evaluated only on the light-slice dataset. 
The agent system, by contrast, is evaluated on both slices and operates iteratively, making up to 30 repair attempts per case. 
Importantly, the study does not report single-iteration success rates for the agent system, as its evaluation focuses on cumulative results across multiple trials.

\toolname's key advantage over the work in \cite{funkte2025} is its context-rich prompting, evident in being able to achieve better performance using a less complicated approach .
\toolname fixed full builds that can consist of multiple file edits, achieving a success rate of \pgfkeysvalueof{o3_P_8_BUILD_SUCCESS_percent}\% using the richest prompt (P8).
Even when the full build does not succeed, \toolname achieves a 78\% fix rate of the individual compilation errors.

In contrast, the agent approach in \cite{funkte2025}  fixes a maximum of 23\% of builds in the full slice and a maximum of 19\% in the light-slice, with the zero-shot approach achieving a maximum success rate of 19\%.
We note that \cite{funkte2025} does not report the average trials for fixes but allow iteration for up to 30 trials. 

To provide a direct comparison of identical breaking dependency updates, we analyze the intersection of breaking dependency updates evaluated by both studies.
The intersection between the 140 (``full slice") breaking updates presented by \cite{funkte2025} and the 103 breaking updates analyzed here yields 97 common breaking dependency updates.
When evaluating \toolname with o3-mini using the P$_8$ configuration on these 97 shared cases, \toolname achieves a 27\% (26/97) success rate, compared to the agent approach's 20\% (19/97) performance on the same breaking dependency updates.
This controlled comparison on the same failing builds eliminates dataset variability and demonstrates that our contextualized prompting strategy outperforms the agent-based approach.

To sum up, the key strength and novelty of \toolname is contextualize the breaking update problem in the prompt with essential information: by considering the line that is causing the errors, structuring the API diffs to the parts relevant to the project under investigation, and utilizing chain-of-thought prompting.

\subsection{Unsuccessful Repairs and Test Failures}
\label{unsuccessful}

Our work aims at repairing compilation failures.
Even when full builds are not repaired, \toolname often generates partial repairs that provide practical assistance.
For example, with the o3-mini model, \toolname fixes 78\% of individual errors and 41\% of erroneous files in otherwise failing builds.
This suggests that many builds fail only due to a small number of remaining issues, and \toolname can significantly reduce developer effort by handling the bulk of the errors automatically.

Yet, once compilation failures are resolved, we have sometimes test failures, indicating that semantic changes have occurred in the library.
For example, in project \texttt{ pay-adminusers}, the dependency update of \texttt{logback-classic} from version \texttt{1.2.11} to \texttt{1.4.5} fails due to compilation error \texttt{\seqsplit{cannot access org.slf4j. spi.LoggingEventAware class file for org.slf4j.spi.LoggingEventAware not found}}.
The compilation error is fixed by replacing the class \texttt{ch.qos.logback.classic.Logger} with the generic slf4j interface \texttt{\seqsplit{org.slf4j.Logger}}.
However, during testing, the method \texttt{doAppend()} was never invoked, resulting in the error \texttt{\seqsplit{Wanted but not invoked:\- mockLogAppender.doAppend()}}.
This is an example where the updates by the LLM fix the compilation failure, but result in a test failure. 

There are two ways in which we can interpret a test failure outcome after automated fixing by LLMs: (1) the LLM-generated code has introduced logical errors that pass compilation but fail functional tests, or (2) some unmodified parts of the code interact with the updated dependency, which has changed semantics.
Distinguishing between these scenarios is a challenging and important direction for future work.

\subsection{Recent Related Industry Tools}
Over the past year, in parallel with this research, several industrial tools have emerged that use large language models and agents to support software development tasks, including dependency updates and related software maintenance tasks. Some of these tools are ``general-purpose development agents'' while some specifically focus on dependency updates.

General-purpose coding agents that can be assigned various tasks across the software development process have become increasingly available. 
Such agents can be standalone CLI tools (e.g., Co-pilot CLI \footnote{https://github.com/features/copilot/cli} or Gemini CLI \footnote{https://geminicli.com/}) or can be directly integrated into development platforms such as GitHub (e.g. GitHub Copilot \footnote{https://github.com/features/copilot}) or Integrated Development Environments (IDEs) such as VSCode (e.g., VSCode Copilot \footnote{https://github.com/features/copilot/ai-code-editor} or Cursor \footnote{https://cursor.com/agents}).
Once prompted with a task, the agents can run in the background to complete the assigned task.
These agents can be assigned issues to solve, whether the issue is about fixing a bug or implementing a new feature.  
For GitHub-integrated tools, the agent may create a PR with the required changes based on the prompt or assigned issue. If it is integrated into an IDE, it can start generating or editing files directly in the editor.
While these agents do not specifically focus on fixing breaking dependency updates, the task can be viewed as a specific instance of program repair that these agents can handle.
Generally speaking, these industrial agents are designed as end-to-end automation pipelines that integrate multiple stages of repair, such as fault diagnosis, patch generation, testing, and deployment—within real development workflows.
Accordingly, while developers can use them to fix breaking dependency updates, we do not have empirical evidence about their effectiveness for this specific task. 

On the other hand, there are some recent industrial tools that are more closely related to breaking dependency updates.
For example, Dynatrace \footnote{https://www.dynatrace.com/} and Develocity \footnote{https://docs.gradle.com/develocity/current/} provide in-depth monitoring of build status. These tools analyze build and test data, with the aim of providing insights into aspects like performance, reliability, and efficiency. 
While these tools do not attempt to fix the breaking update themselves, they can potentially provide richer context that can help the LLM provide a correct fix. Experimenting with the effect of this additional context is an interesting avenue for future work.

The closest to our setting are tools that specialize in fixing breaking dependency updates. For example, FOSSA \footnote{https://fossa.com/products/fossabot/} automatically reviews updates for breaking changes and analyzes the code impact of updates, working alongside tools like Dependabot and Renovate to help developers manage dependencies. Patchwork \footnote{https://docs.patched.codes/patchwork/overview} is another example. These tools support automated repair by performing steps such as comparing dependency versions, generating patches, validating fixes, and potentially opening pull requests with proposed changes. However, similar to general-purpose agents, they are designed as end-to-end automation pipelines that integrate multiple stages of repair, rather than enabling in-depth experimentation and analysis of individual steps.
In contrast, our study evaluates LLM-based repair under scientifically controlled conditions. We focus on empirical evidence for specific repair principles, including (i) the role of targeted contextual information about the breaking change, (ii) fault localization that highlights failing lines and relevant API changes, and (iii) prompt design choices and how they affect repair outcomes. Our goal is to isolate these factors and analyze how they influence repair effectiveness for breaking changes. Thus, rather than providing a full end-to-end automation solution, our work studies the repair step in isolation to understand which design choices contribute to successful outcomes. Our results can inform improvements to relevant academic and industrial tools.

\subsection{Industry versus Academia in Software Research}
The previous section demonstrates that since we began this research, the broader AI-assisted software development ecosystem has advanced quickly, especially with advances in the capabilities of large language models and agent-based systems. This forces us to reflect on these pace differences and their impact. 

Generally speaking, industrial progress is fast paced with rapid updates in response to user feedback and needs.
On the other hand, the requirements of academic research  are different. It requires careful experimental design, dataset construction, controlled evaluation, peer review, and reproducibility.
In other words, industrial tools drive adoption and deliver end-to-end automation, but they are typically evaluated through operational metrics and real-world usage rather than controlled experimental measurements of individual design choices.
As a result, it remains difficult to assess how well their behavior generalizes or to reason about the causes of success and failure.
Research, on the other hand, systematically investigates techniques and provides empirical evidence about what works, under which conditions, and why. These differing objectives naturally lead to different evaluation practices and timelines between industrial development and academic research.

\subsection{Workflow Integration and Future Work}
\label{future-work}

Applications in the field of tools like \toolname require integration. We note that there exist Pull Request (PR) bots, which create pull requests to update dependencies (Dependabot, Renovate). \toolname could be integrated into such systems to further automate the process of dependency update. 
For example, if a PR for a dependency update causes a break in the Continuous Integration (CI) environment, an integrated tool could attempt to fix the breaking dependency update. Such an end-to-end solution to dependency updates would further reduce developers' effort in keeping their dependencies up-to-date, thus making their projects more secure and reliable.

Future work should explore several directions.
First, addressing semantic failures may require incorporating additional context from test suites or integrating program analysis to reason beyond syntax.
Second, as more powerful reasoning-oriented LLMs emerge, they may further improve repair rates, particularly for complex or multi-file repairs.
Finally, while our evaluation focused on Java/Maven projects, the core idea of contextual prompting is generalizable and could be applied to other ecosystems.

\section{Threats to Validity}
\label{sec:threats}

In the following, we discuss internal, construct, and external threats to the validity of our study.

\subsection{Internal Validity}
\label{sec:internal-validity}

A threat identified in the study is the variability in LLM results, which can produce nondeterministic results due to inherent randomness in their operation.
As an effort to limit this threat, we set a temperature of zero during inference, which reduces randomness and ensures that the model always chooses the most likely options.

\subsection{Construct Validity}

Memorization and data leakage are typically a concern in LLM-based techniques.
We reviewed the release dates of the dependencies in the dataset, comparing them to the training cutoff dates of each LLM, and confirmed that all release dates precede the training cut-off. 
As such, there is a potential risk of data leakage, which is similar to that of research based on HumanEval, Defects4j, or SWE-bench. Future work in required to collect breaking updates with new dependency versions released after the models' training cutoff dates. 

\subsection{External Validity}
We used the BUMP dataset, which focuses exclusively on breaking dependency updates in Java projects, which may limit the generalizability of our findings to other programming languages.
In terms of generalization across application domains, the dataset was collected from 153 Java projects, and each project was filtered to ensure that it is not a toy project. 

\section{Related Work}\label{sec:related}

Breaking dependency updates is a common problem in software development.
\citep{brito2018and,xavier2017we,brito2020you} investigate the reasons behind the developer's decision to introduce updates that lead to breaking dependency updates. 
These reasons include introducing new features, refactoring the code, fixing bugs, and addressing security vulnerabilities.
Hejderup et al. analyze how changes in method behavior trigger incompatibilities despite preserving the API contract. Such changes are difficult to detect by static analysis and often only become apparent at runtime~\citep{Hejderup2021CanWT}.
\citep{venturini2023depended} emphasized the challenges developers face in updating client code after dependency updates.
Researchers investigated various approaches to address this issue, where some investigated detecting breaking dependency updates \citep{brito2018apidiff,mujahid2020using}.
\citep{breakinggood} recently introduced an automated approach to explain the breaking in the client code after the version update.
Our approach differs from the previous studies as it uniquely addresses compilation failures caused by breaking updates through fixes generated using LLM. 
\toolname originally exploits contextual information such as APIDiff and erroneous lines to augment prompts, a dimension unexplored in previous studies.

Different approaches to mitigate breaking dependency updates are built on rule-based program transformation techniques.
\citep{dagenais2009semdiff} introduces a novel approach to API evolution by recommending adaptive replacements for deleted methods through analysis of repository history.
\citep{xing2007api} presents Diff-CatchUp, a tool that assists client application migration by automatically identifying broken APIs, suggesting plausible replacements, and providing usage examples based on model differencing and the framework's actual working code.
More recent research focuses on learning migration patterns from existing code samples~\citep{Lamothe2022A3AA}, analyzing changes in customer projects that have already migrated~\citep{Xu2019MeditorIA}, and leveraging documentation~\citep{Ni2021SOARAS}.
These approaches are intended to reduce the manual effort required to update dependencies and increase the likelihood that developers maintain updated dependencies.
Contrary to rule-based systems, our approach dynamically adapts to different types of breaking changes by integrating APIDiff and CoT reasoning.
This eliminates the need for manual engineering of rules or migration examples by leveraging the generalization capabilities of LLMs.

Large Language Models (LLMs) have emerged as a solution to automate API migration and dependency updates.
Almeida et al. explored the use of ChatGPT for these tasks~\citep{Almeida2024AutomaticLM}.
Compared to rule-based approaches, LLMs can handle contextual variations in the code, which makes them more effective in certain transformations~\citep{Nikolov2025HowIG}.
Tools based on LLMs have been developed to address specific migration problems.
RELANCER employs machine learning to predict repairs on deprecated APIs in Jupyter notebooks~\citep{Zhu2021RestoringTE}.
PCART automates compatibility assessment and repair of API parameter changes in Python~\citep{Zhang2024PCARTAR}.
In addition, Liu et al. note that LLMs can repair tests affected by code changes~\citep{Liu2024FixTT}.

Our approach differs from previous research as it focuses on fixing Java compilation errors across build, file, and error levels.

The most closely related work to our study is presented in parallel research by ~\citep{funkte2025}.
The authors explore automated build repair using either zero-shot prompting or agent-based iterative repair loops without explicit prompt engineering, achieving at most 23\% repair on the full-slice dataset.
\toolname improves on this by incorporating contextual information, reaching 27\% success on the same dataset of builds in a single iteration and fixing 78\% of individual compilation errors.

\section{Conclusion}\label{sec:conclude}
In this paper, we have presented \toolname, a novel approach that uses large language models to fix compilation failures caused by breaking dependency updates.
We have provided an empirical analysis of \toolname~using the BUMP dataset, using five different notable models and experimenting with eight different prompt configurations.
Our prompts incorporate contextual information such as the differences between the dependency versions, the lines causing the compilation failure, and providing a set-by-step reasoning in the prompt. 

We have analyzed the results at three granularity levels: at the build level, the file level, and the individual compilation error level.
Our results demonstrate the promising use of LLMs for fixing breaking dependency updates. 
In particular, the o3-mini model is able to successfully repair 27\% of failed builds, fixing 41\% of errors at the file level, and completely removing 78\% of individual errors.
The inclusion of APIDiff and CoT prompts improves performance, showing the importance of contextualized prompt designs to maximize the efficiency of LLMs to fix breaking dependency updates. 

Overall, our results demonstrate the potential for LLMs to fix breaking dependency updates. This is an essential automation step to help developers with keeping their projects' dependencies up-to-date.

\section*{Declarations}

\decl{Funding:}{This work was supported by the CHAINS project funded by Swedish Foundation for Strategic Research (SSF), the WebInspector project funded by Swedish Research Council (VR), as well as by the Wallenberg Autonomous Systems and Software Program (WASP) funded by the Knut and Alice Wallenberg Foundation. Some computation was enabled by resources provided by the National Academic Infrastructure for Supercomputing in Sweden (NAISS).}

\decl{Conflict of Interest:}{ None}

\decl{Ethical approval}{The authors declare that they have no known competing financial interests or personal relationships that could have appeared to influence the work reported in this paper. All data and code related to this paper is available at \byamurl}

\decl{Author Contributions}{ \textbf{Frank Reyes} contributed to the experimental design, data collection, data analysis, and writing.
\textbf{May Mahmoud} contributed to the experimental design, data collection, data analysis, and writing. 
\textbf{Federico Bono} contributed to the experimental design and data collection. 
\textbf{Sarah Nadi} contributed to data analysis and writing. 
\textbf{Benoit Baudry} contributed to data analysis and writing. 
\textbf{Martin Monperrus} contributed to data analysis and writing. }

\decl{Generative AI:}{Generative AI was not used for the generation of any part of the content in this paper or fordata analysis. Grammarly, a tool that uses AI, was used for spell checking, grammar correction, and improving writing clarity.}

\decl{Clinical Trial Number}{ Not applicable}

\bibliographystyle{apalike}       
\bibliography{references}   

\end{document}

%% file: file_error_level.tex
\newcommand{\fileerrorleveltab}{%
\begin{table}[t!]
    \centering
    \caption{Partial Repair (RQ2): File Fix Success Rate of \toolname over the failed cases}
    \label{tab:file_error_level}
    \rowcolors{2}{gray!10}{white}
    \resizebox{\textwidth}{!}{%
    \begin{tabular}{lrrrrr}
    \hline\noalign{\smallskip}
    \makecell[l]{\textbf{Prompt}\\ID} & \makecell[l]{\textbf{Deepseek}\\V3} & \makecell[l]{\textbf{Gemini}\\2.0-flash} & \makecell[l]{\textbf{Gpt}\\4o-mini} & \makecell[l]{\textbf{o3}\\mini} & \makecell[l]{\textbf{Qwen2.5}\\32b-instruct} \\
    \noalign{\smallskip}\hline\noalign{\smallskip}
$P_1$ & \pgfkeysvalueof{P_1_deepseek-deepseek-chat_fixed}/\pgfkeysvalueof{P_1_deepseek-deepseek-chat_total_files}(\pgfkeysvalueof{P_1_deepseek-deepseek-chat_fixed_percentage}\%) & \pgfkeysvalueof{P_1_gemini-2.0-flash-001_fixed}/\pgfkeysvalueof{P_1_gemini-2.0-flash-001_total_files}(\pgfkeysvalueof{P_1_gemini-2.0-flash-001_fixed_percentage}\%) & \pgfkeysvalueof{P_1_gpt-4o-mini_fixed}/\pgfkeysvalueof{P_1_gpt-4o-mini_total_files}(\pgfkeysvalueof{P_1_gpt-4o-mini_fixed_percentage}\%) & \pgfkeysvalueof{P_1_o3-mini-2025-01-31_fixed}/\pgfkeysvalueof{P_1_o3-mini-2025-01-31_total_files}(\pgfkeysvalueof{P_1_o3-mini-2025-01-31_fixed_percentage}\%) & \pgfkeysvalueof{P_1_qwen-qwen2.5-32b-instruct_fixed}/\pgfkeysvalueof{P_1_qwen-qwen2.5-32b-instruct_total_files}(\pgfkeysvalueof{P_1_qwen-qwen2.5-32b-instruct_fixed_percentage}\%) \\
$P_2$ & \pgfkeysvalueof{P_2_deepseek-deepseek-chat_fixed}/\pgfkeysvalueof{P_2_deepseek-deepseek-chat_total_files}(\pgfkeysvalueof{P_2_deepseek-deepseek-chat_fixed_percentage}\%) & \pgfkeysvalueof{P_2_gemini-2.0-flash-001_fixed}/\pgfkeysvalueof{P_2_gemini-2.0-flash-001_total_files}(\pgfkeysvalueof{P_2_gemini-2.0-flash-001_fixed_percentage}\%) & \pgfkeysvalueof{P_2_gpt-4o-mini_fixed}/\pgfkeysvalueof{P_2_gpt-4o-mini_total_files}(\pgfkeysvalueof{P_2_gpt-4o-mini_fixed_percentage}\%) & \pgfkeysvalueof{P_2_o3-mini-2025-01-31_fixed}/\pgfkeysvalueof{P_2_o3-mini-2025-01-31_total_files}(\pgfkeysvalueof{P_2_o3-mini-2025-01-31_fixed_percentage}\%) & \pgfkeysvalueof{P_2_qwen-qwen2.5-32b-instruct_fixed}/\pgfkeysvalueof{P_2_qwen-qwen2.5-32b-instruct_total_files}(\pgfkeysvalueof{P_2_qwen-qwen2.5-32b-instruct_fixed_percentage}\%) \\
$P_3$ & \pgfkeysvalueof{P_3_deepseek-deepseek-chat_fixed}/\pgfkeysvalueof{P_3_deepseek-deepseek-chat_total_files}(\pgfkeysvalueof{P_3_deepseek-deepseek-chat_fixed_percentage}\%) & \pgfkeysvalueof{P_3_gemini-2.0-flash-001_fixed}/\pgfkeysvalueof{P_3_gemini-2.0-flash-001_total_files}(\pgfkeysvalueof{P_3_gemini-2.0-flash-001_fixed_percentage}\%) & \pgfkeysvalueof{P_3_gpt-4o-mini_fixed}/\pgfkeysvalueof{P_3_gpt-4o-mini_total_files}(\pgfkeysvalueof{P_3_gpt-4o-mini_fixed_percentage}\%) & \pgfkeysvalueof{P_3_o3-mini-2025-01-31_fixed}/\pgfkeysvalueof{P_3_o3-mini-2025-01-31_total_files}(\pgfkeysvalueof{P_3_o3-mini-2025-01-31_fixed_percentage}\%) & \pgfkeysvalueof{P_3_qwen-qwen2.5-32b-instruct_fixed}/\pgfkeysvalueof{P_3_qwen-qwen2.5-32b-instruct_total_files}(\pgfkeysvalueof{P_3_qwen-qwen2.5-32b-instruct_fixed_percentage}\%) \\
$P_4$ & \pgfkeysvalueof{P_4_deepseek-deepseek-chat_fixed}/\pgfkeysvalueof{P_4_deepseek-deepseek-chat_total_files}(\pgfkeysvalueof{P_4_deepseek-deepseek-chat_fixed_percentage}\%) & \pgfkeysvalueof{P_4_gemini-2.0-flash-001_fixed}/\pgfkeysvalueof{P_4_gemini-2.0-flash-001_total_files}(\pgfkeysvalueof{P_4_gemini-2.0-flash-001_fixed_percentage}\%) & \pgfkeysvalueof{P_4_gpt-4o-mini_fixed}/\pgfkeysvalueof{P_4_gpt-4o-mini_total_files}(\pgfkeysvalueof{P_4_gpt-4o-mini_fixed_percentage}\%) & \textbf{\pgfkeysvalueof{P_4_o3-mini-2025-01-31_fixed}/\pgfkeysvalueof{P_4_o3-mini-2025-01-31_total_files}(\pgfkeysvalueof{P_4_o3-mini-2025-01-31_fixed_percentage}\%)} & \pgfkeysvalueof{P_4_qwen-qwen2.5-32b-instruct_fixed}/\pgfkeysvalueof{P_4_qwen-qwen2.5-32b-instruct_total_files}(\pgfkeysvalueof{P_4_qwen-qwen2.5-32b-instruct_fixed_percentage}\%) \\
$P_5$ & \pgfkeysvalueof{P_5_deepseek-deepseek-chat_fixed}/\pgfkeysvalueof{P_5_deepseek-deepseek-chat_total_files}(\pgfkeysvalueof{P_5_deepseek-deepseek-chat_fixed_percentage}\%) & \pgfkeysvalueof{P_5_gemini-2.0-flash-001_fixed}/\pgfkeysvalueof{P_5_gemini-2.0-flash-001_total_files}(\pgfkeysvalueof{P_5_gemini-2.0-flash-001_fixed_percentage}\%) & \pgfkeysvalueof{P_5_gpt-4o-mini_fixed}/\pgfkeysvalueof{P_5_gpt-4o-mini_total_files}(\pgfkeysvalueof{P_5_gpt-4o-mini_fixed_percentage}\%) & \pgfkeysvalueof{P_5_o3-mini-2025-01-31_fixed}/\pgfkeysvalueof{P_5_o3-mini-2025-01-31_total_files}(\pgfkeysvalueof{P_5_o3-mini-2025-01-31_fixed_percentage}\%) & \pgfkeysvalueof{P_5_qwen-qwen2.5-32b-instruct_fixed}/\pgfkeysvalueof{P_5_qwen-qwen2.5-32b-instruct_total_files}(\pgfkeysvalueof{P_5_qwen-qwen2.5-32b-instruct_fixed_percentage}\%) \\
$P_6$ & \pgfkeysvalueof{P_6_deepseek-deepseek-chat_fixed}/\pgfkeysvalueof{P_6_deepseek-deepseek-chat_total_files}(\pgfkeysvalueof{P_6_deepseek-deepseek-chat_fixed_percentage}\%) & \pgfkeysvalueof{P_6_gemini-2.0-flash-001_fixed}/\pgfkeysvalueof{P_6_gemini-2.0-flash-001_total_files}(\pgfkeysvalueof{P_6_gemini-2.0-flash-001_fixed_percentage}\%) & \pgfkeysvalueof{P_6_gpt-4o-mini_fixed}/\pgfkeysvalueof{P_6_gpt-4o-mini_total_files}(\pgfkeysvalueof{P_6_gpt-4o-mini_fixed_percentage}\%) & \pgfkeysvalueof{P_6_o3-mini-2025-01-31_fixed}/\pgfkeysvalueof{P_6_o3-mini-2025-01-31_total_files}(\pgfkeysvalueof{P_6_o3-mini-2025-01-31_fixed_percentage}\%) & \pgfkeysvalueof{P_6_qwen-qwen2.5-32b-instruct_fixed}/\pgfkeysvalueof{P_6_qwen-qwen2.5-32b-instruct_total_files}(\pgfkeysvalueof{P_6_qwen-qwen2.5-32b-instruct_fixed_percentage}\%) \\
$P_7$ & \pgfkeysvalueof{P_7_deepseek-deepseek-chat_fixed}/\pgfkeysvalueof{P_7_deepseek-deepseek-chat_total_files}(\pgfkeysvalueof{P_7_deepseek-deepseek-chat_fixed_percentage}\%) & \pgfkeysvalueof{P_7_gemini-2.0-flash-001_fixed}/\pgfkeysvalueof{P_7_gemini-2.0-flash-001_total_files}(\pgfkeysvalueof{P_7_gemini-2.0-flash-001_fixed_percentage}\%) & \pgfkeysvalueof{P_7_gpt-4o-mini_fixed}/\pgfkeysvalueof{P_7_gpt-4o-mini_total_files}(\pgfkeysvalueof{P_7_gpt-4o-mini_fixed_percentage}\%) & \pgfkeysvalueof{P_7_o3-mini-2025-01-31_fixed}/\pgfkeysvalueof{P_7_o3-mini-2025-01-31_total_files}(\pgfkeysvalueof{P_7_o3-mini-2025-01-31_fixed_percentage}\%) & \pgfkeysvalueof{P_7_qwen-qwen2.5-32b-instruct_fixed}/\pgfkeysvalueof{P_7_qwen-qwen2.5-32b-instruct_total_files}(\pgfkeysvalueof{P_7_qwen-qwen2.5-32b-instruct_fixed_percentage}\%) \\
$P_8$ & \pgfkeysvalueof{P_8_deepseek-deepseek-chat_fixed}/\pgfkeysvalueof{P_8_deepseek-deepseek-chat_total_files}(\pgfkeysvalueof{P_8_deepseek-deepseek-chat_fixed_percentage}\%) & \pgfkeysvalueof{P_8_gemini-2.0-flash-001_fixed}/\pgfkeysvalueof{P_8_gemini-2.0-flash-001_total_files}(\pgfkeysvalueof{P_8_gemini-2.0-flash-001_fixed_percentage}\%) & \pgfkeysvalueof{P_8_gpt-4o-mini_fixed}/\pgfkeysvalueof{P_8_gpt-4o-mini_total_files}(\pgfkeysvalueof{P_8_gpt-4o-mini_fixed_percentage}\%) & \pgfkeysvalueof{P_8_o3-mini-2025-01-31_fixed}/\pgfkeysvalueof{P_8_o3-mini-2025-01-31_total_files}(\pgfkeysvalueof{P_8_o3-mini-2025-01-31_fixed_percentage}\%) & \pgfkeysvalueof{P_8_qwen-qwen2.5-32b-instruct_fixed}/\pgfkeysvalueof{P_8_qwen-qwen2.5-32b-instruct_total_files}(\pgfkeysvalueof{P_8_qwen-qwen2.5-32b-instruct_fixed_percentage}\%) \\

       \noalign{\smallskip}\hline
    \end{tabular}
    }
\end{table}
}

%% file: rq1-tab.tex
\newcommand{\buildsuccesstab}{%
\begin{table}[ht]
\centering
\rowcolors{2}{gray!10}{white}
\caption{Build Success Rate (BSR) of \toolname}
\label{tab:build_success_prompt}
\resizebox{\columnwidth}{!}{%
\begin{tabular}{lrrrrrrrrr}
\toprule
\textbf{\makecell[l]{Prompt\\ID}} & \textbf{\makecell[l]{Deepseek\\V3}} & \textbf{\makecell[l]{Gemini\\2.0-flash}} & \textbf{\makecell[l]{Gpt\\4o-mini}} & \textbf{\makecell[l]{o3-\\mini}} & \textbf{\makecell[l]{Qwen2.5-\\32b-instruct}} \\
  
\midrule
$P_1$ & \pgfkeysvalueof{deepseek_P_1_BUILD_SUCCESS}/\finaldata~(\pgfkeysvalueof{deepseek_P_1_BUILD_SUCCESS_percent}\%) & \pgfkeysvalueof{gemini_P_1_BUILD_SUCCESS}/\finaldata~(\pgfkeysvalueof{gemini_P_1_BUILD_SUCCESS_percent}\%) & \pgfkeysvalueof{gpt_P_1_BUILD_SUCCESS}/\finaldata~(\pgfkeysvalueof{gpt_P_1_BUILD_SUCCESS_percent}\%) & \pgfkeysvalueof{o3_P_1_BUILD_SUCCESS}/\finaldata~(\pgfkeysvalueof{o3_P_1_BUILD_SUCCESS_percent}\%) & \pgfkeysvalueof{qwen_P_1_BUILD_SUCCESS}/\finaldata~(\pgfkeysvalueof{qwen_P_1_BUILD_SUCCESS_percent}\%) \\
$P_2$ & \pgfkeysvalueof{deepseek_P_2_BUILD_SUCCESS}/\finaldata~(\pgfkeysvalueof{deepseek_P_2_BUILD_SUCCESS_percent}\%) & \pgfkeysvalueof{gemini_P_2_BUILD_SUCCESS}/\finaldata~(\pgfkeysvalueof{gemini_P_2_BUILD_SUCCESS_percent}\%) & \pgfkeysvalueof{gpt_P_2_BUILD_SUCCESS}/\finaldata~(\pgfkeysvalueof{gpt_P_2_BUILD_SUCCESS_percent}\%) & \pgfkeysvalueof{o3_P_2_BUILD_SUCCESS}/\finaldata~(\pgfkeysvalueof{o3_P_2_BUILD_SUCCESS_percent}\%) & \pgfkeysvalueof{qwen_P_2_BUILD_SUCCESS}/\finaldata~(\pgfkeysvalueof{qwen_P_2_BUILD_SUCCESS_percent}\%) \\
$P_3$ & \pgfkeysvalueof{deepseek_P_3_BUILD_SUCCESS}/\finaldata~(\pgfkeysvalueof{deepseek_P_3_BUILD_SUCCESS_percent}\%) & \pgfkeysvalueof{gemini_P_3_BUILD_SUCCESS}/\finaldata~(\pgfkeysvalueof{gemini_P_3_BUILD_SUCCESS_percent}\%) & \pgfkeysvalueof{gpt_P_3_BUILD_SUCCESS}/\finaldata~(\pgfkeysvalueof{gpt_P_3_BUILD_SUCCESS_percent}\%) & \pgfkeysvalueof{o3_P_3_BUILD_SUCCESS}/\finaldata~(\pgfkeysvalueof{o3_P_3_BUILD_SUCCESS_percent}\%) & \pgfkeysvalueof{qwen_P_3_BUILD_SUCCESS}/\finaldata~(\pgfkeysvalueof{qwen_P_3_BUILD_SUCCESS_percent}\%) \\
$P_4$ & \pgfkeysvalueof{deepseek_P_4_BUILD_SUCCESS}/\finaldata~(\pgfkeysvalueof{deepseek_P_4_BUILD_SUCCESS_percent}\%) & \pgfkeysvalueof{gemini_P_4_BUILD_SUCCESS}/\finaldata~(\pgfkeysvalueof{gemini_P_4_BUILD_SUCCESS_percent}\%) & \pgfkeysvalueof{gpt_P_4_BUILD_SUCCESS}/\finaldata~(\pgfkeysvalueof{gpt_P_4_BUILD_SUCCESS_percent}\%) & \pgfkeysvalueof{o3_P_4_BUILD_SUCCESS}/\finaldata~(\pgfkeysvalueof{o3_P_4_BUILD_SUCCESS_percent}\%) & \pgfkeysvalueof{qwen_P_4_BUILD_SUCCESS}/\finaldata~(\pgfkeysvalueof{qwen_P_4_BUILD_SUCCESS_percent}\%) \\
$P_5$ & \pgfkeysvalueof{deepseek_P_5_BUILD_SUCCESS}/\finaldata~(\pgfkeysvalueof{deepseek_P_5_BUILD_SUCCESS_percent}\%) & \pgfkeysvalueof{gemini_P_5_BUILD_SUCCESS}/\finaldata~(\pgfkeysvalueof{gemini_P_5_BUILD_SUCCESS_percent}\%) & \pgfkeysvalueof{gpt_P_5_BUILD_SUCCESS}/\finaldata~(\pgfkeysvalueof{gpt_P_5_BUILD_SUCCESS_percent}\%) & \pgfkeysvalueof{o3_P_5_BUILD_SUCCESS}/\finaldata~(\pgfkeysvalueof{o3_P_5_BUILD_SUCCESS_percent}\%) & \pgfkeysvalueof{qwen_P_5_BUILD_SUCCESS}/\finaldata~(\pgfkeysvalueof{qwen_P_5_BUILD_SUCCESS_percent}\%) \\
$P_6$ & \pgfkeysvalueof{deepseek_P_6_BUILD_SUCCESS}/\finaldata~(\pgfkeysvalueof{deepseek_P_6_BUILD_SUCCESS_percent}\%) & \pgfkeysvalueof{gemini_P_6_BUILD_SUCCESS}/\finaldata~(\pgfkeysvalueof{gemini_P_6_BUILD_SUCCESS_percent}\%) & \pgfkeysvalueof{gpt_P_6_BUILD_SUCCESS}/\finaldata~(\pgfkeysvalueof{gpt_P_6_BUILD_SUCCESS_percent}\%) & \pgfkeysvalueof{o3_P_6_BUILD_SUCCESS}/\finaldata~(\pgfkeysvalueof{o3_P_6_BUILD_SUCCESS_percent}\%) & \pgfkeysvalueof{qwen_P_6_BUILD_SUCCESS}/\finaldata~(\pgfkeysvalueof{qwen_P_6_BUILD_SUCCESS_percent}\%) \\
$P_7$ & \pgfkeysvalueof{deepseek_P_7_BUILD_SUCCESS}/\finaldata~(\pgfkeysvalueof{deepseek_P_7_BUILD_SUCCESS_percent}\%) & \pgfkeysvalueof{gemini_P_7_BUILD_SUCCESS}/\finaldata~(\pgfkeysvalueof{gemini_P_7_BUILD_SUCCESS_percent}\%) & \pgfkeysvalueof{gpt_P_7_BUILD_SUCCESS}/\finaldata~(\pgfkeysvalueof{gpt_P_7_BUILD_SUCCESS_percent}\%) & \pgfkeysvalueof{o3_P_7_BUILD_SUCCESS}/\finaldata~(\pgfkeysvalueof{o3_P_7_BUILD_SUCCESS_percent}\%) & \pgfkeysvalueof{qwen_P_7_BUILD_SUCCESS}/\finaldata~(\pgfkeysvalueof{qwen_P_7_BUILD_SUCCESS_percent}\%) \\
$P_8$ & \pgfkeysvalueof{deepseek_P_8_BUILD_SUCCESS}/\finaldata~(\pgfkeysvalueof{deepseek_P_8_BUILD_SUCCESS_percent}\%) & \pgfkeysvalueof{gemini_P_8_BUILD_SUCCESS}/\finaldata~(\pgfkeysvalueof{gemini_P_8_BUILD_SUCCESS_percent}\%) & \pgfkeysvalueof{gpt_P_8_BUILD_SUCCESS}/\finaldata~(\pgfkeysvalueof{gpt_P_8_BUILD_SUCCESS_percent}\%) & \textbf{\pgfkeysvalueof{o3_P_8_BUILD_SUCCESS}/\finaldata~(\pgfkeysvalueof{o3_P_8_BUILD_SUCCESS_percent}\%)} & \pgfkeysvalueof{qwen_P_8_BUILD_SUCCESS}/\finaldata~(\pgfkeysvalueof{qwen_P_8_BUILD_SUCCESS_percent}\%) \\

\bottomrule
\end{tabular}}
\end{table}
}

%% file: rq1_pgfkeys.tex
\pgfkeyssetvalue{total_commits}{103}
\pgfkeyssetvalue{gemini_P_8_COMPILATION_FAILURE}{65}
\pgfkeyssetvalue{gemini_P_8_COMPILATION_FAILURE_percent}{63}
\pgfkeyssetvalue{gemini_P_8_BUILD_SUCCESS}{18}
\pgfkeyssetvalue{gemini_P_8_BUILD_SUCCESS_percent}{17}
\pgfkeyssetvalue{gemini_P_8_TEST_FAILURE}{12}
\pgfkeyssetvalue{gemini_P_8_TEST_FAILURE_percent}{12}
\pgfkeyssetvalue{gemini_P_8_ERROR_MODEL_RESPONSE}{6}
\pgfkeyssetvalue{gemini_P_8_ERROR_MODEL_RESPONSE_percent}{6}
\pgfkeyssetvalue{gemini_P_8_NOT_REPAIRED}{2}
\pgfkeyssetvalue{gemini_P_8_NOT_REPAIRED_percent}{2}
\pgfkeyssetvalue{o3_P_8_COMPILATION_FAILURE}{56}
\pgfkeyssetvalue{o3_P_8_COMPILATION_FAILURE_percent}{54}
\pgfkeyssetvalue{o3_P_8_BUILD_SUCCESS}{28}
\pgfkeyssetvalue{o3_P_8_BUILD_SUCCESS_percent}{27}
\pgfkeyssetvalue{o3_P_8_TEST_FAILURE}{17}
\pgfkeyssetvalue{o3_P_8_TEST_FAILURE_percent}{17}
\pgfkeyssetvalue{o3_P_8_ERROR_MODEL_RESPONSE}{1}
\pgfkeyssetvalue{o3_P_8_ERROR_MODEL_RESPONSE_percent}{1}
\pgfkeyssetvalue{o3_P_8_NOT_REPAIRED}{1}
\pgfkeyssetvalue{o3_P_8_NOT_REPAIRED_percent}{1}
\pgfkeyssetvalue{deepseek_P_8_COMPILATION_FAILURE}{63}
\pgfkeyssetvalue{deepseek_P_8_COMPILATION_FAILURE_percent}{61}
\pgfkeyssetvalue{deepseek_P_8_BUILD_SUCCESS}{19}
\pgfkeyssetvalue{deepseek_P_8_BUILD_SUCCESS_percent}{18}
\pgfkeyssetvalue{deepseek_P_8_ERROR_MODEL_RESPONSE}{12}
\pgfkeyssetvalue{deepseek_P_8_ERROR_MODEL_RESPONSE_percent}{12}
\pgfkeyssetvalue{deepseek_P_8_TEST_FAILURE}{8}
\pgfkeyssetvalue{deepseek_P_8_TEST_FAILURE_percent}{8}
\pgfkeyssetvalue{deepseek_P_8_NOT_REPAIRED}{1}
\pgfkeyssetvalue{deepseek_P_8_NOT_REPAIRED_percent}{1}
\pgfkeyssetvalue{gpt_P_8_COMPILATION_FAILURE}{75}
\pgfkeyssetvalue{gpt_P_8_COMPILATION_FAILURE_percent}{73}
\pgfkeyssetvalue{gpt_P_8_BUILD_SUCCESS}{15}
\pgfkeyssetvalue{gpt_P_8_BUILD_SUCCESS_percent}{15}
\pgfkeyssetvalue{gpt_P_8_TEST_FAILURE}{6}
\pgfkeyssetvalue{gpt_P_8_TEST_FAILURE_percent}{6}
\pgfkeyssetvalue{gpt_P_8_ERROR_MODEL_RESPONSE}{5}
\pgfkeyssetvalue{gpt_P_8_ERROR_MODEL_RESPONSE_percent}{5}
\pgfkeyssetvalue{gpt_P_8_NOT_REPAIRED}{2}
\pgfkeyssetvalue{gpt_P_8_NOT_REPAIRED_percent}{2}
\pgfkeyssetvalue{qwen_P_8_COMPILATION_FAILURE}{81}
\pgfkeyssetvalue{qwen_P_8_COMPILATION_FAILURE_percent}{79}
\pgfkeyssetvalue{qwen_P_8_BUILD_SUCCESS}{11}
\pgfkeyssetvalue{qwen_P_8_BUILD_SUCCESS_percent}{11}
\pgfkeyssetvalue{qwen_P_8_ERROR_MODEL_RESPONSE}{7}
\pgfkeyssetvalue{qwen_P_8_ERROR_MODEL_RESPONSE_percent}{7}
\pgfkeyssetvalue{qwen_P_8_TEST_FAILURE}{3}
\pgfkeyssetvalue{qwen_P_8_TEST_FAILURE_percent}{3}
\pgfkeyssetvalue{qwen_P_8_NOT_REPAIRED}{1}
\pgfkeyssetvalue{qwen_P_8_NOT_REPAIRED_percent}{1}
\pgfkeyssetvalue{gemini_P_6_COMPILATION_FAILURE}{57}
\pgfkeyssetvalue{gemini_P_6_COMPILATION_FAILURE_percent}{55}
\pgfkeyssetvalue{gemini_P_6_BUILD_SUCCESS}{17}
\pgfkeyssetvalue{gemini_P_6_BUILD_SUCCESS_percent}{17}
\pgfkeyssetvalue{gemini_P_6_TEST_FAILURE}{12}
\pgfkeyssetvalue{gemini_P_6_TEST_FAILURE_percent}{12}
\pgfkeyssetvalue{gemini_P_6_NOT_REPAIRED}{11}
\pgfkeyssetvalue{gemini_P_6_NOT_REPAIRED_percent}{11}
\pgfkeyssetvalue{gemini_P_6_ERROR_MODEL_RESPONSE}{6}
\pgfkeyssetvalue{gemini_P_6_ERROR_MODEL_RESPONSE_percent}{6}
\pgfkeyssetvalue{o3_P_6_COMPILATION_FAILURE}{67}
\pgfkeyssetvalue{o3_P_6_COMPILATION_FAILURE_percent}{65}
\pgfkeyssetvalue{o3_P_6_BUILD_SUCCESS}{22}
\pgfkeyssetvalue{o3_P_6_BUILD_SUCCESS_percent}{21}
\pgfkeyssetvalue{o3_P_6_TEST_FAILURE}{13}
\pgfkeyssetvalue{o3_P_6_TEST_FAILURE_percent}{13}
\pgfkeyssetvalue{o3_P_6_NOT_REPAIRED}{1}
\pgfkeyssetvalue{o3_P_6_NOT_REPAIRED_percent}{1}
\pgfkeyssetvalue{deepseek_P_6_COMPILATION_FAILURE}{64}
\pgfkeyssetvalue{deepseek_P_6_COMPILATION_FAILURE_percent}{62}
\pgfkeyssetvalue{deepseek_P_6_BUILD_SUCCESS}{22}
\pgfkeyssetvalue{deepseek_P_6_BUILD_SUCCESS_percent}{21}
\pgfkeyssetvalue{deepseek_P_6_ERROR_MODEL_RESPONSE}{8}
\pgfkeyssetvalue{deepseek_P_6_ERROR_MODEL_RESPONSE_percent}{8}
\pgfkeyssetvalue{deepseek_P_6_TEST_FAILURE}{7}
\pgfkeyssetvalue{deepseek_P_6_TEST_FAILURE_percent}{7}
\pgfkeyssetvalue{deepseek_P_6_NOT_REPAIRED}{2}
\pgfkeyssetvalue{deepseek_P_6_NOT_REPAIRED_percent}{2}
\pgfkeyssetvalue{gpt_P_6_COMPILATION_FAILURE}{66}
\pgfkeyssetvalue{gpt_P_6_COMPILATION_FAILURE_percent}{64}
\pgfkeyssetvalue{gpt_P_6_BUILD_SUCCESS}{16}
\pgfkeyssetvalue{gpt_P_6_BUILD_SUCCESS_percent}{16}
\pgfkeyssetvalue{gpt_P_6_NOT_REPAIRED}{12}
\pgfkeyssetvalue{gpt_P_6_NOT_REPAIRED_percent}{12}
\pgfkeyssetvalue{gpt_P_6_TEST_FAILURE}{9}
\pgfkeyssetvalue{gpt_P_6_TEST_FAILURE_percent}{9}
\pgfkeyssetvalue{qwen_P_6_COMPILATION_FAILURE}{79}
\pgfkeyssetvalue{qwen_P_6_COMPILATION_FAILURE_percent}{77}
\pgfkeyssetvalue{qwen_P_6_BUILD_SUCCESS}{12}
\pgfkeyssetvalue{qwen_P_6_BUILD_SUCCESS_percent}{12}
\pgfkeyssetvalue{qwen_P_6_TEST_FAILURE}{6}
\pgfkeyssetvalue{qwen_P_6_TEST_FAILURE_percent}{6}
\pgfkeyssetvalue{qwen_P_6_ERROR_MODEL_RESPONSE}{5}
\pgfkeyssetvalue{qwen_P_6_ERROR_MODEL_RESPONSE_percent}{5}
\pgfkeyssetvalue{qwen_P_6_NOT_REPAIRED}{1}
\pgfkeyssetvalue{qwen_P_6_NOT_REPAIRED_percent}{1}
\pgfkeyssetvalue{gemini_P_2_COMPILATION_FAILURE}{58}
\pgfkeyssetvalue{gemini_P_2_COMPILATION_FAILURE_percent}{56}
\pgfkeyssetvalue{gemini_P_2_BUILD_SUCCESS}{18}
\pgfkeyssetvalue{gemini_P_2_BUILD_SUCCESS_percent}{17}
\pgfkeyssetvalue{gemini_P_2_TEST_FAILURE}{13}
\pgfkeyssetvalue{gemini_P_2_TEST_FAILURE_percent}{13}
\pgfkeyssetvalue{gemini_P_2_NOT_REPAIRED}{8}
\pgfkeyssetvalue{gemini_P_2_NOT_REPAIRED_percent}{8}
\pgfkeyssetvalue{gemini_P_2_ERROR_MODEL_RESPONSE}{6}
\pgfkeyssetvalue{gemini_P_2_ERROR_MODEL_RESPONSE_percent}{6}
\pgfkeyssetvalue{o3_P_2_COMPILATION_FAILURE}{64}
\pgfkeyssetvalue{o3_P_2_COMPILATION_FAILURE_percent}{62}
\pgfkeyssetvalue{o3_P_2_BUILD_SUCCESS}{24}
\pgfkeyssetvalue{o3_P_2_BUILD_SUCCESS_percent}{23}
\pgfkeyssetvalue{o3_P_2_TEST_FAILURE}{13}
\pgfkeyssetvalue{o3_P_2_TEST_FAILURE_percent}{13}
\pgfkeyssetvalue{o3_P_2_ERROR_MODEL_RESPONSE}{1}
\pgfkeyssetvalue{o3_P_2_ERROR_MODEL_RESPONSE_percent}{1}
\pgfkeyssetvalue{o3_P_2_NOT_REPAIRED}{1}
\pgfkeyssetvalue{o3_P_2_NOT_REPAIRED_percent}{1}
\pgfkeyssetvalue{deepseek_P_2_COMPILATION_FAILURE}{67}
\pgfkeyssetvalue{deepseek_P_2_COMPILATION_FAILURE_percent}{65}
\pgfkeyssetvalue{deepseek_P_2_BUILD_SUCCESS}{15}
\pgfkeyssetvalue{deepseek_P_2_BUILD_SUCCESS_percent}{15}
\pgfkeyssetvalue{deepseek_P_2_ERROR_MODEL_RESPONSE}{11}
\pgfkeyssetvalue{deepseek_P_2_ERROR_MODEL_RESPONSE_percent}{11}
\pgfkeyssetvalue{deepseek_P_2_TEST_FAILURE}{8}
\pgfkeyssetvalue{deepseek_P_2_TEST_FAILURE_percent}{8}
\pgfkeyssetvalue{deepseek_P_2_NOT_REPAIRED}{2}
\pgfkeyssetvalue{deepseek_P_2_NOT_REPAIRED_percent}{2}
\pgfkeyssetvalue{gpt_P_2_COMPILATION_FAILURE}{62}
\pgfkeyssetvalue{gpt_P_2_COMPILATION_FAILURE_percent}{60}
\pgfkeyssetvalue{gpt_P_2_BUILD_SUCCESS}{18}
\pgfkeyssetvalue{gpt_P_2_BUILD_SUCCESS_percent}{17}
\pgfkeyssetvalue{gpt_P_2_TEST_FAILURE}{14}
\pgfkeyssetvalue{gpt_P_2_TEST_FAILURE_percent}{14}
\pgfkeyssetvalue{gpt_P_2_NOT_REPAIRED}{8}
\pgfkeyssetvalue{gpt_P_2_NOT_REPAIRED_percent}{8}
\pgfkeyssetvalue{gpt_P_2_ERROR_MODEL_RESPONSE}{1}
\pgfkeyssetvalue{gpt_P_2_ERROR_MODEL_RESPONSE_percent}{1}
\pgfkeyssetvalue{qwen_P_2_COMPILATION_FAILURE}{78}
\pgfkeyssetvalue{qwen_P_2_COMPILATION_FAILURE_percent}{76}
\pgfkeyssetvalue{qwen_P_2_BUILD_SUCCESS}{11}
\pgfkeyssetvalue{qwen_P_2_BUILD_SUCCESS_percent}{11}
\pgfkeyssetvalue{qwen_P_2_ERROR_MODEL_RESPONSE}{7}
\pgfkeyssetvalue{qwen_P_2_ERROR_MODEL_RESPONSE_percent}{7}
\pgfkeyssetvalue{qwen_P_2_TEST_FAILURE}{6}
\pgfkeyssetvalue{qwen_P_2_TEST_FAILURE_percent}{6}
\pgfkeyssetvalue{qwen_P_2_NOT_REPAIRED}{1}
\pgfkeyssetvalue{qwen_P_2_NOT_REPAIRED_percent}{1}
\pgfkeyssetvalue{gemini_P_7_ERROR_MODEL_RESPONSE}{64}
\pgfkeyssetvalue{gemini_P_7_ERROR_MODEL_RESPONSE_percent}{62}
\pgfkeyssetvalue{gemini_P_7_COMPILATION_FAILURE}{24}
\pgfkeyssetvalue{gemini_P_7_COMPILATION_FAILURE_percent}{23}
\pgfkeyssetvalue{gemini_P_7_BUILD_SUCCESS}{7}
\pgfkeyssetvalue{gemini_P_7_BUILD_SUCCESS_percent}{7}
\pgfkeyssetvalue{gemini_P_7_TEST_FAILURE}{4}
\pgfkeyssetvalue{gemini_P_7_TEST_FAILURE_percent}{4}
\pgfkeyssetvalue{gemini_P_7_NOT_REPAIRED}{4}
\pgfkeyssetvalue{gemini_P_7_NOT_REPAIRED_percent}{4}
\pgfkeyssetvalue{o3_P_7_COMPILATION_FAILURE}{48}
\pgfkeyssetvalue{o3_P_7_COMPILATION_FAILURE_percent}{47}
\pgfkeyssetvalue{o3_P_7_BUILD_SUCCESS}{25}
\pgfkeyssetvalue{o3_P_7_BUILD_SUCCESS_percent}{24}
\pgfkeyssetvalue{o3_P_7_TEST_FAILURE}{17}
\pgfkeyssetvalue{o3_P_7_TEST_FAILURE_percent}{17}
\pgfkeyssetvalue{o3_P_7_ERROR_MODEL_RESPONSE}{12}
\pgfkeyssetvalue{o3_P_7_ERROR_MODEL_RESPONSE_percent}{12}
\pgfkeyssetvalue{o3_P_7_NOT_REPAIRED}{1}
\pgfkeyssetvalue{o3_P_7_NOT_REPAIRED_percent}{1}
\pgfkeyssetvalue{deepseek_P_7_COMPILATION_FAILURE}{54}
\pgfkeyssetvalue{deepseek_P_7_COMPILATION_FAILURE_percent}{52}
\pgfkeyssetvalue{deepseek_P_7_BUILD_SUCCESS}{20}
\pgfkeyssetvalue{deepseek_P_7_BUILD_SUCCESS_percent}{19}
\pgfkeyssetvalue{deepseek_P_7_ERROR_MODEL_RESPONSE}{16}
\pgfkeyssetvalue{deepseek_P_7_ERROR_MODEL_RESPONSE_percent}{16}
\pgfkeyssetvalue{deepseek_P_7_TEST_FAILURE}{10}
\pgfkeyssetvalue{deepseek_P_7_TEST_FAILURE_percent}{10}
\pgfkeyssetvalue{deepseek_P_7_NOT_REPAIRED}{3}
\pgfkeyssetvalue{deepseek_P_7_NOT_REPAIRED_percent}{3}
\pgfkeyssetvalue{gpt_P_7_COMPILATION_FAILURE}{73}
\pgfkeyssetvalue{gpt_P_7_COMPILATION_FAILURE_percent}{71}
\pgfkeyssetvalue{gpt_P_7_BUILD_SUCCESS}{14}
\pgfkeyssetvalue{gpt_P_7_BUILD_SUCCESS_percent}{14}
\pgfkeyssetvalue{gpt_P_7_NOT_REPAIRED}{6}
\pgfkeyssetvalue{gpt_P_7_NOT_REPAIRED_percent}{6}
\pgfkeyssetvalue{gpt_P_7_TEST_FAILURE}{6}
\pgfkeyssetvalue{gpt_P_7_TEST_FAILURE_percent}{6}
\pgfkeyssetvalue{gpt_P_7_ERROR_MODEL_RESPONSE}{4}
\pgfkeyssetvalue{gpt_P_7_ERROR_MODEL_RESPONSE_percent}{4}
\pgfkeyssetvalue{qwen_P_7_COMPILATION_FAILURE}{78}
\pgfkeyssetvalue{qwen_P_7_COMPILATION_FAILURE_percent}{76}
\pgfkeyssetvalue{qwen_P_7_BUILD_SUCCESS}{13}
\pgfkeyssetvalue{qwen_P_7_BUILD_SUCCESS_percent}{13}
\pgfkeyssetvalue{qwen_P_7_ERROR_MODEL_RESPONSE}{9}
\pgfkeyssetvalue{qwen_P_7_ERROR_MODEL_RESPONSE_percent}{9}
\pgfkeyssetvalue{qwen_P_7_TEST_FAILURE}{2}
\pgfkeyssetvalue{qwen_P_7_TEST_FAILURE_percent}{2}
\pgfkeyssetvalue{qwen_P_7_NOT_REPAIRED}{1}
\pgfkeyssetvalue{qwen_P_7_NOT_REPAIRED_percent}{1}
\pgfkeyssetvalue{gemini_P_5_COMPILATION_FAILURE}{60}
\pgfkeyssetvalue{gemini_P_5_COMPILATION_FAILURE_percent}{58}
\pgfkeyssetvalue{gemini_P_5_BUILD_SUCCESS}{16}
\pgfkeyssetvalue{gemini_P_5_BUILD_SUCCESS_percent}{16}
\pgfkeyssetvalue{gemini_P_5_TEST_FAILURE}{13}
\pgfkeyssetvalue{gemini_P_5_TEST_FAILURE_percent}{13}
\pgfkeyssetvalue{gemini_P_5_NOT_REPAIRED}{8}
\pgfkeyssetvalue{gemini_P_5_NOT_REPAIRED_percent}{8}
\pgfkeyssetvalue{gemini_P_5_ERROR_MODEL_RESPONSE}{6}
\pgfkeyssetvalue{gemini_P_5_ERROR_MODEL_RESPONSE_percent}{6}
\pgfkeyssetvalue{o3_P_5_COMPILATION_FAILURE}{68}
\pgfkeyssetvalue{o3_P_5_COMPILATION_FAILURE_percent}{66}
\pgfkeyssetvalue{o3_P_5_BUILD_SUCCESS}{19}
\pgfkeyssetvalue{o3_P_5_BUILD_SUCCESS_percent}{18}
\pgfkeyssetvalue{o3_P_5_TEST_FAILURE}{15}
\pgfkeyssetvalue{o3_P_5_TEST_FAILURE_percent}{15}
\pgfkeyssetvalue{o3_P_5_NOT_REPAIRED}{1}
\pgfkeyssetvalue{o3_P_5_NOT_REPAIRED_percent}{1}
\pgfkeyssetvalue{deepseek_P_5_COMPILATION_FAILURE}{58}
\pgfkeyssetvalue{deepseek_P_5_COMPILATION_FAILURE_percent}{56}
\pgfkeyssetvalue{deepseek_P_5_BUILD_SUCCESS}{20}
\pgfkeyssetvalue{deepseek_P_5_BUILD_SUCCESS_percent}{19}
\pgfkeyssetvalue{deepseek_P_5_TEST_FAILURE}{13}
\pgfkeyssetvalue{deepseek_P_5_TEST_FAILURE_percent}{13}
\pgfkeyssetvalue{deepseek_P_5_NOT_REPAIRED}{7}
\pgfkeyssetvalue{deepseek_P_5_NOT_REPAIRED_percent}{7}
\pgfkeyssetvalue{deepseek_P_5_ERROR_MODEL_RESPONSE}{5}
\pgfkeyssetvalue{deepseek_P_5_ERROR_MODEL_RESPONSE_percent}{5}
\pgfkeyssetvalue{gpt_P_5_COMPILATION_FAILURE}{71}
\pgfkeyssetvalue{gpt_P_5_COMPILATION_FAILURE_percent}{69}
\pgfkeyssetvalue{gpt_P_5_BUILD_SUCCESS}{14}
\pgfkeyssetvalue{gpt_P_5_BUILD_SUCCESS_percent}{14}
\pgfkeyssetvalue{gpt_P_5_NOT_REPAIRED}{12}
\pgfkeyssetvalue{gpt_P_5_NOT_REPAIRED_percent}{12}
\pgfkeyssetvalue{gpt_P_5_TEST_FAILURE}{6}
\pgfkeyssetvalue{gpt_P_5_TEST_FAILURE_percent}{6}
\pgfkeyssetvalue{qwen_P_5_COMPILATION_FAILURE}{81}
\pgfkeyssetvalue{qwen_P_5_COMPILATION_FAILURE_percent}{79}
\pgfkeyssetvalue{qwen_P_5_BUILD_SUCCESS}{11}
\pgfkeyssetvalue{qwen_P_5_BUILD_SUCCESS_percent}{11}
\pgfkeyssetvalue{qwen_P_5_ERROR_MODEL_RESPONSE}{6}
\pgfkeyssetvalue{qwen_P_5_ERROR_MODEL_RESPONSE_percent}{6}
\pgfkeyssetvalue{qwen_P_5_TEST_FAILURE}{4}
\pgfkeyssetvalue{qwen_P_5_TEST_FAILURE_percent}{4}
\pgfkeyssetvalue{qwen_P_5_NOT_REPAIRED}{1}
\pgfkeyssetvalue{qwen_P_5_NOT_REPAIRED_percent}{1}
\pgfkeyssetvalue{gemini_P_1_COMPILATION_FAILURE}{62}
\pgfkeyssetvalue{gemini_P_1_COMPILATION_FAILURE_percent}{60}
\pgfkeyssetvalue{gemini_P_1_BUILD_SUCCESS}{15}
\pgfkeyssetvalue{gemini_P_1_BUILD_SUCCESS_percent}{15}
\pgfkeyssetvalue{gemini_P_1_NOT_REPAIRED}{10}
\pgfkeyssetvalue{gemini_P_1_NOT_REPAIRED_percent}{10}
\pgfkeyssetvalue{gemini_P_1_TEST_FAILURE}{10}
\pgfkeyssetvalue{gemini_P_1_TEST_FAILURE_percent}{10}
\pgfkeyssetvalue{gemini_P_1_ERROR_MODEL_RESPONSE}{6}
\pgfkeyssetvalue{gemini_P_1_ERROR_MODEL_RESPONSE_percent}{6}
\pgfkeyssetvalue{o3_P_1_COMPILATION_FAILURE}{64}
\pgfkeyssetvalue{o3_P_1_COMPILATION_FAILURE_percent}{62}
\pgfkeyssetvalue{o3_P_1_BUILD_SUCCESS}{24}
\pgfkeyssetvalue{o3_P_1_BUILD_SUCCESS_percent}{23}
\pgfkeyssetvalue{o3_P_1_TEST_FAILURE}{14}
\pgfkeyssetvalue{o3_P_1_TEST_FAILURE_percent}{14}
\pgfkeyssetvalue{o3_P_1_NOT_REPAIRED}{1}
\pgfkeyssetvalue{o3_P_1_NOT_REPAIRED_percent}{1}
\pgfkeyssetvalue{deepseek_P_1_COMPILATION_FAILURE}{63}
\pgfkeyssetvalue{deepseek_P_1_COMPILATION_FAILURE_percent}{61}
\pgfkeyssetvalue{deepseek_P_1_BUILD_SUCCESS}{16}
\pgfkeyssetvalue{deepseek_P_1_BUILD_SUCCESS_percent}{16}
\pgfkeyssetvalue{deepseek_P_1_TEST_FAILURE}{10}
\pgfkeyssetvalue{deepseek_P_1_TEST_FAILURE_percent}{10}
\pgfkeyssetvalue{deepseek_P_1_ERROR_MODEL_RESPONSE}{10}
\pgfkeyssetvalue{deepseek_P_1_ERROR_MODEL_RESPONSE_percent}{10}
\pgfkeyssetvalue{deepseek_P_1_NOT_REPAIRED}{4}
\pgfkeyssetvalue{deepseek_P_1_NOT_REPAIRED_percent}{4}
\pgfkeyssetvalue{gpt_P_1_COMPILATION_FAILURE}{63}
\pgfkeyssetvalue{gpt_P_1_COMPILATION_FAILURE_percent}{61}
\pgfkeyssetvalue{gpt_P_1_BUILD_SUCCESS}{15}
\pgfkeyssetvalue{gpt_P_1_BUILD_SUCCESS_percent}{15}
\pgfkeyssetvalue{gpt_P_1_TEST_FAILURE}{12}
\pgfkeyssetvalue{gpt_P_1_TEST_FAILURE_percent}{12}
\pgfkeyssetvalue{gpt_P_1_NOT_REPAIRED}{9}
\pgfkeyssetvalue{gpt_P_1_NOT_REPAIRED_percent}{9}
\pgfkeyssetvalue{gpt_P_1_ERROR_MODEL_RESPONSE}{4}
\pgfkeyssetvalue{gpt_P_1_ERROR_MODEL_RESPONSE_percent}{4}
\pgfkeyssetvalue{qwen_P_1_COMPILATION_FAILURE}{74}
\pgfkeyssetvalue{qwen_P_1_COMPILATION_FAILURE_percent}{72}
\pgfkeyssetvalue{qwen_P_1_TEST_FAILURE}{11}
\pgfkeyssetvalue{qwen_P_1_TEST_FAILURE_percent}{11}
\pgfkeyssetvalue{qwen_P_1_BUILD_SUCCESS}{9}
\pgfkeyssetvalue{qwen_P_1_BUILD_SUCCESS_percent}{9}
\pgfkeyssetvalue{qwen_P_1_ERROR_MODEL_RESPONSE}{8}
\pgfkeyssetvalue{qwen_P_1_ERROR_MODEL_RESPONSE_percent}{8}
\pgfkeyssetvalue{qwen_P_1_NOT_REPAIRED}{1}
\pgfkeyssetvalue{qwen_P_1_NOT_REPAIRED_percent}{1}
\pgfkeyssetvalue{gemini_P_4_COMPILATION_FAILURE}{67}
\pgfkeyssetvalue{gemini_P_4_COMPILATION_FAILURE_percent}{65}
\pgfkeyssetvalue{gemini_P_4_BUILD_SUCCESS}{17}
\pgfkeyssetvalue{gemini_P_4_BUILD_SUCCESS_percent}{17}
\pgfkeyssetvalue{gemini_P_4_TEST_FAILURE}{9}
\pgfkeyssetvalue{gemini_P_4_TEST_FAILURE_percent}{9}
\pgfkeyssetvalue{gemini_P_4_ERROR_MODEL_RESPONSE}{7}
\pgfkeyssetvalue{gemini_P_4_ERROR_MODEL_RESPONSE_percent}{7}
\pgfkeyssetvalue{gemini_P_4_NOT_REPAIRED}{3}
\pgfkeyssetvalue{gemini_P_4_NOT_REPAIRED_percent}{3}
\pgfkeyssetvalue{o3_P_4_COMPILATION_FAILURE}{52}
\pgfkeyssetvalue{o3_P_4_COMPILATION_FAILURE_percent}{50}
\pgfkeyssetvalue{o3_P_4_BUILD_SUCCESS}{27}
\pgfkeyssetvalue{o3_P_4_BUILD_SUCCESS_percent}{26}
\pgfkeyssetvalue{o3_P_4_TEST_FAILURE}{21}
\pgfkeyssetvalue{o3_P_4_TEST_FAILURE_percent}{20}
\pgfkeyssetvalue{o3_P_4_ERROR_MODEL_RESPONSE}{2}
\pgfkeyssetvalue{o3_P_4_ERROR_MODEL_RESPONSE_percent}{2}
\pgfkeyssetvalue{o3_P_4_NOT_REPAIRED}{1}
\pgfkeyssetvalue{o3_P_4_NOT_REPAIRED_percent}{1}
\pgfkeyssetvalue{deepseek_P_4_COMPILATION_FAILURE}{61}
\pgfkeyssetvalue{deepseek_P_4_COMPILATION_FAILURE_percent}{59}
\pgfkeyssetvalue{deepseek_P_4_BUILD_SUCCESS}{14}
\pgfkeyssetvalue{deepseek_P_4_BUILD_SUCCESS_percent}{14}
\pgfkeyssetvalue{deepseek_P_4_ERROR_MODEL_RESPONSE}{14}
\pgfkeyssetvalue{deepseek_P_4_ERROR_MODEL_RESPONSE_percent}{14}
\pgfkeyssetvalue{deepseek_P_4_TEST_FAILURE}{13}
\pgfkeyssetvalue{deepseek_P_4_TEST_FAILURE_percent}{13}
\pgfkeyssetvalue{deepseek_P_4_NOT_REPAIRED}{1}
\pgfkeyssetvalue{deepseek_P_4_NOT_REPAIRED_percent}{1}
\pgfkeyssetvalue{gpt_P_4_COMPILATION_FAILURE}{77}
\pgfkeyssetvalue{gpt_P_4_COMPILATION_FAILURE_percent}{75}
\pgfkeyssetvalue{gpt_P_4_BUILD_SUCCESS}{16}
\pgfkeyssetvalue{gpt_P_4_BUILD_SUCCESS_percent}{16}
\pgfkeyssetvalue{gpt_P_4_TEST_FAILURE}{5}
\pgfkeyssetvalue{gpt_P_4_TEST_FAILURE_percent}{5}
\pgfkeyssetvalue{gpt_P_4_ERROR_MODEL_RESPONSE}{4}
\pgfkeyssetvalue{gpt_P_4_ERROR_MODEL_RESPONSE_percent}{4}
\pgfkeyssetvalue{gpt_P_4_NOT_REPAIRED}{1}
\pgfkeyssetvalue{gpt_P_4_NOT_REPAIRED_percent}{1}
\pgfkeyssetvalue{qwen_P_4_COMPILATION_FAILURE}{78}
\pgfkeyssetvalue{qwen_P_4_COMPILATION_FAILURE_percent}{76}
\pgfkeyssetvalue{qwen_P_4_BUILD_SUCCESS}{11}
\pgfkeyssetvalue{qwen_P_4_BUILD_SUCCESS_percent}{11}
\pgfkeyssetvalue{qwen_P_4_ERROR_MODEL_RESPONSE}{8}
\pgfkeyssetvalue{qwen_P_4_ERROR_MODEL_RESPONSE_percent}{8}
\pgfkeyssetvalue{qwen_P_4_TEST_FAILURE}{5}
\pgfkeyssetvalue{qwen_P_4_TEST_FAILURE_percent}{5}
\pgfkeyssetvalue{qwen_P_4_NOT_REPAIRED}{1}
\pgfkeyssetvalue{qwen_P_4_NOT_REPAIRED_percent}{1}
\pgfkeyssetvalue{gemini_P_3_COMPILATION_FAILURE}{61}
\pgfkeyssetvalue{gemini_P_3_COMPILATION_FAILURE_percent}{59}
\pgfkeyssetvalue{gemini_P_3_BUILD_SUCCESS}{21}
\pgfkeyssetvalue{gemini_P_3_BUILD_SUCCESS_percent}{20}
\pgfkeyssetvalue{gemini_P_3_TEST_FAILURE}{12}
\pgfkeyssetvalue{gemini_P_3_TEST_FAILURE_percent}{12}
\pgfkeyssetvalue{gemini_P_3_ERROR_MODEL_RESPONSE}{6}
\pgfkeyssetvalue{gemini_P_3_ERROR_MODEL_RESPONSE_percent}{6}
\pgfkeyssetvalue{gemini_P_3_NOT_REPAIRED}{3}
\pgfkeyssetvalue{gemini_P_3_NOT_REPAIRED_percent}{3}
\pgfkeyssetvalue{o3_P_3_COMPILATION_FAILURE}{58}
\pgfkeyssetvalue{o3_P_3_COMPILATION_FAILURE_percent}{56}
\pgfkeyssetvalue{o3_P_3_BUILD_SUCCESS}{26}
\pgfkeyssetvalue{o3_P_3_BUILD_SUCCESS_percent}{25}
\pgfkeyssetvalue{o3_P_3_TEST_FAILURE}{16}
\pgfkeyssetvalue{o3_P_3_TEST_FAILURE_percent}{16}
\pgfkeyssetvalue{o3_P_3_ERROR_MODEL_RESPONSE}{2}
\pgfkeyssetvalue{o3_P_3_ERROR_MODEL_RESPONSE_percent}{2}
\pgfkeyssetvalue{o3_P_3_NOT_REPAIRED}{1}
\pgfkeyssetvalue{o3_P_3_NOT_REPAIRED_percent}{1}
\pgfkeyssetvalue{deepseek_P_3_COMPILATION_FAILURE}{53}
\pgfkeyssetvalue{deepseek_P_3_COMPILATION_FAILURE_percent}{51}
\pgfkeyssetvalue{deepseek_P_3_BUILD_SUCCESS}{19}
\pgfkeyssetvalue{deepseek_P_3_BUILD_SUCCESS_percent}{18}
\pgfkeyssetvalue{deepseek_P_3_ERROR_MODEL_RESPONSE}{16}
\pgfkeyssetvalue{deepseek_P_3_ERROR_MODEL_RESPONSE_percent}{16}
\pgfkeyssetvalue{deepseek_P_3_TEST_FAILURE}{14}
\pgfkeyssetvalue{deepseek_P_3_TEST_FAILURE_percent}{14}
\pgfkeyssetvalue{deepseek_P_3_NOT_REPAIRED}{1}
\pgfkeyssetvalue{deepseek_P_3_NOT_REPAIRED_percent}{1}
\pgfkeyssetvalue{gpt_P_3_COMPILATION_FAILURE}{81}
\pgfkeyssetvalue{gpt_P_3_COMPILATION_FAILURE_percent}{79}
\pgfkeyssetvalue{gpt_P_3_BUILD_SUCCESS}{13}
\pgfkeyssetvalue{gpt_P_3_BUILD_SUCCESS_percent}{13}
\pgfkeyssetvalue{gpt_P_3_TEST_FAILURE}{4}
\pgfkeyssetvalue{gpt_P_3_TEST_FAILURE_percent}{4}
\pgfkeyssetvalue{gpt_P_3_ERROR_MODEL_RESPONSE}{3}
\pgfkeyssetvalue{gpt_P_3_ERROR_MODEL_RESPONSE_percent}{3}
\pgfkeyssetvalue{gpt_P_3_NOT_REPAIRED}{2}
\pgfkeyssetvalue{gpt_P_3_NOT_REPAIRED_percent}{2}
\pgfkeyssetvalue{qwen_P_3_COMPILATION_FAILURE}{82}
\pgfkeyssetvalue{qwen_P_3_COMPILATION_FAILURE_percent}{80}
\pgfkeyssetvalue{qwen_P_3_BUILD_SUCCESS}{9}
\pgfkeyssetvalue{qwen_P_3_BUILD_SUCCESS_percent}{9}
\pgfkeyssetvalue{qwen_P_3_ERROR_MODEL_RESPONSE}{7}
\pgfkeyssetvalue{qwen_P_3_ERROR_MODEL_RESPONSE_percent}{7}
\pgfkeyssetvalue{qwen_P_3_TEST_FAILURE}{4}
\pgfkeyssetvalue{qwen_P_3_TEST_FAILURE_percent}{4}
\pgfkeyssetvalue{qwen_P_3_NOT_REPAIRED}{1}
\pgfkeyssetvalue{qwen_P_3_NOT_REPAIRED_percent}{1}

%% file: rq3-values.tex
\pgfkeyssetvalue{P8_gemini-2.0-flash-001_mean_relative_fixed}{2.40}
\pgfkeyssetvalue{P8_gemini-2.0-flash-001_median_relative_fixed}{40.00}
\pgfkeyssetvalue{P8_gemini-2.0-flash-001_max_relative_fixed}{100.00}
\pgfkeyssetvalue{P8_gemini-2.0-flash-001_min_relative_fixed}{-2900.00}
\pgfkeyssetvalue{P8_o3-mini-2025-01-31_mean_relative_fixed}{-44.63}
\pgfkeyssetvalue{P8_o3-mini-2025-01-31_median_relative_fixed}{80.00}
\pgfkeyssetvalue{P8_o3-mini-2025-01-31_max_relative_fixed}{100.00}
\pgfkeyssetvalue{P8_o3-mini-2025-01-31_min_relative_fixed}{-4400.00}
\pgfkeyssetvalue{P8_deepseek-deepseek-chat_mean_relative_fixed}{11.48}
\pgfkeyssetvalue{P8_deepseek-deepseek-chat_median_relative_fixed}{80.00}
\pgfkeyssetvalue{P8_deepseek-deepseek-chat_max_relative_fixed}{100.00}
\pgfkeyssetvalue{P8_deepseek-deepseek-chat_min_relative_fixed}{-2900.00}
\pgfkeyssetvalue{P8_gpt-4o-mini_mean_relative_fixed}{-7.08}
\pgfkeyssetvalue{P8_gpt-4o-mini_median_relative_fixed}{42.86}
\pgfkeyssetvalue{P8_gpt-4o-mini_max_relative_fixed}{100.00}
\pgfkeyssetvalue{P8_gpt-4o-mini_min_relative_fixed}{-2900.00}
\pgfkeyssetvalue{P8_qwen-qwen2.5-32b-instruct_mean_relative_fixed}{-136.05}
\pgfkeyssetvalue{P8_qwen-qwen2.5-32b-instruct_median_relative_fixed}{0.00}
\pgfkeyssetvalue{P8_qwen-qwen2.5-32b-instruct_max_relative_fixed}{100.00}
\pgfkeyssetvalue{P8_qwen-qwen2.5-32b-instruct_min_relative_fixed}{-2900.00}
\pgfkeyssetvalue{P6_gemini-2.0-flash-001_mean_relative_fixed}{5.29}
\pgfkeyssetvalue{P6_gemini-2.0-flash-001_median_relative_fixed}{50.00}
\pgfkeyssetvalue{P6_gemini-2.0-flash-001_max_relative_fixed}{100.00}
\pgfkeyssetvalue{P6_gemini-2.0-flash-001_min_relative_fixed}{-2900.00}
\pgfkeyssetvalue{P6_o3-mini-2025-01-31_mean_relative_fixed}{-74.17}
\pgfkeyssetvalue{P6_o3-mini-2025-01-31_median_relative_fixed}{80.00}
\pgfkeyssetvalue{P6_o3-mini-2025-01-31_max_relative_fixed}{100.00}
\pgfkeyssetvalue{P6_o3-mini-2025-01-31_min_relative_fixed}{-7200.00}
\pgfkeyssetvalue{P6_deepseek-deepseek-chat_mean_relative_fixed}{-18.95}
\pgfkeyssetvalue{P6_deepseek-deepseek-chat_median_relative_fixed}{38.00}
\pgfkeyssetvalue{P6_deepseek-deepseek-chat_max_relative_fixed}{100.00}
\pgfkeyssetvalue{P6_deepseek-deepseek-chat_min_relative_fixed}{-2900.00}
\pgfkeyssetvalue{P6_gpt-4o-mini_mean_relative_fixed}{11.78}
\pgfkeyssetvalue{P6_gpt-4o-mini_median_relative_fixed}{40.00}
\pgfkeyssetvalue{P6_gpt-4o-mini_max_relative_fixed}{100.00}
\pgfkeyssetvalue{P6_gpt-4o-mini_min_relative_fixed}{-2900.00}
\pgfkeyssetvalue{P6_qwen-qwen2.5-32b-instruct_mean_relative_fixed}{-542.02}
\pgfkeyssetvalue{P6_qwen-qwen2.5-32b-instruct_median_relative_fixed}{0.00}
\pgfkeyssetvalue{P6_qwen-qwen2.5-32b-instruct_max_relative_fixed}{100.00}
\pgfkeyssetvalue{P6_qwen-qwen2.5-32b-instruct_min_relative_fixed}{-8200.00}
\pgfkeyssetvalue{P2_gemini-2.0-flash-001_mean_relative_fixed}{5.76}
\pgfkeyssetvalue{P2_gemini-2.0-flash-001_median_relative_fixed}{79.17}
\pgfkeyssetvalue{P2_gemini-2.0-flash-001_max_relative_fixed}{100.00}
\pgfkeyssetvalue{P2_gemini-2.0-flash-001_min_relative_fixed}{-2900.00}
\pgfkeyssetvalue{P2_o3-mini-2025-01-31_mean_relative_fixed}{-81.36}
\pgfkeyssetvalue{P2_o3-mini-2025-01-31_median_relative_fixed}{80.00}
\pgfkeyssetvalue{P2_o3-mini-2025-01-31_max_relative_fixed}{100.00}
\pgfkeyssetvalue{P2_o3-mini-2025-01-31_min_relative_fixed}{-9900.00}
\pgfkeyssetvalue{P2_deepseek-deepseek-chat_mean_relative_fixed}{-1.59}
\pgfkeyssetvalue{P2_deepseek-deepseek-chat_median_relative_fixed}{66.67}
\pgfkeyssetvalue{P2_deepseek-deepseek-chat_max_relative_fixed}{100.00}
\pgfkeyssetvalue{P2_deepseek-deepseek-chat_min_relative_fixed}{-2900.00}
\pgfkeyssetvalue{P2_gpt-4o-mini_mean_relative_fixed}{8.67}
\pgfkeyssetvalue{P2_gpt-4o-mini_median_relative_fixed}{50.00}
\pgfkeyssetvalue{P2_gpt-4o-mini_max_relative_fixed}{100.00}
\pgfkeyssetvalue{P2_gpt-4o-mini_min_relative_fixed}{-2900.00}
\pgfkeyssetvalue{P2_qwen-qwen2.5-32b-instruct_mean_relative_fixed}{-186.88}
\pgfkeyssetvalue{P2_qwen-qwen2.5-32b-instruct_median_relative_fixed}{0.00}
\pgfkeyssetvalue{P2_qwen-qwen2.5-32b-instruct_max_relative_fixed}{100.00}
\pgfkeyssetvalue{P2_qwen-qwen2.5-32b-instruct_min_relative_fixed}{-5100.00}
\pgfkeyssetvalue{P7_gemini-2.0-flash-001_mean_relative_fixed}{9.44}
\pgfkeyssetvalue{P7_gemini-2.0-flash-001_median_relative_fixed}{40.00}
\pgfkeyssetvalue{P7_gemini-2.0-flash-001_max_relative_fixed}{100.00}
\pgfkeyssetvalue{P7_gemini-2.0-flash-001_min_relative_fixed}{-2900.00}
\pgfkeyssetvalue{P7_o3-mini-2025-01-31_mean_relative_fixed}{-109.91}
\pgfkeyssetvalue{P7_o3-mini-2025-01-31_median_relative_fixed}{85.71}
\pgfkeyssetvalue{P7_o3-mini-2025-01-31_max_relative_fixed}{100.00}
\pgfkeyssetvalue{P7_o3-mini-2025-01-31_min_relative_fixed}{-9900.00}
\pgfkeyssetvalue{P7_deepseek-deepseek-chat_mean_relative_fixed}{-2.64}
\pgfkeyssetvalue{P7_deepseek-deepseek-chat_median_relative_fixed}{60.00}
\pgfkeyssetvalue{P7_deepseek-deepseek-chat_max_relative_fixed}{100.00}
\pgfkeyssetvalue{P7_deepseek-deepseek-chat_min_relative_fixed}{-2900.00}
\pgfkeyssetvalue{P7_gpt-4o-mini_mean_relative_fixed}{0.56}
\pgfkeyssetvalue{P7_gpt-4o-mini_median_relative_fixed}{50.00}
\pgfkeyssetvalue{P7_gpt-4o-mini_max_relative_fixed}{100.00}
\pgfkeyssetvalue{P7_gpt-4o-mini_min_relative_fixed}{-2900.00}
\pgfkeyssetvalue{P7_qwen-qwen2.5-32b-instruct_mean_relative_fixed}{-211.98}
\pgfkeyssetvalue{P7_qwen-qwen2.5-32b-instruct_median_relative_fixed}{0.00}
\pgfkeyssetvalue{P7_qwen-qwen2.5-32b-instruct_max_relative_fixed}{100.00}
\pgfkeyssetvalue{P7_qwen-qwen2.5-32b-instruct_min_relative_fixed}{-3600.00}
\pgfkeyssetvalue{P5_gemini-2.0-flash-001_mean_relative_fixed}{-9.40}
\pgfkeyssetvalue{P5_gemini-2.0-flash-001_median_relative_fixed}{20.00}
\pgfkeyssetvalue{P5_gemini-2.0-flash-001_max_relative_fixed}{100.00}
\pgfkeyssetvalue{P5_gemini-2.0-flash-001_min_relative_fixed}{-2900.00}
\pgfkeyssetvalue{P5_o3-mini-2025-01-31_mean_relative_fixed}{6.38}
\pgfkeyssetvalue{P5_o3-mini-2025-01-31_median_relative_fixed}{79.17}
\pgfkeyssetvalue{P5_o3-mini-2025-01-31_max_relative_fixed}{100.00}
\pgfkeyssetvalue{P5_o3-mini-2025-01-31_min_relative_fixed}{-2900.00}
\pgfkeyssetvalue{P5_deepseek-deepseek-chat_mean_relative_fixed}{-22.73}
\pgfkeyssetvalue{P5_deepseek-deepseek-chat_median_relative_fixed}{50.00}
\pgfkeyssetvalue{P5_deepseek-deepseek-chat_max_relative_fixed}{100.00}
\pgfkeyssetvalue{P5_deepseek-deepseek-chat_min_relative_fixed}{-2900.00}
\pgfkeyssetvalue{P5_gpt-4o-mini_mean_relative_fixed}{6.28}
\pgfkeyssetvalue{P5_gpt-4o-mini_median_relative_fixed}{20.83}
\pgfkeyssetvalue{P5_gpt-4o-mini_max_relative_fixed}{100.00}
\pgfkeyssetvalue{P5_gpt-4o-mini_min_relative_fixed}{-2900.00}
\pgfkeyssetvalue{P5_qwen-qwen2.5-32b-instruct_mean_relative_fixed}{-276.71}
\pgfkeyssetvalue{P5_qwen-qwen2.5-32b-instruct_median_relative_fixed}{0.00}
\pgfkeyssetvalue{P5_qwen-qwen2.5-32b-instruct_max_relative_fixed}{100.00}
\pgfkeyssetvalue{P5_qwen-qwen2.5-32b-instruct_min_relative_fixed}{-5000.00}
\pgfkeyssetvalue{P1_gemini-2.0-flash-001_mean_relative_fixed}{2.66}
\pgfkeyssetvalue{P1_gemini-2.0-flash-001_median_relative_fixed}{50.00}
\pgfkeyssetvalue{P1_gemini-2.0-flash-001_max_relative_fixed}{100.00}
\pgfkeyssetvalue{P1_gemini-2.0-flash-001_min_relative_fixed}{-2900.00}
\pgfkeyssetvalue{P1_o3-mini-2025-01-31_mean_relative_fixed}{-84.12}
\pgfkeyssetvalue{P1_o3-mini-2025-01-31_median_relative_fixed}{80.00}
\pgfkeyssetvalue{P1_o3-mini-2025-01-31_max_relative_fixed}{100.00}
\pgfkeyssetvalue{P1_o3-mini-2025-01-31_min_relative_fixed}{-8600.00}
\pgfkeyssetvalue{P1_deepseek-deepseek-chat_mean_relative_fixed}{-8.62}
\pgfkeyssetvalue{P1_deepseek-deepseek-chat_median_relative_fixed}{50.00}
\pgfkeyssetvalue{P1_deepseek-deepseek-chat_max_relative_fixed}{100.00}
\pgfkeyssetvalue{P1_deepseek-deepseek-chat_min_relative_fixed}{-2900.00}
\pgfkeyssetvalue{P1_gpt-4o-mini_mean_relative_fixed}{6.86}
\pgfkeyssetvalue{P1_gpt-4o-mini_median_relative_fixed}{25.00}
\pgfkeyssetvalue{P1_gpt-4o-mini_max_relative_fixed}{100.00}
\pgfkeyssetvalue{P1_gpt-4o-mini_min_relative_fixed}{-2900.00}
\pgfkeyssetvalue{P1_qwen-qwen2.5-32b-instruct_mean_relative_fixed}{-359.38}
\pgfkeyssetvalue{P1_qwen-qwen2.5-32b-instruct_median_relative_fixed}{0.00}
\pgfkeyssetvalue{P1_qwen-qwen2.5-32b-instruct_max_relative_fixed}{100.00}
\pgfkeyssetvalue{P1_qwen-qwen2.5-32b-instruct_min_relative_fixed}{-7400.00}
\pgfkeyssetvalue{P4_gemini-2.0-flash-001_mean_relative_fixed}{-3.51}
\pgfkeyssetvalue{P4_gemini-2.0-flash-001_median_relative_fixed}{40.00}
\pgfkeyssetvalue{P4_gemini-2.0-flash-001_max_relative_fixed}{100.00}
\pgfkeyssetvalue{P4_gemini-2.0-flash-001_min_relative_fixed}{-2900.00}
\pgfkeyssetvalue{P4_o3-mini-2025-01-31_mean_relative_fixed}{-79.39}
\pgfkeyssetvalue{P4_o3-mini-2025-01-31_median_relative_fixed}{93.33}
\pgfkeyssetvalue{P4_o3-mini-2025-01-31_max_relative_fixed}{100.00}
\pgfkeyssetvalue{P4_o3-mini-2025-01-31_min_relative_fixed}{-9800.00}
\pgfkeyssetvalue{P4_deepseek-deepseek-chat_mean_relative_fixed}{3.06}
\pgfkeyssetvalue{P4_deepseek-deepseek-chat_median_relative_fixed}{45.00}
\pgfkeyssetvalue{P4_deepseek-deepseek-chat_max_relative_fixed}{100.00}
\pgfkeyssetvalue{P4_deepseek-deepseek-chat_min_relative_fixed}{-2900.00}
\pgfkeyssetvalue{P4_gpt-4o-mini_mean_relative_fixed}{-15.25}
\pgfkeyssetvalue{P4_gpt-4o-mini_median_relative_fixed}{33.33}
\pgfkeyssetvalue{P4_gpt-4o-mini_max_relative_fixed}{100.00}
\pgfkeyssetvalue{P4_gpt-4o-mini_min_relative_fixed}{-2900.00}
\pgfkeyssetvalue{P4_qwen-qwen2.5-32b-instruct_mean_relative_fixed}{-138.92}
\pgfkeyssetvalue{P4_qwen-qwen2.5-32b-instruct_median_relative_fixed}{0.00}
\pgfkeyssetvalue{P4_qwen-qwen2.5-32b-instruct_max_relative_fixed}{100.00}
\pgfkeyssetvalue{P4_qwen-qwen2.5-32b-instruct_min_relative_fixed}{-2900.00}
\pgfkeyssetvalue{P3_gemini-2.0-flash-001_mean_relative_fixed}{6.12}
\pgfkeyssetvalue{P3_gemini-2.0-flash-001_median_relative_fixed}{50.00}
\pgfkeyssetvalue{P3_gemini-2.0-flash-001_max_relative_fixed}{100.00}
\pgfkeyssetvalue{P3_gemini-2.0-flash-001_min_relative_fixed}{-2900.00}
\pgfkeyssetvalue{P3_o3-mini-2025-01-31_mean_relative_fixed}{-20.31}
\pgfkeyssetvalue{P3_o3-mini-2025-01-31_median_relative_fixed}{66.67}
\pgfkeyssetvalue{P3_o3-mini-2025-01-31_max_relative_fixed}{100.00}
\pgfkeyssetvalue{P3_o3-mini-2025-01-31_min_relative_fixed}{-2900.00}
\pgfkeyssetvalue{P3_deepseek-deepseek-chat_mean_relative_fixed}{-4.31}
\pgfkeyssetvalue{P3_deepseek-deepseek-chat_median_relative_fixed}{60.00}
\pgfkeyssetvalue{P3_deepseek-deepseek-chat_max_relative_fixed}{100.00}
\pgfkeyssetvalue{P3_deepseek-deepseek-chat_min_relative_fixed}{-2900.00}
\pgfkeyssetvalue{P3_gpt-4o-mini_mean_relative_fixed}{4.19}
\pgfkeyssetvalue{P3_gpt-4o-mini_median_relative_fixed}{33.33}
\pgfkeyssetvalue{P3_gpt-4o-mini_max_relative_fixed}{100.00}
\pgfkeyssetvalue{P3_gpt-4o-mini_min_relative_fixed}{-2900.00}
\pgfkeyssetvalue{P3_qwen-qwen2.5-32b-instruct_mean_relative_fixed}{-139.49}
\pgfkeyssetvalue{P3_qwen-qwen2.5-32b-instruct_median_relative_fixed}{0.00}
\pgfkeyssetvalue{P3_qwen-qwen2.5-32b-instruct_max_relative_fixed}{100.00}
\pgfkeyssetvalue{P3_qwen-qwen2.5-32b-instruct_min_relative_fixed}{-2900.00}

%% file: rq3-tab2.tex
\newcommand{\relativeerrortab}{%
\begin{table}[t!]
\centering
\rowcolors{2}{gray!10}{white}
\caption{Median of Relative Error Fixed Ratio of \toolname}
\label{tab:releative_error_prompt}
\resizebox{\textwidth}{!}{%

\begin{tabular}{lrrrrr}
\toprule
   \makecell[l]{\textbf{Prompt}\\ID} & \makecell[l]{\textbf{Deepseek}\\V3} & \makecell[l]{\textbf{Gemini}\\2.0-flash} & \makecell[l]{\textbf{Gpt}\\4o-mini} & \makecell[l]{\textbf{o3}\\mini} & \makecell[l]{\textbf{Qwen2.5}\\32b-instruct} \\
  \midrule
$P_1$ & \pgfkeysvalueof{P1_deepseek-deepseek-chat_median_relative_fixed}\% & \pgfkeysvalueof{P1_gemini-2.0-flash-001_median_relative_fixed}\% & \pgfkeysvalueof{P1_gpt-4o-mini_median_relative_fixed}\% & \pgfkeysvalueof{P1_o3-mini-2025-01-31_median_relative_fixed}\% & \pgfkeysvalueof{P1_qwen-qwen2.5-32b-instruct_median_relative_fixed}\% \\
$P_2$ & \pgfkeysvalueof{P2_deepseek-deepseek-chat_median_relative_fixed}\% & \pgfkeysvalueof{P2_gemini-2.0-flash-001_median_relative_fixed}\% & \pgfkeysvalueof{P2_gpt-4o-mini_median_relative_fixed}\% & \pgfkeysvalueof{P2_o3-mini-2025-01-31_median_relative_fixed}\% & \pgfkeysvalueof{P2_qwen-qwen2.5-32b-instruct_median_relative_fixed}\% \\
$P_3$ & \pgfkeysvalueof{P3_deepseek-deepseek-chat_median_relative_fixed}\% & \pgfkeysvalueof{P3_gemini-2.0-flash-001_median_relative_fixed}\% & \pgfkeysvalueof{P3_gpt-4o-mini_median_relative_fixed}\% & \pgfkeysvalueof{P3_o3-mini-2025-01-31_median_relative_fixed}\% & \pgfkeysvalueof{P3_qwen-qwen2.5-32b-instruct_median_relative_fixed}\% \\
$P_4$ & \pgfkeysvalueof{P4_deepseek-deepseek-chat_median_relative_fixed}\% & \pgfkeysvalueof{P4_gemini-2.0-flash-001_median_relative_fixed}\% & \pgfkeysvalueof{P4_gpt-4o-mini_median_relative_fixed}\% & \textbf{\pgfkeysvalueof{P4_o3-mini-2025-01-31_median_relative_fixed}}\% & \pgfkeysvalueof{P4_qwen-qwen2.5-32b-instruct_median_relative_fixed}\% \\
$P_5$ & \pgfkeysvalueof{P5_deepseek-deepseek-chat_median_relative_fixed}\% & \pgfkeysvalueof{P5_gemini-2.0-flash-001_median_relative_fixed}\% & \pgfkeysvalueof{P5_gpt-4o-mini_median_relative_fixed}\% & \pgfkeysvalueof{P5_o3-mini-2025-01-31_median_relative_fixed}\% & \pgfkeysvalueof{P5_qwen-qwen2.5-32b-instruct_median_relative_fixed}\% \\
$P_6$ & \pgfkeysvalueof{P6_deepseek-deepseek-chat_median_relative_fixed}\% & \pgfkeysvalueof{P6_gemini-2.0-flash-001_median_relative_fixed}\% & \pgfkeysvalueof{P6_gpt-4o-mini_median_relative_fixed}\% & \pgfkeysvalueof{P6_o3-mini-2025-01-31_median_relative_fixed}\% & \pgfkeysvalueof{P6_qwen-qwen2.5-32b-instruct_median_relative_fixed}\% \\
$P_7$ & \pgfkeysvalueof{P7_deepseek-deepseek-chat_median_relative_fixed}\% & \pgfkeysvalueof{P7_gemini-2.0-flash-001_median_relative_fixed}\% & \pgfkeysvalueof{P7_gpt-4o-mini_median_relative_fixed}\% & \pgfkeysvalueof{P7_o3-mini-2025-01-31_median_relative_fixed}\% & \pgfkeysvalueof{P7_qwen-qwen2.5-32b-instruct_median_relative_fixed}\% \\
$P_8$ & \pgfkeysvalueof{P8_deepseek-deepseek-chat_median_relative_fixed}\% & \pgfkeysvalueof{P8_gemini-2.0-flash-001_median_relative_fixed}\% & \pgfkeysvalueof{P8_gpt-4o-mini_median_relative_fixed}\% & \pgfkeysvalueof{P8_o3-mini-2025-01-31_median_relative_fixed}\% & \pgfkeysvalueof{P8_qwen-qwen2.5-32b-instruct_median_relative_fixed}\% \\
\bottomrule
\end{tabular}}
\end{table}
}

%% file: rq2-pgfkeys_file-level.tex
\pgfkeyssetvalue{P_8_gemini-2.0-flash-001_total_files}{250}
\pgfkeyssetvalue{P_8_gemini-2.0-flash-001_fixed}{76}
\pgfkeyssetvalue{P_8_gemini-2.0-flash-001_fixed_percentage}{30}
\pgfkeyssetvalue{P_8_o3-mini-2025-01-31_total_files}{246}
\pgfkeyssetvalue{P_8_o3-mini-2025-01-31_fixed}{92}
\pgfkeyssetvalue{P_8_o3-mini-2025-01-31_fixed_percentage}{37}
\pgfkeyssetvalue{P_8_deepseek-deepseek-chat_total_files}{248}
\pgfkeyssetvalue{P_8_deepseek-deepseek-chat_fixed}{71}
\pgfkeyssetvalue{P_8_deepseek-deepseek-chat_fixed_percentage}{29}
\pgfkeyssetvalue{P_8_gpt-4o-mini_total_files}{256}
\pgfkeyssetvalue{P_8_gpt-4o-mini_fixed}{53}
\pgfkeyssetvalue{P_8_gpt-4o-mini_fixed_percentage}{21}
\pgfkeyssetvalue{P_8_qwen-qwen2.5-32b-instruct_total_files}{267}
\pgfkeyssetvalue{P_8_qwen-qwen2.5-32b-instruct_fixed}{54}
\pgfkeyssetvalue{P_8_qwen-qwen2.5-32b-instruct_fixed_percentage}{20}
\pgfkeyssetvalue{P_6_gemini-2.0-flash-001_total_files}{243}
\pgfkeyssetvalue{P_6_gemini-2.0-flash-001_fixed}{39}
\pgfkeyssetvalue{P_6_gemini-2.0-flash-001_fixed_percentage}{16}
\pgfkeyssetvalue{P_6_o3-mini-2025-01-31_total_files}{243}
\pgfkeyssetvalue{P_6_o3-mini-2025-01-31_fixed}{74}
\pgfkeyssetvalue{P_6_o3-mini-2025-01-31_fixed_percentage}{30}
\pgfkeyssetvalue{P_6_deepseek-deepseek-chat_total_files}{248}
\pgfkeyssetvalue{P_6_deepseek-deepseek-chat_fixed}{42}
\pgfkeyssetvalue{P_6_deepseek-deepseek-chat_fixed_percentage}{17}
\pgfkeyssetvalue{P_6_gpt-4o-mini_total_files}{254}
\pgfkeyssetvalue{P_6_gpt-4o-mini_fixed}{40}
\pgfkeyssetvalue{P_6_gpt-4o-mini_fixed_percentage}{16}
\pgfkeyssetvalue{P_6_qwen-qwen2.5-32b-instruct_total_files}{263}
\pgfkeyssetvalue{P_6_qwen-qwen2.5-32b-instruct_fixed}{62}
\pgfkeyssetvalue{P_6_qwen-qwen2.5-32b-instruct_fixed_percentage}{24}
\pgfkeyssetvalue{P_2_gemini-2.0-flash-001_total_files}{248}
\pgfkeyssetvalue{P_2_gemini-2.0-flash-001_fixed}{54}
\pgfkeyssetvalue{P_2_gemini-2.0-flash-001_fixed_percentage}{22}
\pgfkeyssetvalue{P_2_o3-mini-2025-01-31_total_files}{241}
\pgfkeyssetvalue{P_2_o3-mini-2025-01-31_fixed}{66}
\pgfkeyssetvalue{P_2_o3-mini-2025-01-31_fixed_percentage}{27}
\pgfkeyssetvalue{P_2_deepseek-deepseek-chat_total_files}{254}
\pgfkeyssetvalue{P_2_deepseek-deepseek-chat_fixed}{61}
\pgfkeyssetvalue{P_2_deepseek-deepseek-chat_fixed_percentage}{24}
\pgfkeyssetvalue{P_2_gpt-4o-mini_total_files}{253}
\pgfkeyssetvalue{P_2_gpt-4o-mini_fixed}{46}
\pgfkeyssetvalue{P_2_gpt-4o-mini_fixed_percentage}{18}
\pgfkeyssetvalue{P_2_qwen-qwen2.5-32b-instruct_total_files}{263}
\pgfkeyssetvalue{P_2_qwen-qwen2.5-32b-instruct_fixed}{44}
\pgfkeyssetvalue{P_2_qwen-qwen2.5-32b-instruct_fixed_percentage}{17}
\pgfkeyssetvalue{P_7_gemini-2.0-flash-001_total_files}{253}
\pgfkeyssetvalue{P_7_gemini-2.0-flash-001_fixed}{71}
\pgfkeyssetvalue{P_7_gemini-2.0-flash-001_fixed_percentage}{28}
\pgfkeyssetvalue{P_7_o3-mini-2025-01-31_total_files}{238}
\pgfkeyssetvalue{P_7_o3-mini-2025-01-31_fixed}{88}
\pgfkeyssetvalue{P_7_o3-mini-2025-01-31_fixed_percentage}{37}
\pgfkeyssetvalue{P_7_deepseek-deepseek-chat_total_files}{244}
\pgfkeyssetvalue{P_7_deepseek-deepseek-chat_fixed}{59}
\pgfkeyssetvalue{P_7_deepseek-deepseek-chat_fixed_percentage}{24}
\pgfkeyssetvalue{P_7_gpt-4o-mini_total_files}{260}
\pgfkeyssetvalue{P_7_gpt-4o-mini_fixed}{57}
\pgfkeyssetvalue{P_7_gpt-4o-mini_fixed_percentage}{22}
\pgfkeyssetvalue{P_7_qwen-qwen2.5-32b-instruct_total_files}{264}
\pgfkeyssetvalue{P_7_qwen-qwen2.5-32b-instruct_fixed}{61}
\pgfkeyssetvalue{P_7_qwen-qwen2.5-32b-instruct_fixed_percentage}{23}
\pgfkeyssetvalue{P_5_gemini-2.0-flash-001_total_files}{260}
\pgfkeyssetvalue{P_5_gemini-2.0-flash-001_fixed}{59}
\pgfkeyssetvalue{P_5_gemini-2.0-flash-001_fixed_percentage}{23}
\pgfkeyssetvalue{P_5_o3-mini-2025-01-31_total_files}{252}
\pgfkeyssetvalue{P_5_o3-mini-2025-01-31_fixed}{72}
\pgfkeyssetvalue{P_5_o3-mini-2025-01-31_fixed_percentage}{29}
\pgfkeyssetvalue{P_5_deepseek-deepseek-chat_total_files}{251}
\pgfkeyssetvalue{P_5_deepseek-deepseek-chat_fixed}{52}
\pgfkeyssetvalue{P_5_deepseek-deepseek-chat_fixed_percentage}{21}
\pgfkeyssetvalue{P_5_gpt-4o-mini_total_files}{254}
\pgfkeyssetvalue{P_5_gpt-4o-mini_fixed}{36}
\pgfkeyssetvalue{P_5_gpt-4o-mini_fixed_percentage}{14}
\pgfkeyssetvalue{P_5_qwen-qwen2.5-32b-instruct_total_files}{268}
\pgfkeyssetvalue{P_5_qwen-qwen2.5-32b-instruct_fixed}{63}
\pgfkeyssetvalue{P_5_qwen-qwen2.5-32b-instruct_fixed_percentage}{24}
\pgfkeyssetvalue{P_1_gemini-2.0-flash-001_total_files}{251}
\pgfkeyssetvalue{P_1_gemini-2.0-flash-001_fixed}{48}
\pgfkeyssetvalue{P_1_gemini-2.0-flash-001_fixed_percentage}{19}
\pgfkeyssetvalue{P_1_o3-mini-2025-01-31_total_files}{242}
\pgfkeyssetvalue{P_1_o3-mini-2025-01-31_fixed}{75}
\pgfkeyssetvalue{P_1_o3-mini-2025-01-31_fixed_percentage}{31}
\pgfkeyssetvalue{P_1_deepseek-deepseek-chat_total_files}{252}
\pgfkeyssetvalue{P_1_deepseek-deepseek-chat_fixed}{48}
\pgfkeyssetvalue{P_1_deepseek-deepseek-chat_fixed_percentage}{19}
\pgfkeyssetvalue{P_1_gpt-4o-mini_total_files}{256}
\pgfkeyssetvalue{P_1_gpt-4o-mini_fixed}{40}
\pgfkeyssetvalue{P_1_gpt-4o-mini_fixed_percentage}{16}
\pgfkeyssetvalue{P_1_qwen-qwen2.5-32b-instruct_total_files}{262}
\pgfkeyssetvalue{P_1_qwen-qwen2.5-32b-instruct_fixed}{41}
\pgfkeyssetvalue{P_1_qwen-qwen2.5-32b-instruct_fixed_percentage}{16}
\pgfkeyssetvalue{P_4_gemini-2.0-flash-001_total_files}{251}
\pgfkeyssetvalue{P_4_gemini-2.0-flash-001_fixed}{87}
\pgfkeyssetvalue{P_4_gemini-2.0-flash-001_fixed_percentage}{35}
\pgfkeyssetvalue{P_4_o3-mini-2025-01-31_total_files}{239}
\pgfkeyssetvalue{P_4_o3-mini-2025-01-31_fixed}{97}
\pgfkeyssetvalue{P_4_o3-mini-2025-01-31_fixed_percentage}{41}
\pgfkeyssetvalue{P_4_deepseek-deepseek-chat_total_files}{255}
\pgfkeyssetvalue{P_4_deepseek-deepseek-chat_fixed}{64}
\pgfkeyssetvalue{P_4_deepseek-deepseek-chat_fixed_percentage}{25}
\pgfkeyssetvalue{P_4_gpt-4o-mini_total_files}{255}
\pgfkeyssetvalue{P_4_gpt-4o-mini_fixed}{53}
\pgfkeyssetvalue{P_4_gpt-4o-mini_fixed_percentage}{21}
\pgfkeyssetvalue{P_4_qwen-qwen2.5-32b-instruct_total_files}{263}
\pgfkeyssetvalue{P_4_qwen-qwen2.5-32b-instruct_fixed}{46}
\pgfkeyssetvalue{P_4_qwen-qwen2.5-32b-instruct_fixed_percentage}{17}
\pgfkeyssetvalue{P_3_gemini-2.0-flash-001_total_files}{246}
\pgfkeyssetvalue{P_3_gemini-2.0-flash-001_fixed}{82}
\pgfkeyssetvalue{P_3_gemini-2.0-flash-001_fixed_percentage}{33}
\pgfkeyssetvalue{P_3_o3-mini-2025-01-31_total_files}{246}
\pgfkeyssetvalue{P_3_o3-mini-2025-01-31_fixed}{89}
\pgfkeyssetvalue{P_3_o3-mini-2025-01-31_fixed_percentage}{36}
\pgfkeyssetvalue{P_3_deepseek-deepseek-chat_total_files}{249}
\pgfkeyssetvalue{P_3_deepseek-deepseek-chat_fixed}{62}
\pgfkeyssetvalue{P_3_deepseek-deepseek-chat_fixed_percentage}{25}
\pgfkeyssetvalue{P_3_gpt-4o-mini_total_files}{262}
\pgfkeyssetvalue{P_3_gpt-4o-mini_fixed}{53}
\pgfkeyssetvalue{P_3_gpt-4o-mini_fixed_percentage}{20}
\pgfkeyssetvalue{P_3_qwen-qwen2.5-32b-instruct_total_files}{265}
\pgfkeyssetvalue{P_3_qwen-qwen2.5-32b-instruct_fixed}{61}
\pgfkeyssetvalue{P_3_qwen-qwen2.5-32b-instruct_fixed_percentage}{23}

%% file: rq2-pgfkeys_error-level.tex
\pgfkeyssetvalue{P_8_gemini-2.0-flash-001_total_errors}{959}
\pgfkeyssetvalue{P_8_gemini-2.0-flash-001_errors_fixed}{714}
\pgfkeyssetvalue{P_8_gemini-2.0-flash-001_fixed_errors_percentage}{74}
\pgfkeyssetvalue{P_8_o3-mini-2025-01-31_total_errors}{955}
\pgfkeyssetvalue{P_8_o3-mini-2025-01-31_errors_fixed}{741}
\pgfkeyssetvalue{P_8_o3-mini-2025-01-31_fixed_errors_percentage}{78}
\pgfkeyssetvalue{P_8_deepseek-deepseek-chat_total_errors}{934}
\pgfkeyssetvalue{P_8_deepseek-deepseek-chat_errors_fixed}{679}
\pgfkeyssetvalue{P_8_deepseek-deepseek-chat_fixed_errors_percentage}{73}
\pgfkeyssetvalue{P_8_gpt-4o-mini_total_errors}{973}
\pgfkeyssetvalue{P_8_gpt-4o-mini_errors_fixed}{680}
\pgfkeyssetvalue{P_8_gpt-4o-mini_fixed_errors_percentage}{70}
\pgfkeyssetvalue{P_8_qwen-qwen2.5-32b-instruct_total_errors}{997}
\pgfkeyssetvalue{P_8_qwen-qwen2.5-32b-instruct_errors_fixed}{602}
\pgfkeyssetvalue{P_8_qwen-qwen2.5-32b-instruct_fixed_errors_percentage}{60}
\pgfkeyssetvalue{P_6_gemini-2.0-flash-001_total_errors}{943}
\pgfkeyssetvalue{P_6_gemini-2.0-flash-001_errors_fixed}{654}
\pgfkeyssetvalue{P_6_gemini-2.0-flash-001_fixed_errors_percentage}{69}
\pgfkeyssetvalue{P_6_o3-mini-2025-01-31_total_errors}{944}
\pgfkeyssetvalue{P_6_o3-mini-2025-01-31_errors_fixed}{712}
\pgfkeyssetvalue{P_6_o3-mini-2025-01-31_fixed_errors_percentage}{75}
\pgfkeyssetvalue{P_6_deepseek-deepseek-chat_total_errors}{933}
\pgfkeyssetvalue{P_6_deepseek-deepseek-chat_errors_fixed}{534}
\pgfkeyssetvalue{P_6_deepseek-deepseek-chat_fixed_errors_percentage}{57}
\pgfkeyssetvalue{P_6_gpt-4o-mini_total_errors}{973}
\pgfkeyssetvalue{P_6_gpt-4o-mini_errors_fixed}{536}
\pgfkeyssetvalue{P_6_gpt-4o-mini_fixed_errors_percentage}{55}
\pgfkeyssetvalue{P_6_qwen-qwen2.5-32b-instruct_total_errors}{988}
\pgfkeyssetvalue{P_6_qwen-qwen2.5-32b-instruct_errors_fixed}{712}
\pgfkeyssetvalue{P_6_qwen-qwen2.5-32b-instruct_fixed_errors_percentage}{72}
\pgfkeyssetvalue{P_2_gemini-2.0-flash-001_total_errors}{955}
\pgfkeyssetvalue{P_2_gemini-2.0-flash-001_errors_fixed}{661}
\pgfkeyssetvalue{P_2_gemini-2.0-flash-001_fixed_errors_percentage}{69}
\pgfkeyssetvalue{P_2_o3-mini-2025-01-31_total_errors}{916}
\pgfkeyssetvalue{P_2_o3-mini-2025-01-31_errors_fixed}{683}
\pgfkeyssetvalue{P_2_o3-mini-2025-01-31_fixed_errors_percentage}{75}
\pgfkeyssetvalue{P_2_deepseek-deepseek-chat_total_errors}{941}
\pgfkeyssetvalue{P_2_deepseek-deepseek-chat_errors_fixed}{645}
\pgfkeyssetvalue{P_2_deepseek-deepseek-chat_fixed_errors_percentage}{69}
\pgfkeyssetvalue{P_2_gpt-4o-mini_total_errors}{966}
\pgfkeyssetvalue{P_2_gpt-4o-mini_errors_fixed}{548}
\pgfkeyssetvalue{P_2_gpt-4o-mini_fixed_errors_percentage}{57}
\pgfkeyssetvalue{P_2_qwen-qwen2.5-32b-instruct_total_errors}{983}
\pgfkeyssetvalue{P_2_qwen-qwen2.5-32b-instruct_errors_fixed}{491}
\pgfkeyssetvalue{P_2_qwen-qwen2.5-32b-instruct_fixed_errors_percentage}{50}
\pgfkeyssetvalue{P_7_gemini-2.0-flash-001_total_errors}{962}
\pgfkeyssetvalue{P_7_gemini-2.0-flash-001_errors_fixed}{711}
\pgfkeyssetvalue{P_7_gemini-2.0-flash-001_fixed_errors_percentage}{74}
\pgfkeyssetvalue{P_7_o3-mini-2025-01-31_total_errors}{937}
\pgfkeyssetvalue{P_7_o3-mini-2025-01-31_errors_fixed}{723}
\pgfkeyssetvalue{P_7_o3-mini-2025-01-31_fixed_errors_percentage}{77}
\pgfkeyssetvalue{P_7_deepseek-deepseek-chat_total_errors}{921}
\pgfkeyssetvalue{P_7_deepseek-deepseek-chat_errors_fixed}{664}
\pgfkeyssetvalue{P_7_deepseek-deepseek-chat_fixed_errors_percentage}{72}
\pgfkeyssetvalue{P_7_gpt-4o-mini_total_errors}{976}
\pgfkeyssetvalue{P_7_gpt-4o-mini_errors_fixed}{696}
\pgfkeyssetvalue{P_7_gpt-4o-mini_fixed_errors_percentage}{71}
\pgfkeyssetvalue{P_7_qwen-qwen2.5-32b-instruct_total_errors}{986}
\pgfkeyssetvalue{P_7_qwen-qwen2.5-32b-instruct_errors_fixed}{669}
\pgfkeyssetvalue{P_7_qwen-qwen2.5-32b-instruct_fixed_errors_percentage}{68}
\pgfkeyssetvalue{P_5_gemini-2.0-flash-001_total_errors}{978}
\pgfkeyssetvalue{P_5_gemini-2.0-flash-001_errors_fixed}{684}
\pgfkeyssetvalue{P_5_gemini-2.0-flash-001_fixed_errors_percentage}{70}
\pgfkeyssetvalue{P_5_o3-mini-2025-01-31_total_errors}{964}
\pgfkeyssetvalue{P_5_o3-mini-2025-01-31_errors_fixed}{731}
\pgfkeyssetvalue{P_5_o3-mini-2025-01-31_fixed_errors_percentage}{76}
\pgfkeyssetvalue{P_5_deepseek-deepseek-chat_total_errors}{937}
\pgfkeyssetvalue{P_5_deepseek-deepseek-chat_errors_fixed}{555}
\pgfkeyssetvalue{P_5_deepseek-deepseek-chat_fixed_errors_percentage}{59}
\pgfkeyssetvalue{P_5_gpt-4o-mini_total_errors}{962}
\pgfkeyssetvalue{P_5_gpt-4o-mini_errors_fixed}{531}
\pgfkeyssetvalue{P_5_gpt-4o-mini_fixed_errors_percentage}{55}
\pgfkeyssetvalue{P_5_qwen-qwen2.5-32b-instruct_total_errors}{994}
\pgfkeyssetvalue{P_5_qwen-qwen2.5-32b-instruct_errors_fixed}{720}
\pgfkeyssetvalue{P_5_qwen-qwen2.5-32b-instruct_fixed_errors_percentage}{72}
\pgfkeyssetvalue{P_1_gemini-2.0-flash-001_total_errors}{959}
\pgfkeyssetvalue{P_1_gemini-2.0-flash-001_errors_fixed}{680}
\pgfkeyssetvalue{P_1_gemini-2.0-flash-001_fixed_errors_percentage}{71}
\pgfkeyssetvalue{P_1_o3-mini-2025-01-31_total_errors}{941}
\pgfkeyssetvalue{P_1_o3-mini-2025-01-31_errors_fixed}{705}
\pgfkeyssetvalue{P_1_o3-mini-2025-01-31_fixed_errors_percentage}{75}
\pgfkeyssetvalue{P_1_deepseek-deepseek-chat_total_errors}{938}
\pgfkeyssetvalue{P_1_deepseek-deepseek-chat_errors_fixed}{548}
\pgfkeyssetvalue{P_1_deepseek-deepseek-chat_fixed_errors_percentage}{58}
\pgfkeyssetvalue{P_1_gpt-4o-mini_total_errors}{965}
\pgfkeyssetvalue{P_1_gpt-4o-mini_errors_fixed}{490}
\pgfkeyssetvalue{P_1_gpt-4o-mini_fixed_errors_percentage}{51}
\pgfkeyssetvalue{P_1_qwen-qwen2.5-32b-instruct_total_errors}{979}
\pgfkeyssetvalue{P_1_qwen-qwen2.5-32b-instruct_errors_fixed}{529}
\pgfkeyssetvalue{P_1_qwen-qwen2.5-32b-instruct_fixed_errors_percentage}{54}
\pgfkeyssetvalue{P_4_gemini-2.0-flash-001_total_errors}{936}
\pgfkeyssetvalue{P_4_gemini-2.0-flash-001_errors_fixed}{687}
\pgfkeyssetvalue{P_4_gemini-2.0-flash-001_fixed_errors_percentage}{73}
\pgfkeyssetvalue{P_4_o3-mini-2025-01-31_total_errors}{938}
\pgfkeyssetvalue{P_4_o3-mini-2025-01-31_errors_fixed}{726}
\pgfkeyssetvalue{P_4_o3-mini-2025-01-31_fixed_errors_percentage}{77}
\pgfkeyssetvalue{P_4_deepseek-deepseek-chat_total_errors}{942}
\pgfkeyssetvalue{P_4_deepseek-deepseek-chat_errors_fixed}{614}
\pgfkeyssetvalue{P_4_deepseek-deepseek-chat_fixed_errors_percentage}{65}
\pgfkeyssetvalue{P_4_gpt-4o-mini_total_errors}{964}
\pgfkeyssetvalue{P_4_gpt-4o-mini_errors_fixed}{670}
\pgfkeyssetvalue{P_4_gpt-4o-mini_fixed_errors_percentage}{70}
\pgfkeyssetvalue{P_4_qwen-qwen2.5-32b-instruct_total_errors}{983}
\pgfkeyssetvalue{P_4_qwen-qwen2.5-32b-instruct_errors_fixed}{622}
\pgfkeyssetvalue{P_4_qwen-qwen2.5-32b-instruct_fixed_errors_percentage}{63}
\pgfkeyssetvalue{P_3_gemini-2.0-flash-001_total_errors}{931}
\pgfkeyssetvalue{P_3_gemini-2.0-flash-001_errors_fixed}{679}
\pgfkeyssetvalue{P_3_gemini-2.0-flash-001_fixed_errors_percentage}{73}
\pgfkeyssetvalue{P_3_o3-mini-2025-01-31_total_errors}{955}
\pgfkeyssetvalue{P_3_o3-mini-2025-01-31_errors_fixed}{736}
\pgfkeyssetvalue{P_3_o3-mini-2025-01-31_fixed_errors_percentage}{77}
\pgfkeyssetvalue{P_3_deepseek-deepseek-chat_total_errors}{935}
\pgfkeyssetvalue{P_3_deepseek-deepseek-chat_errors_fixed}{553}
\pgfkeyssetvalue{P_3_deepseek-deepseek-chat_fixed_errors_percentage}{59}
\pgfkeyssetvalue{P_3_gpt-4o-mini_total_errors}{979}
\pgfkeyssetvalue{P_3_gpt-4o-mini_errors_fixed}{696}
\pgfkeyssetvalue{P_3_gpt-4o-mini_fixed_errors_percentage}{71}
\pgfkeyssetvalue{P_3_qwen-qwen2.5-32b-instruct_total_errors}{987}
\pgfkeyssetvalue{P_3_qwen-qwen2.5-32b-instruct_errors_fixed}{710}
\pgfkeyssetvalue{P_3_qwen-qwen2.5-32b-instruct_fixed_errors_percentage}{72}

%% file: error_level.tex
\newcommand{\errorleveltab}{%
\begin{table}[t!]
    \centering
    \caption{Partial Repair (RQ2): Compilation Error Fix Rate of \toolname over failed cases}
    \label{tab:error_level}
    \rowcolors{2}{gray!10}{white}
    \resizebox{\textwidth}{!}{%
    \begin{tabular}{lrrrrr}
    \toprule
    \makecell[l]{\textbf{Prompt}\\ID} & \makecell[l]{\textbf{Deepseek}\\V3} & \makecell[l]{\textbf{Gemini}\\2.0-flash} & \makecell[l]{\textbf{Gpt}\\4o-mini} & \makecell[l]{\textbf{o3}\\mini} & \makecell[l]{\textbf{Qwen2.5}\\32b-instruct} \\
    \midrule

   $P_1$ & \pgfkeysvalueof{P_1_deepseek-deepseek-chat_errors_fixed}/\pgfkeysvalueof{P_1_deepseek-deepseek-chat_total_errors}(\pgfkeysvalueof{P_1_deepseek-deepseek-chat_fixed_errors_percentage}\%) & \pgfkeysvalueof{P_1_gemini-2.0-flash-001_errors_fixed}/\pgfkeysvalueof{P_1_gemini-2.0-flash-001_total_errors}(\pgfkeysvalueof{P_1_gemini-2.0-flash-001_fixed_errors_percentage}\%) & \pgfkeysvalueof{P_1_gpt-4o-mini_errors_fixed}/\pgfkeysvalueof{P_1_gpt-4o-mini_total_errors}(\pgfkeysvalueof{P_1_gpt-4o-mini_fixed_errors_percentage}\%) & \pgfkeysvalueof{P_1_o3-mini-2025-01-31_errors_fixed}/\pgfkeysvalueof{P_1_o3-mini-2025-01-31_total_errors}(\pgfkeysvalueof{P_1_o3-mini-2025-01-31_fixed_errors_percentage}\%) & \pgfkeysvalueof{P_1_qwen-qwen2.5-32b-instruct_errors_fixed}/\pgfkeysvalueof{P_1_qwen-qwen2.5-32b-instruct_total_errors}(\pgfkeysvalueof{P_1_qwen-qwen2.5-32b-instruct_fixed_errors_percentage}\%) \\
$P_2$ & \pgfkeysvalueof{P_2_deepseek-deepseek-chat_errors_fixed}/\pgfkeysvalueof{P_2_deepseek-deepseek-chat_total_errors}(\pgfkeysvalueof{P_2_deepseek-deepseek-chat_fixed_errors_percentage}\%) & \pgfkeysvalueof{P_2_gemini-2.0-flash-001_errors_fixed}/\pgfkeysvalueof{P_2_gemini-2.0-flash-001_total_errors}(\pgfkeysvalueof{P_2_gemini-2.0-flash-001_fixed_errors_percentage}\%) & \pgfkeysvalueof{P_2_gpt-4o-mini_errors_fixed}/\pgfkeysvalueof{P_2_gpt-4o-mini_total_errors}(\pgfkeysvalueof{P_2_gpt-4o-mini_fixed_errors_percentage}\%) & \pgfkeysvalueof{P_2_o3-mini-2025-01-31_errors_fixed}/\pgfkeysvalueof{P_2_o3-mini-2025-01-31_total_errors}(\pgfkeysvalueof{P_2_o3-mini-2025-01-31_fixed_errors_percentage}\%) & \pgfkeysvalueof{P_2_qwen-qwen2.5-32b-instruct_errors_fixed}/\pgfkeysvalueof{P_2_qwen-qwen2.5-32b-instruct_total_errors}(\pgfkeysvalueof{P_2_qwen-qwen2.5-32b-instruct_fixed_errors_percentage}\%) \\
$P_3$ & \pgfkeysvalueof{P_3_deepseek-deepseek-chat_errors_fixed}/\pgfkeysvalueof{P_3_deepseek-deepseek-chat_total_errors}(\pgfkeysvalueof{P_3_deepseek-deepseek-chat_fixed_errors_percentage}\%) & \pgfkeysvalueof{P_3_gemini-2.0-flash-001_errors_fixed}/\pgfkeysvalueof{P_3_gemini-2.0-flash-001_total_errors}(\pgfkeysvalueof{P_3_gemini-2.0-flash-001_fixed_errors_percentage}\%) & \pgfkeysvalueof{P_3_gpt-4o-mini_errors_fixed}/\pgfkeysvalueof{P_3_gpt-4o-mini_total_errors}(\pgfkeysvalueof{P_3_gpt-4o-mini_fixed_errors_percentage}\%) & \pgfkeysvalueof{P_3_o3-mini-2025-01-31_errors_fixed}/\pgfkeysvalueof{P_3_o3-mini-2025-01-31_total_errors}(\pgfkeysvalueof{P_3_o3-mini-2025-01-31_fixed_errors_percentage}\%) & \pgfkeysvalueof{P_3_qwen-qwen2.5-32b-instruct_errors_fixed}/\pgfkeysvalueof{P_3_qwen-qwen2.5-32b-instruct_total_errors}(\pgfkeysvalueof{P_3_qwen-qwen2.5-32b-instruct_fixed_errors_percentage}\%) \\
$P_4$ & \pgfkeysvalueof{P_4_deepseek-deepseek-chat_errors_fixed}/\pgfkeysvalueof{P_4_deepseek-deepseek-chat_total_errors}(\pgfkeysvalueof{P_4_deepseek-deepseek-chat_fixed_errors_percentage}\%) & \pgfkeysvalueof{P_4_gemini-2.0-flash-001_errors_fixed}/\pgfkeysvalueof{P_4_gemini-2.0-flash-001_total_errors}(\pgfkeysvalueof{P_4_gemini-2.0-flash-001_fixed_errors_percentage}\%) & \pgfkeysvalueof{P_4_gpt-4o-mini_errors_fixed}/\pgfkeysvalueof{P_4_gpt-4o-mini_total_errors}(\pgfkeysvalueof{P_4_gpt-4o-mini_fixed_errors_percentage}\%) & \pgfkeysvalueof{P_4_o3-mini-2025-01-31_errors_fixed}/\pgfkeysvalueof{P_4_o3-mini-2025-01-31_total_errors}(\pgfkeysvalueof{P_4_o3-mini-2025-01-31_fixed_errors_percentage}\%) & \pgfkeysvalueof{P_4_qwen-qwen2.5-32b-instruct_errors_fixed}/\pgfkeysvalueof{P_4_qwen-qwen2.5-32b-instruct_total_errors}(\pgfkeysvalueof{P_4_qwen-qwen2.5-32b-instruct_fixed_errors_percentage}\%) \\
$P_5$ & \pgfkeysvalueof{P_5_deepseek-deepseek-chat_errors_fixed}/\pgfkeysvalueof{P_5_deepseek-deepseek-chat_total_errors}(\pgfkeysvalueof{P_5_deepseek-deepseek-chat_fixed_errors_percentage}\%) & \pgfkeysvalueof{P_5_gemini-2.0-flash-001_errors_fixed}/\pgfkeysvalueof{P_5_gemini-2.0-flash-001_total_errors}(\pgfkeysvalueof{P_5_gemini-2.0-flash-001_fixed_errors_percentage}\%) & \pgfkeysvalueof{P_5_gpt-4o-mini_errors_fixed}/\pgfkeysvalueof{P_5_gpt-4o-mini_total_errors}(\pgfkeysvalueof{P_5_gpt-4o-mini_fixed_errors_percentage}\%) & \pgfkeysvalueof{P_5_o3-mini-2025-01-31_errors_fixed}/\pgfkeysvalueof{P_5_o3-mini-2025-01-31_total_errors}(\pgfkeysvalueof{P_5_o3-mini-2025-01-31_fixed_errors_percentage}\%) & \pgfkeysvalueof{P_5_qwen-qwen2.5-32b-instruct_errors_fixed}/\pgfkeysvalueof{P_5_qwen-qwen2.5-32b-instruct_total_errors}(\pgfkeysvalueof{P_5_qwen-qwen2.5-32b-instruct_fixed_errors_percentage}\%) \\
$P_6$ & \pgfkeysvalueof{P_6_deepseek-deepseek-chat_errors_fixed}/\pgfkeysvalueof{P_6_deepseek-deepseek-chat_total_errors}(\pgfkeysvalueof{P_6_deepseek-deepseek-chat_fixed_errors_percentage}\%) & \pgfkeysvalueof{P_6_gemini-2.0-flash-001_errors_fixed}/\pgfkeysvalueof{P_6_gemini-2.0-flash-001_total_errors}(\pgfkeysvalueof{P_6_gemini-2.0-flash-001_fixed_errors_percentage}\%) & \pgfkeysvalueof{P_6_gpt-4o-mini_errors_fixed}/\pgfkeysvalueof{P_6_gpt-4o-mini_total_errors}(\pgfkeysvalueof{P_6_gpt-4o-mini_fixed_errors_percentage}\%) & \pgfkeysvalueof{P_6_o3-mini-2025-01-31_errors_fixed}/\pgfkeysvalueof{P_6_o3-mini-2025-01-31_total_errors}(\pgfkeysvalueof{P_6_o3-mini-2025-01-31_fixed_errors_percentage}\%) & \pgfkeysvalueof{P_6_qwen-qwen2.5-32b-instruct_errors_fixed}/\pgfkeysvalueof{P_6_qwen-qwen2.5-32b-instruct_total_errors}(\pgfkeysvalueof{P_6_qwen-qwen2.5-32b-instruct_fixed_errors_percentage}\%) \\
$P_7$ & \pgfkeysvalueof{P_7_deepseek-deepseek-chat_errors_fixed}/\pgfkeysvalueof{P_7_deepseek-deepseek-chat_total_errors}(\pgfkeysvalueof{P_7_deepseek-deepseek-chat_fixed_errors_percentage}\%) & \pgfkeysvalueof{P_7_gemini-2.0-flash-001_errors_fixed}/\pgfkeysvalueof{P_7_gemini-2.0-flash-001_total_errors}(\pgfkeysvalueof{P_7_gemini-2.0-flash-001_fixed_errors_percentage}\%) & \pgfkeysvalueof{P_7_gpt-4o-mini_errors_fixed}/\pgfkeysvalueof{P_7_gpt-4o-mini_total_errors}(\pgfkeysvalueof{P_7_gpt-4o-mini_fixed_errors_percentage}\%) & \pgfkeysvalueof{P_7_o3-mini-2025-01-31_errors_fixed}/\pgfkeysvalueof{P_7_o3-mini-2025-01-31_total_errors}(\pgfkeysvalueof{P_7_o3-mini-2025-01-31_fixed_errors_percentage}\%) & \pgfkeysvalueof{P_7_qwen-qwen2.5-32b-instruct_errors_fixed}/\pgfkeysvalueof{P_7_qwen-qwen2.5-32b-instruct_total_errors}(\pgfkeysvalueof{P_7_qwen-qwen2.5-32b-instruct_fixed_errors_percentage}\%) \\
$P_8$ & \pgfkeysvalueof{P_8_deepseek-deepseek-chat_errors_fixed}/\pgfkeysvalueof{P_8_deepseek-deepseek-chat_total_errors}(\pgfkeysvalueof{P_8_deepseek-deepseek-chat_fixed_errors_percentage}\%) & \pgfkeysvalueof{P_8_gemini-2.0-flash-001_errors_fixed}/\pgfkeysvalueof{P_8_gemini-2.0-flash-001_total_errors}(\pgfkeysvalueof{P_8_gemini-2.0-flash-001_fixed_errors_percentage}\%) & \pgfkeysvalueof{P_8_gpt-4o-mini_errors_fixed}/\pgfkeysvalueof{P_8_gpt-4o-mini_total_errors}(\pgfkeysvalueof{P_8_gpt-4o-mini_fixed_errors_percentage}\%) & \textbf{\pgfkeysvalueof{P_8_o3-mini-2025-01-31_errors_fixed}/\pgfkeysvalueof{P_8_o3-mini-2025-01-31_total_errors}(\pgfkeysvalueof{P_8_o3-mini-2025-01-31_fixed_errors_percentage}\%)} & \pgfkeysvalueof{P_8_qwen-qwen2.5-32b-instruct_errors_fixed}/\pgfkeysvalueof{P_8_qwen-qwen2.5-32b-instruct_total_errors}(\pgfkeysvalueof{P_8_qwen-qwen2.5-32b-instruct_fixed_errors_percentage}\%) \\

       \bottomrule
    \end{tabular}}
\end{table}
}

%% file: rq3_pgfkeys_output.tex
\pgfkeyssetvalue{P8_gemini_total_prefix}{1004}
\pgfkeyssetvalue{P8_gemini_total_fixed}{759}
\pgfkeyssetvalue{P8_gemini_total_new}{393}
\pgfkeyssetvalue{P8_gemini_relative_fixed_percentage}{36}
\pgfkeyssetvalue{P8_gemini_relative_nominator}{366}
\pgfkeyssetvalue{P8_o3-mini_total_prefix}{1004}
\pgfkeyssetvalue{P8_o3-mini_total_fixed}{790}
\pgfkeyssetvalue{P8_o3-mini_total_new}{445}
\pgfkeyssetvalue{P8_o3-mini_relative_fixed_percentage}{34}
\pgfkeyssetvalue{P8_o3-mini_relative_nominator}{345}
\pgfkeyssetvalue{P8_deepseek_total_prefix}{1004}
\pgfkeyssetvalue{P8_deepseek_total_fixed}{749}
\pgfkeyssetvalue{P8_deepseek_total_new}{202}
\pgfkeyssetvalue{P8_deepseek_relative_fixed_percentage}{54}
\pgfkeyssetvalue{P8_deepseek_relative_nominator}{547}
\pgfkeyssetvalue{P8_gpt_total_prefix}{1004}
\pgfkeyssetvalue{P8_gpt_total_fixed}{711}
\pgfkeyssetvalue{P8_gpt_total_new}{364}
\pgfkeyssetvalue{P8_gpt_relative_fixed_percentage}{35}
\pgfkeyssetvalue{P8_gpt_relative_nominator}{347}
\pgfkeyssetvalue{P8_qwen_total_prefix}{1004}
\pgfkeyssetvalue{P8_qwen_total_fixed}{609}
\pgfkeyssetvalue{P8_qwen_total_new}{522}
\pgfkeyssetvalue{P8_qwen_relative_fixed_percentage}{9}
\pgfkeyssetvalue{P8_qwen_relative_nominator}{87}
\pgfkeyssetvalue{P6_gemini_total_prefix}{1004}
\pgfkeyssetvalue{P6_gemini_total_fixed}{715}
\pgfkeyssetvalue{P6_gemini_total_new}{441}
\pgfkeyssetvalue{P6_gemini_relative_fixed_percentage}{27}
\pgfkeyssetvalue{P6_gemini_relative_nominator}{274}
\pgfkeyssetvalue{P6_o3-mini_total_prefix}{1004}
\pgfkeyssetvalue{P6_o3-mini_total_fixed}{772}
\pgfkeyssetvalue{P6_o3-mini_total_new}{399}
\pgfkeyssetvalue{P6_o3-mini_relative_fixed_percentage}{37}
\pgfkeyssetvalue{P6_o3-mini_relative_nominator}{373}
\pgfkeyssetvalue{P6_deepseek_total_prefix}{1004}
\pgfkeyssetvalue{P6_deepseek_total_fixed}{605}
\pgfkeyssetvalue{P6_deepseek_total_new}{307}
\pgfkeyssetvalue{P6_deepseek_relative_fixed_percentage}{30}
\pgfkeyssetvalue{P6_deepseek_relative_nominator}{298}
\pgfkeyssetvalue{P6_gpt_total_prefix}{1004}
\pgfkeyssetvalue{P6_gpt_total_fixed}{567}
\pgfkeyssetvalue{P6_gpt_total_new}{285}
\pgfkeyssetvalue{P6_gpt_relative_fixed_percentage}{28}
\pgfkeyssetvalue{P6_gpt_relative_nominator}{282}
\pgfkeyssetvalue{P6_qwen_total_prefix}{1004}
\pgfkeyssetvalue{P6_qwen_total_fixed}{728}
\pgfkeyssetvalue{P6_qwen_total_new}{1125}
\pgfkeyssetvalue{P6_qwen_relative_fixed_percentage}{-40}
\pgfkeyssetvalue{P6_qwen_relative_nominator}{-397}
\pgfkeyssetvalue{P2_gemini_total_prefix}{1004}
\pgfkeyssetvalue{P2_gemini_total_fixed}{710}
\pgfkeyssetvalue{P2_gemini_total_new}{333}
\pgfkeyssetvalue{P2_gemini_relative_fixed_percentage}{38}
\pgfkeyssetvalue{P2_gemini_relative_nominator}{377}
\pgfkeyssetvalue{P2_o3-mini_total_prefix}{1004}
\pgfkeyssetvalue{P2_o3-mini_total_fixed}{771}
\pgfkeyssetvalue{P2_o3-mini_total_new}{347}
\pgfkeyssetvalue{P2_o3-mini_relative_fixed_percentage}{42}
\pgfkeyssetvalue{P2_o3-mini_relative_nominator}{424}
\pgfkeyssetvalue{P2_deepseek_total_prefix}{1004}
\pgfkeyssetvalue{P2_deepseek_total_fixed}{708}
\pgfkeyssetvalue{P2_deepseek_total_new}{290}
\pgfkeyssetvalue{P2_deepseek_relative_fixed_percentage}{42}
\pgfkeyssetvalue{P2_deepseek_relative_nominator}{418}
\pgfkeyssetvalue{P2_gpt_total_prefix}{1004}
\pgfkeyssetvalue{P2_gpt_total_fixed}{586}
\pgfkeyssetvalue{P2_gpt_total_new}{302}
\pgfkeyssetvalue{P2_gpt_relative_fixed_percentage}{28}
\pgfkeyssetvalue{P2_gpt_relative_nominator}{284}
\pgfkeyssetvalue{P2_qwen_total_prefix}{1004}
\pgfkeyssetvalue{P2_qwen_total_fixed}{512}
\pgfkeyssetvalue{P2_qwen_total_new}{759}
\pgfkeyssetvalue{P2_qwen_relative_fixed_percentage}{-25}
\pgfkeyssetvalue{P2_qwen_relative_nominator}{-247}
\pgfkeyssetvalue{P7_gemini_total_prefix}{1004}
\pgfkeyssetvalue{P7_gemini_total_fixed}{753}
\pgfkeyssetvalue{P7_gemini_total_new}{367}
\pgfkeyssetvalue{P7_gemini_relative_fixed_percentage}{38}
\pgfkeyssetvalue{P7_gemini_relative_nominator}{386}
\pgfkeyssetvalue{P7_o3-mini_total_prefix}{1004}
\pgfkeyssetvalue{P7_o3-mini_total_fixed}{790}
\pgfkeyssetvalue{P7_o3-mini_total_new}{525}
\pgfkeyssetvalue{P7_o3-mini_relative_fixed_percentage}{26}
\pgfkeyssetvalue{P7_o3-mini_relative_nominator}{265}
\pgfkeyssetvalue{P7_deepseek_total_prefix}{1004}
\pgfkeyssetvalue{P7_deepseek_total_fixed}{747}
\pgfkeyssetvalue{P7_deepseek_total_new}{296}
\pgfkeyssetvalue{P7_deepseek_relative_fixed_percentage}{45}
\pgfkeyssetvalue{P7_deepseek_relative_nominator}{451}
\pgfkeyssetvalue{P7_gpt_total_prefix}{1004}
\pgfkeyssetvalue{P7_gpt_total_fixed}{724}
\pgfkeyssetvalue{P7_gpt_total_new}{346}
\pgfkeyssetvalue{P7_gpt_relative_fixed_percentage}{38}
\pgfkeyssetvalue{P7_gpt_relative_nominator}{378}
\pgfkeyssetvalue{P7_qwen_total_prefix}{1004}
\pgfkeyssetvalue{P7_qwen_total_fixed}{687}
\pgfkeyssetvalue{P7_qwen_total_new}{606}
\pgfkeyssetvalue{P7_qwen_relative_fixed_percentage}{8}
\pgfkeyssetvalue{P7_qwen_relative_nominator}{81}
\pgfkeyssetvalue{P5_gemini_total_prefix}{1004}
\pgfkeyssetvalue{P5_gemini_total_fixed}{710}
\pgfkeyssetvalue{P5_gemini_total_new}{406}
\pgfkeyssetvalue{P5_gemini_relative_fixed_percentage}{30}
\pgfkeyssetvalue{P5_gemini_relative_nominator}{304}
\pgfkeyssetvalue{P5_o3-mini_total_prefix}{1004}
\pgfkeyssetvalue{P5_o3-mini_total_fixed}{771}
\pgfkeyssetvalue{P5_o3-mini_total_new}{267}
\pgfkeyssetvalue{P5_o3-mini_relative_fixed_percentage}{50}
\pgfkeyssetvalue{P5_o3-mini_relative_nominator}{504}
\pgfkeyssetvalue{P5_deepseek_total_prefix}{1004}
\pgfkeyssetvalue{P5_deepseek_total_fixed}{622}
\pgfkeyssetvalue{P5_deepseek_total_new}{278}
\pgfkeyssetvalue{P5_deepseek_relative_fixed_percentage}{34}
\pgfkeyssetvalue{P5_deepseek_relative_nominator}{344}
\pgfkeyssetvalue{P5_gpt_total_prefix}{1004}
\pgfkeyssetvalue{P5_gpt_total_fixed}{573}
\pgfkeyssetvalue{P5_gpt_total_new}{283}
\pgfkeyssetvalue{P5_gpt_relative_fixed_percentage}{29}
\pgfkeyssetvalue{P5_gpt_relative_nominator}{290}
\pgfkeyssetvalue{P5_qwen_total_prefix}{1004}
\pgfkeyssetvalue{P5_qwen_total_fixed}{730}
\pgfkeyssetvalue{P5_qwen_total_new}{919}
\pgfkeyssetvalue{P5_qwen_relative_fixed_percentage}{-19}
\pgfkeyssetvalue{P5_qwen_relative_nominator}{-189}
\pgfkeyssetvalue{P1_gemini_total_prefix}{1004}
\pgfkeyssetvalue{P1_gemini_total_fixed}{725}
\pgfkeyssetvalue{P1_gemini_total_new}{505}
\pgfkeyssetvalue{P1_gemini_relative_fixed_percentage}{22}
\pgfkeyssetvalue{P1_gemini_relative_nominator}{220}
\pgfkeyssetvalue{P1_o3-mini_total_prefix}{1004}
\pgfkeyssetvalue{P1_o3-mini_total_fixed}{768}
\pgfkeyssetvalue{P1_o3-mini_total_new}{399}
\pgfkeyssetvalue{P1_o3-mini_relative_fixed_percentage}{37}
\pgfkeyssetvalue{P1_o3-mini_relative_nominator}{369}
\pgfkeyssetvalue{P1_deepseek_total_prefix}{1004}
\pgfkeyssetvalue{P1_deepseek_total_fixed}{614}
\pgfkeyssetvalue{P1_deepseek_total_new}{264}
\pgfkeyssetvalue{P1_deepseek_relative_fixed_percentage}{35}
\pgfkeyssetvalue{P1_deepseek_relative_nominator}{350}
\pgfkeyssetvalue{P1_gpt_total_prefix}{1004}
\pgfkeyssetvalue{P1_gpt_total_fixed}{529}
\pgfkeyssetvalue{P1_gpt_total_new}{289}
\pgfkeyssetvalue{P1_gpt_relative_fixed_percentage}{24}
\pgfkeyssetvalue{P1_gpt_relative_nominator}{240}
\pgfkeyssetvalue{P1_qwen_total_prefix}{1004}
\pgfkeyssetvalue{P1_qwen_total_fixed}{554}
\pgfkeyssetvalue{P1_qwen_total_new}{887}
\pgfkeyssetvalue{P1_qwen_relative_fixed_percentage}{-33}
\pgfkeyssetvalue{P1_qwen_relative_nominator}{-333}
\pgfkeyssetvalue{P4_gemini_total_prefix}{1004}
\pgfkeyssetvalue{P4_gemini_total_fixed}{755}
\pgfkeyssetvalue{P4_gemini_total_new}{293}
\pgfkeyssetvalue{P4_gemini_relative_fixed_percentage}{46}
\pgfkeyssetvalue{P4_gemini_relative_nominator}{462}
\pgfkeyssetvalue{P4_o3-mini_total_prefix}{1004}
\pgfkeyssetvalue{P4_o3-mini_total_fixed}{792}
\pgfkeyssetvalue{P4_o3-mini_total_new}{395}
\pgfkeyssetvalue{P4_o3-mini_relative_fixed_percentage}{40}
\pgfkeyssetvalue{P4_o3-mini_relative_nominator}{397}
\pgfkeyssetvalue{P4_deepseek_total_prefix}{1004}
\pgfkeyssetvalue{P4_deepseek_total_fixed}{676}
\pgfkeyssetvalue{P4_deepseek_total_new}{183}
\pgfkeyssetvalue{P4_deepseek_relative_fixed_percentage}{49}
\pgfkeyssetvalue{P4_deepseek_relative_nominator}{493}
\pgfkeyssetvalue{P4_gpt_total_prefix}{1004}
\pgfkeyssetvalue{P4_gpt_total_fixed}{710}
\pgfkeyssetvalue{P4_gpt_total_new}{347}
\pgfkeyssetvalue{P4_gpt_relative_fixed_percentage}{36}
\pgfkeyssetvalue{P4_gpt_relative_nominator}{363}
\pgfkeyssetvalue{P4_qwen_total_prefix}{1004}
\pgfkeyssetvalue{P4_qwen_total_fixed}{643}
\pgfkeyssetvalue{P4_qwen_total_new}{553}
\pgfkeyssetvalue{P4_qwen_relative_fixed_percentage}{9}
\pgfkeyssetvalue{P4_qwen_relative_nominator}{90}
\pgfkeyssetvalue{P3_gemini_total_prefix}{1004}
\pgfkeyssetvalue{P3_gemini_total_fixed}{752}
\pgfkeyssetvalue{P3_gemini_total_new}{278}
\pgfkeyssetvalue{P3_gemini_relative_fixed_percentage}{47}
\pgfkeyssetvalue{P3_gemini_relative_nominator}{474}
\pgfkeyssetvalue{P3_o3-mini_total_prefix}{1004}
\pgfkeyssetvalue{P3_o3-mini_total_fixed}{785}
\pgfkeyssetvalue{P3_o3-mini_total_new}{506}
\pgfkeyssetvalue{P3_o3-mini_relative_fixed_percentage}{28}
\pgfkeyssetvalue{P3_o3-mini_relative_nominator}{279}
\pgfkeyssetvalue{P3_deepseek_total_prefix}{1004}
\pgfkeyssetvalue{P3_deepseek_total_fixed}{622}
\pgfkeyssetvalue{P3_deepseek_total_new}{172}
\pgfkeyssetvalue{P3_deepseek_relative_fixed_percentage}{45}
\pgfkeyssetvalue{P3_deepseek_relative_nominator}{450}
\pgfkeyssetvalue{P3_gpt_total_prefix}{1004}
\pgfkeyssetvalue{P3_gpt_total_fixed}{721}
\pgfkeyssetvalue{P3_gpt_total_new}{326}
\pgfkeyssetvalue{P3_gpt_relative_fixed_percentage}{39}
\pgfkeyssetvalue{P3_gpt_relative_nominator}{395}
\pgfkeyssetvalue{P3_qwen_total_prefix}{1004}
\pgfkeyssetvalue{P3_qwen_total_fixed}{727}
\pgfkeyssetvalue{P3_qwen_total_new}{651}
\pgfkeyssetvalue{P3_qwen_relative_fixed_percentage}{8}
\pgfkeyssetvalue{P3_qwen_relative_nominator}{76}